%% file: paper.tex
\documentclass[nofootinbib,showkeys,twocolumn,a4paper,10pt,aps,prd,preprintnumbers,superscriptaddress,floatfix]{revtex4-2}
\input{preamble}

\def\pmbanner{}
\begin{document}
\title{\pmbanner{\LARGE \texorpdfstring{Letter of Intent: AN Underground Belayed In-Shaft (ANUBIS) detector\\
\includegraphics[width=0.1\linewidth]{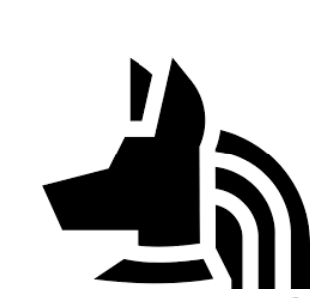}\\
{\large\textbf{ANUBIS Collaboration}}
\\[-0.2em]
{\normalsize\textit{E-mail: }\href{mailto:anubis-publications@cern.ch}{anubis-publications@cern.ch}}
\\[-0.2em]
}{Letter of Intent: AN Underground Belayed In-Shaft (ANUBIS) detector}}}

\begin{abstract}
\textit{Abstract:}
Long-lived particles (LLPs) arise naturally in many extensions of the Standard Model, notably in models with dark matter candidates or motivated by baryogenesis.
LLPs produced at or above the electroweak scale with decay lengths of tens of metres escape the general-purpose LHC detectors and cannot be produced at sufficient rates in forward and beam-dump experiments.
This Letter of Intent proposes the ANUBIS detector, which would instrument the ceiling of the ATLAS cavern and the bottom of its access shafts with about 1600~m$^2$ of resistive plate chambers in two tracking layers about 1~m apart, using the air-filled volume above the ATLAS detector as decay volume.
The ANUBIS detector exploits the full ATLAS event information to suppress backgrounds, and exchanges triggers with the ATLAS experiment.
With 3~ab$^{-1}$ at the High-Luminosity LHC, it would improve the sensitivity to exotic Higgs-boson decays into long-lived scalars by two to three orders of magnitude, probe heavy neutral leptons of all three flavours down to mixings of about $2\times10^{-8}$, and extend the reach for axion-like particles.
The \textsl{pro}ANUBIS demonstrator, operated in the ATLAS cavern in 2024 and 2025, has validated the detector technology.
The detector design, trigger concept, integration with the ATLAS experiment and a timeline towards installation in LHC Long Shutdown 4 are presented.
%
\end{abstract}

\input{authorList}
\def\andname{\hspace*{-0.5em}}
\maketitle

\hbox{}
\clearpage

\section*{Executive summary}
\input{sections/executive_summary}

\clearpage
\tableofcontents
\clearpage

\section{Introduction}
\label{sec:intro}
\input{sections/introduction}

\section{Detector location}
\label{sec:location}
\input{sections/location}

\section{Discovery potential}
\label{sec:physics}
\input{sections/physics}

\section{Backgrounds}
\label{sec:backgrounds}
\input{sections/backgrounds}

\section{Detector performance and layout}
\label{sec:performance}
\input{sections/performance}

\section{\NoCaseChange{\proanubis} demonstrator: results}
\label{sec:proanubis}
\input{sections/proanubis}

\section{Engineering challenges}
\label{sec:engineering}
\input{sections/structure}

\section{Detector design}
\label{sec:design}
\input{sections/detector}

\section{Trigger design}
\label{sec:trigger}
\input{sections/trigger}


\section{Interface between \anubis and \atlas}
\label{sec:interfacing}
\input{sections/interface}

\section{Timeline}
\label{sec:timeline}
\input{sections/timeline}

\section{Preliminary costing}
\label{sec:costing}
\input{sections/costing}

\section{Conclusion and outlook} 
\label{sec:Summary}
\input{sections/conclusion}

\begin{acknowledgments}
The authors would like to express their gratitude to: the ATLAS Collaboration for hosting \proanubis, facilitating the data taking and data analysis, and provision of consumables; the ATLAS RPC community for their guidance with constructing the RPCs for \proanubis and the chamber cages, and maintenance/troubleshooting; and the ATLAS Technical Coordination team for help with installing, connection of services, maintenance, and safety; to the colleagues in the CODEX-b collaboration for lending front-end board electronics to allow for a timely installation of the \proanubis demonstrator; to the PBC BSM group and especially group conveners Felix Kahlh\"ofer and Torben Ferber for their guidance and efficient moderation of the scientific discussions, and finally, to colleagues in the wider LLP community for the scrutiny of our results and thoughtful suggestions. Particular thanks go to: Ben Allanach, Cristiano Alpigiani, David Curtin, Albert De Roeck, Miriam Diamond, Erez Etzion, Simon Knapen, Gaurav Kumar, Steven Lowette, Maxim Ovchinnikov, Runze Ren, Heather Russell, Richard Shaw, and Gustavo Uribe.

This Letter of Intent is dedicated to the memory of Prof.~Henry Lubatti, who pioneered the field of dedicated LLP detectors at the LHC.
\end{acknowledgments}

\clearpage
\bibliographystyle{JHEP}
\bibliography{bibs/anubis}

\end{document}

%% file: preamble.tex
\pdfoutput=1
\usepackage[T1]{fontenc}   
\usepackage{amsmath,amssymb,graphicx,bm,color,slashed,array,multirow}
\usepackage[dvipsnames]{xcolor}
\usepackage[font=footnotesize,labelfont=bf,skip=5pt,justification=RaggedRight]{caption}
\usepackage{subcaption}
\usepackage[framemethod=default]{mdframed}
\usepackage{tikz}
\usepackage{hyperref}
\hypersetup{
    colorlinks=true,
    linkcolor=blue,
    citecolor=blue,
    filecolor=magenta,
    urlcolor=cyan,
    pdftitle={Letter of Intent: AN Underground Belayed In-Shaft (ANUBIS) search experiment},
    pdfpagemode=UseOutlines
}
\usepackage{xspace}
\usepackage{siunitx}
\usepackage{enumitem}
\usepackage{booktabs}
\usepackage{verbatim}
\setlist{itemsep=0mm}

\graphicspath{{fig/}}

\input{style}

%% file: style.tex
\newcommand{\eg}{\mbox{\itshape e.g.}\xspace}
\newcommand{\ie}{\mbox{\itshape i.e.}\xspace}

\newcommand{\proanubis}{\mbox{\textsl{pro}ANUBIS}\xspace}
\newcommand{\anubis}{\mbox{ANUBIS}\xspace}
\newcommand{\setanubis}{\mbox{\textsc{SET-ANUBIS}}\xspace}

\newcommand{\atlas}{\mbox{ATLAS}\xspace}
\newcommand{\cms}{\mbox{CMS}\xspace}
\newcommand{\mathusla}{\mbox{MATHUSLA}\xspace}
\newcommand{\codex}{\mbox{CODEX-b}\xspace}

\newcommand{\geant}{\mbox{\textsc{Geant4}}\xspace}

\newcommand{\mum}{\ensuremath{\mu\textnormal{m}}\xspace}
\newcommand{\mm}{\ensuremath{\textnormal{mm}}\xspace}

\newcommand{\metre}{\ensuremath{\textnormal{m}}\xspace}

\newcommand{\MeV}{\ensuremath{\textnormal{MeV}}\xspace}
\newcommand{\GeV}{\ensuremath{\textnormal{GeV}}\xspace}
\newcommand{\TeV}{\ensuremath{\textnormal{TeV}}\xspace}

\newcommand{\pb}{\ensuremath{\rm pb}\xspace}

\newcommand{\invfb}{\ensuremath{{\rm fb}^{-1}}\xspace}

\newcommand{\invab}{\ensuremath{{\rm ab}^{-1}}\xspace}

\newcommand{\met}{\ensuremath{E_T^\text{miss}}\xspace} 
\newcommand{\pt}{\ensuremath{p_T}\xspace} 
\newcommand{\order}[1]{\ensuremath{\mathcal{O}(#1)}\xspace} 
\newcommand{\leff}{\ensuremath{L_\text{eff}}\xspace} 

\newcommand{\eps}{\varepsilon} 
\newcommand{\acc}{\ensuremath{\mathcal A}\xspace} 

\newcommand{\br}{\ensuremath{\mathcal Br}\xspace}

\newcommand{\hllhc}{\mbox{HL-LHC}\xspace}
\providecommand{\unit}[1]{\,\mathrm{#1}}

\newif\ifshowoutline \showoutlinetrue
\definecolor{outlinecol}{RGB}{30,80,160}
\newenvironment{outline}
  {\ifshowoutline
     \begin{mdframed}[linecolor=outlinecol,linewidth=0.8pt,
                      backgroundcolor=outlinecol!5,
                      innertopmargin=4pt,innerbottommargin=6pt]
     \color{outlinecol}\small\textbf{Outline:}
   \else\setbox0\vbox\bgroup\fi}
  {\ifshowoutline\end{mdframed}\else\egroup\fi}

\newcommand{\tabcell}[2][0.18\textwidth]{\parbox[t]{#1}{\raggedright #2\strut}}

%% file: authorList.tex
%
%
%
%
\author{Giulio~Aielli}
\affiliation{INFN and University Roma Tor Vergata, Rome, Italy}
\affiliation{Department of Basic Sciences, Istinye University, Istanbul, Turkiye}
\author{Martin~Bauer}
\affiliation{IPPP, Dept of Physics, Durham University, Durham, United Kingdom}
\author{Oleg~Brandt}
\affiliation{Cavendish Laboratory, University of Cambridge, Cambridge, United Kingdom}
\author{Jackson~Burzynski}
\affiliation{Homer L. Dodge Department of Physics and Astronomy, University of Oklahoma, Norman OK; United States of America}
\author{Paolo~Camarri}
\affiliation{INFN and University Roma Tor Vergata, Rome, Italy}
\author{Roberto~Cardarelli}
\affiliation{INFN and University Roma Tor Vergata, Rome, Italy}
\author{Lucio~Cerrito}
\affiliation{INFN and University Roma Tor Vergata, Rome, Italy}
\author{Serkant~Cetin}
\affiliation{Department of Basic Sciences, Istinye University, Istanbul, Turkiye}
\author{Ming~Chung~Chu}
\affiliation{Department of Physics, The Chinese University of Hong Kong, Hong Kong SAR, China}
\author{Anna~Di~Ciaccio}
\affiliation{INFN and University Roma Tor Vergata, Rome, Italy}
\author{Abdualazem~F.~M.~Ebrahim}
\affiliation{Department of Physics, Nanjing University, Nanjing, China}
\author{Cayetano~Fernandez~Ruiz}
\affiliation{Cavendish Laboratory, University of Cambridge, Cambridge, United Kingdom}
\author{Luis~Roberto~Flores~Castillo}
\affiliation{Department of Physics, The Chinese University of Hong Kong, Hong Kong SAR, China}
\author{Christopher~Lester}
\affiliation{Cavendish Laboratory, University of Cambridge, Cambridge, United Kingdom}
\author{Kam-Biu~Luk}
\affiliation{Department of Physics, Hong Kong University of Science and Technology, Hong Kong SAR, China}
\author{Vasiliki~A.~Mitsou}
\affiliation{Instituto de F\'isica Corpuscular (IFIC), CSIC -- Universitat de Val\`encia, Paterna (Valencia), Spain}
\author{Patrick~Marcel~Morscheck}
\affiliation{Cavendish Laboratory, University of Cambridge, Cambridge, United Kingdom}
\author{Luca~Pizzimento}
\affiliation{Department of Physics, University of Hong Kong, Hong Kong SAR, China}
\author{Tina~Potter}
\affiliation{Cavendish Laboratory, University of Cambridge, Cambridge, United Kingdom}
\author{Kirill~Prokofiev}
\affiliation{Department of Physics, Hong Kong University of Science and Technology, Hong Kong SAR, China}
\author{Michael~Revering}
\affiliation{Cavendish Laboratory, University of Cambridge, Cambridge, United Kingdom}
\author{Th\'eo~Reymermier}
\affiliation{Institut de Physique des 2 Infinis (IP2I), Lyon, France}
\author{Anatoli~Romaniouk}
\affiliation{Institute of Astro and Particle Physics, University of Innsbruck, Innsbruck, Austria}
\affiliation{Department of Basic Sciences, Istinye University, Istanbul, Turkiye}
\author{Aashaq~Shah}
\affiliation{Cavendish Laboratory, University of Cambridge, Cambridge, United Kingdom}
\author{Sinem~Simsek}
\affiliation{Department of Basic Sciences, Istinye University, Istanbul, Turkiye}
\author{John~Stupak}
\affiliation{Homer L. Dodge Department of Physics and Astronomy, University of Oklahoma, Norman OK; United States of America}
\author{Yongjie~Sun}
\affiliation{Department of Modern Physics, University of Science and Technology of China, Hefei, China}
\author{Paul~Swallow}
\affiliation{Cavendish Laboratory, University of Cambridge, Cambridge, United Kingdom}
\affiliation{Formerly at: Cavendish Laboratory, University of Cambridge, Cambridge, United Kingdom}
\author{Yanjun~Tu}
\affiliation{Department of Physics, University of Hong Kong, Hong Kong SAR, China}
\author{Monica~Verducci}
\affiliation{University and INFN Pisa, Pisa, Italy}
\author{Oscar~Vives}
\affiliation{Instituto de F\'isica Corpuscular (IFIC), CSIC -- Universitat de Val\`encia, Paterna (Valencia), Spain}
\author{Ngoc~Khanh~Vu}
\affiliation{Department of Physics, Nanjing University, Nanjing, China}
\author{Julian~Wack}
\affiliation{Cavendish Laboratory, University of Cambridge, Cambridge, United Kingdom}
\author{Zeren~Simon~Wang}
\affiliation{School of Physics, Hefei University of Technology, Hefei 230601, China}
\author{Michael~K.~Wilkinson}
\affiliation{Homer L. Dodge Department of Physics and Astronomy, University of Oklahoma, Norman OK; United States of America}
\author{Yusheng~Wu}
\affiliation{Department of Modern Physics, University of Science and Technology of China, Hefei, China}
\author{Rui~Zhang}
\affiliation{Department of Physics, Nanjing University, Nanjing, China}
\author{and Yu~Zhang}
\affiliation{School of Physics, Hefei University of Technology, Hefei 230601, China}
\author{for the ANUBIS Collaboration}
\noaffiliation


%% file: sections/executive_summary.tex
{\em The physics case.}
Long-lived particles (LLPs) arise naturally in many extensions of the Standard Model (SM), notably in models with dark matter candidates or motivated by baryogenesis.
Neutral LLPs produced at or above the electroweak scale with decay lengths of $\order{10~\metre}$ and beyond escape the general-purpose LHC detectors, and cannot be produced at sufficient rates in forward and beam-dump experiments.
A dedicated transverse detector close to an LHC interaction point is required to cover this region of parameter space at the High-Luminosity LHC (HL-LHC).

{\em The \anubis detector.}
The \anubis detector would instrument the ceiling of the \atlas cavern UX15 and the bottom of the PX14 and PX16 access shafts with two layers of resistive plate chambers (RPCs) about 1~\metre apart, each covering about 1600~m$^2$ (Sections~\ref{sec:location} and~\ref{sec:performance}).
The air-filled volume between the \atlas detector and the cavern ceiling serves as decay volume, so that the \atlas and \anubis detectors together form a continuous decay volume from the interaction point up to about 23~\metre.
The \atlas detector acts as a passive shield against SM particles, and its event information provides an active veto that suppresses the backgrounds from interactions of SM LLPs with material (Section~\ref{sec:backgrounds}).

{\em Discovery potential.}
With 3~\invab at the HL-LHC, the \anubis detector would probe exotic decays of the Higgs boson into long-lived scalars down to branching ratios of $\order{10^{-5}}$--$\order{10^{-6}}$, two to three orders of magnitude beyond the latest \atlas and \cms results, and extend their $c\tau$ reach by three to four orders of magnitude (Section~\ref{sec:physics}).
It would reach mixings of heavy neutral leptons of all three flavours of about $2\times10^{-8}$ at masses of about 6~GeV, where it is the only proposal with sensitivity to small couplings, and improve on the current limits on axion-like particles by up to two orders of magnitude.
For the studied benchmarks, it surpasses the projections of the other transverse proposals \codex and \mathusla{}40, and it is complementary to forward and beam-dump experiments.

{\em Technology and design.}
The \proanubis demonstrator recorded 104~\invfb in 2024 and 73~\invfb in 2025 close to the ceiling of the \atlas cavern, and showed that the RPC technology meets the efficiency and timing requirements of the \anubis detector (Section~\ref{sec:proanubis}).
The detector design builds on the RPC technology of the \atlas Phase~II upgrade, combined with a CO$_2$-rich gas mixture with a low global warming potential (GWP), \ie free of SF$_6$ and of R134a or at least with an R134a fraction of at most 15\%, and a sensitive front end (Section~\ref{sec:design}).
A trigger concept based on coincidences between two RPC doublets retains more than 90\% of the signal at an accidental rate below 100~Hz, within the latency envelope of the \atlas Level-0 (L0) trigger (Section~\ref{sec:trigger}).
A baseline mechanical design for the installation on the cavern vault and in the shaft plugs has been developed (Section~\ref{sec:engineering}).

{\em Relationship with the \atlas experiment.}
The \anubis detector is conceived as an integral extension of the \atlas experiment (Section~\ref{sec:interfacing}).
It relies on the full \atlas event information to reach its sensitivity, and the two detectors would exchange triggers.
In turn, it extends the \atlas muon system by two tracking layers about 23~\metre above the beamline, improving the reconstruction of muons and of displaced vertices in the outer muon spectrometer, and enabling time-of-flight measurements with a precision of $\order{0.5\%}$ on the velocity.
The \anubis Collaboration is fully committed to fitting into the existing constraints without any destructive interference with the \atlas experiment.
Its services are designed to be largely self-sufficient, with few and well-defined interfaces to the \atlas infrastructure.
Several \anubis institutes have contributed to the Phase~II upgrade of the \atlas RPC system, and the two communities collaborate on R\&D of common interest, such as low-GWP RPC gas mixtures.

{\em Timeline.}
The next step is to formalise the \anubis detector as an \atlas Phase~III upgrade project, starting with an Initial Design Review in spring 2027, followed by a Technical Design Report (TDR) for the detector and electronics in summer 2028 and its approval by the CERN Research Board by the end of 2028, and a separate TDR for the mechanics and services in 2031 (Section~\ref{sec:timeline}).
A Slice Test of $\order{10}$ tracking modules would be used to develop the data-acquisition, detector-control and trigger interfaces during Run~4.
It would be installed on the cavern ceiling at the end of the third LHC Long Shutdown (LS3) if, and only if, the LS3 schedule allows, and otherwise on the HO gantry on the A side of the \atlas cavern.
The tracking modules would be produced from 2031 until the end of 2034 and installed during the fourth Long Shutdown (LS4), from December 2033 to 2035, for data taking from the start of Run~5 in 2036.
The installation of the services and of the mechanical support system could already start in the year-end technical stops of Run~4, if this can be accommodated.
Beyond the HL-LHC, the \anubis detector is a proof of concept for transverse LLP detectors proposed for the electron--positron stage of the Future Circular Collider, FCC-ee, and it could be upcycled should a future FCC-ee detector be hosted in the UX15 cavern.

{\em Collaboration.}
The \anubis Collaboration currently comprises 38 authors from 15 institutes in Europe, Asia and North America, and is open to new members.

%% file: sections/introduction.tex
The Standard Model (SM) of particle physics has been tested to remarkable precision, yet it leaves fundamental questions unanswered, such as the nature of dark matter~\cite{Hall:2010jx,Cheung:2010gk,Zurek:2013wia}, the origin of the baryon asymmetry of the Universe~\cite{Barry:2013nva}, and the naturalness of the electroweak scale~\cite{Giudice:1998bp,Burdman:2006tz}.
Many of the extensions of the SM proposed to address these questions, including hidden sectors~\cite{Han:2007ae,Strassler:2006im}, supersymmetry~\cite{Barbier:2004ez,Giudice:1998bp} and neutral-naturalness models with exotic Higgs-boson decays~\cite{Chacko:2005pe,Cai:2008au}, predict long-lived particles (LLPs) with macroscopic lifetimes, $\tau\gtrsim10$~ps.
Such particles may have escaped detection so far not because they are too heavy or too feebly produced, but because they decay far away from the interaction point.

\begin{figure}[ht]
    \centering
    \includegraphics[width=0.90\linewidth]{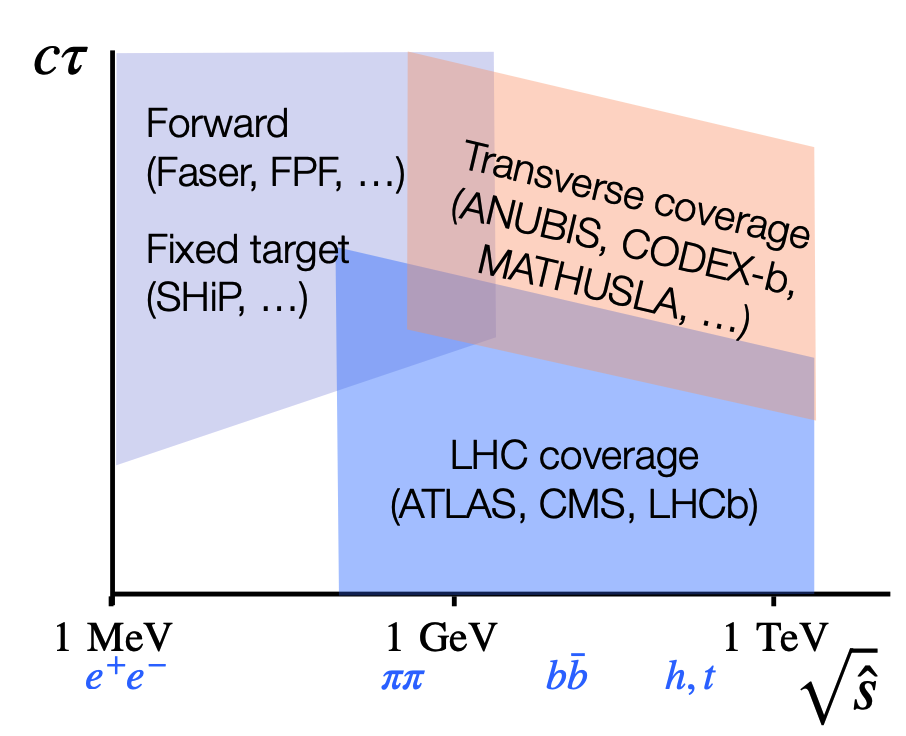}
    \caption{Schematic representation of the coverage of different experiments as a function of the proper lifetime, $c\tau$, and the partonic centre-of-mass energy, $\sqrt{\hat{s}}$.}
    \label{fig:LLP_Complimentarity}
\end{figure}

The general-purpose detectors at the Large Hadron Collider (LHC)~\cite{Evans:1129806}, \atlas~\cite{Aad:1129811} and \cms~\cite{Chatrchyan:1129810}, were optimised for particles that decay promptly, while LHCb~\cite{LHCb:2008vvz} targets $b$-hadron decays with lifetimes of $\order{1~\text{ps}}$~\cite{bib:pdg2024}.
Their instrumented volumes, trigger strategies and backgrounds limit their sensitivity to neutral LLPs with decay lengths beyond a few metres~\cite{Alimena:2019zri,CMS:2021sch,CMS:2021juv,ATLAS:2018tup,satterthwaite_2022_292b9-eck73}.
Dedicated LLP experiments fall into two classes, depending on the region of LLP parameter space they cover, as illustrated in Figure~\ref{fig:LLP_Complimentarity}~\cite{PBC:2025sny}.
{\em Forward} and beam-dump experiments, such as FASER~\cite{FASER:2018eoc} and SHiP~\cite{Ahdida:2654870}, efficiently probe LLPs produced at small partonic centre-of-mass energies, $\sqrt{\hat s}=\order{1~\GeV}$, \eg in meson decays~\cite{Antel:2023hkf}, but their sensitivity drops rapidly for $\sqrt{\hat s}\gtrsim10~\GeV$.
{\em Transverse} experiments, located at large angles to the beam close to an LHC interaction point, are required to cover LLPs produced at the electroweak scale, $\sqrt{\hat s}\approx100~\GeV$, and above, such as in decays of the Higgs boson.
Proposed transverse experiments include \mathusla~\cite{Curtin:2018mvb,MATHUSLA:2020uve,MATHUSLA:2025zyt}, \codex~\cite{CODEX-b:2019jve}, GRENDEL~\cite{citron2026}, and \anubis~\cite{bib:anubis_orig,Shah:2024fpl,Brandt:2025fdj}.

This Letter of Intent proposes the \anubis (AN Underground Belayed In-Shaft) detector, which would instrument the ceiling of the \atlas experimental cavern and the bottom of its access shafts with tracking detectors.
Using established and cost-effective resistive plate chamber (RPC) technology, the \anubis detector would be sensitive to LLPs with proper lifetimes $c\tau$ from $\order{0.1~\metre}$ up to $\order{10^5~\metre}$.

A defining feature of the \anubis detector is its close integration with the \atlas experiment.
Events recorded by the \anubis detector are synchronised with those of the \atlas experiment, so that the \atlas detector serves both as a passive shield against collision backgrounds and as an active veto through event-level selections based on reconstructed tracks, jets and missing transverse momentum.
Conversely, the \anubis detector would extend the \atlas sensitivity to neutral LLPs decaying beyond its instrumented volume, improve the reconstruction of muons crossing both detectors, and provide an external trigger for the \atlas readout, enabling the recording of events that might otherwise evade the standard trigger selections.
Joint analyses could combine displaced vertices reconstructed in the \anubis detector with prompt objects and missing transverse momentum measured by the \atlas detector, helping to identify the LLP production mechanism and to characterise a potential signal.
These complementary capabilities provide a strong rationale for the \anubis detector to become an \atlas Phase~III upgrade project (Section~\ref{sec:interfacing}).

The \proanubis demonstrator was installed close to the ceiling of the \atlas cavern in 2023~\cite{Shah:2024fpl,ANUBIS:2025mme}, commissioned~\cite{ANUBIS:2025inn}, and decommissioned in 2026.
It recorded 104~\invfb and 73~\invfb of $pp$ collision data at $\sqrt s=13.6~\TeV$ in 2024 and 2025, respectively.
The \proanubis results show that the proposed RPC technology meets the efficiency and timing requirements of the \anubis detector, and that \proanubis events can be synchronised with those recorded by the \atlas experiment~\cite{ANUBIS:2026jxg}.
Further studies combining \proanubis and \atlas data are under way.

Transverse LLP detectors have also been proposed for the Future Circular Collider (FCC)~\cite{Benedikt:2928193}, such as LAYCAST~\cite{laycast} and HECATE~\cite{hecate}, which would instrument the FCC-ee experimental caverns or their surroundings with large-area tracking detectors.
The \anubis detector would therefore also serve as a proof of concept for these experiments, developing the techniques and expertise required and informing their design.

This Letter of Intent is organised as follows.
The location of the \anubis detector and its discovery potential are discussed in Sections~\ref{sec:location} and~\ref{sec:physics}, followed by the backgrounds in Section~\ref{sec:backgrounds}.
The detector performance requirements, the results of the \proanubis demonstrator, the engineering challenges, the detector design and the trigger design are presented in Sections~\ref{sec:performance}--\ref{sec:trigger}.
The interface between the \anubis and \atlas detectors, the timeline and the preliminary costing are discussed in Sections~\ref{sec:interfacing}--\ref{sec:costing}, before concluding in Section~\ref{sec:Summary}.

%% file: sections/location.tex
\begin{figure}[tb]
    \centering
    \begin{tikzpicture}
        \node[inner sep=0] (image) at (0,0) {\includegraphics[width=0.75\linewidth]{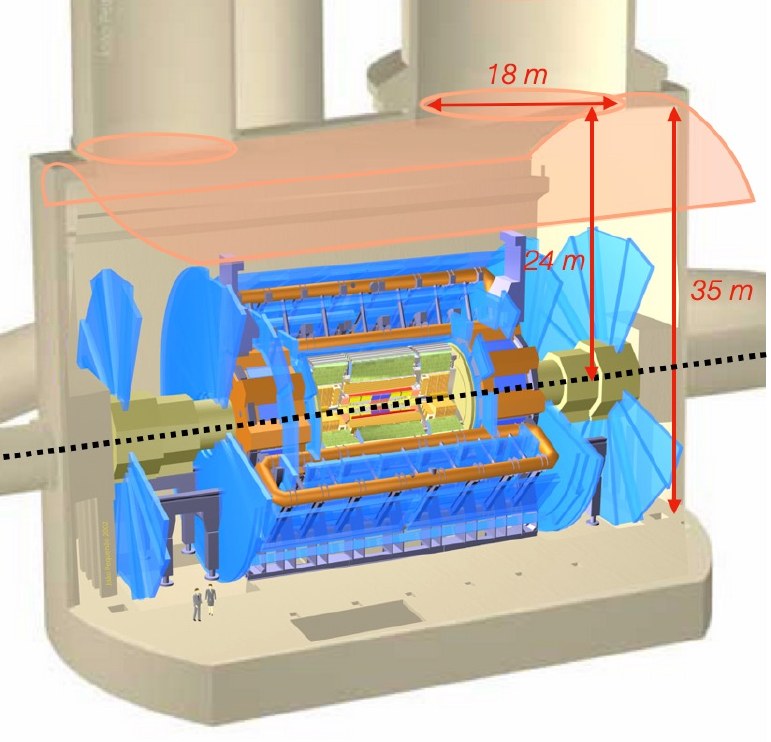}};
        \node[inner sep=0] at (1.1,2.9) {{\color{red}{\textbf{PX14}}}};
        \node[inner sep=0] at (-2,2.9) {{\color{red}{\textbf{PX16}}}};
    \end{tikzpicture}
    \caption{Layout of the \atlas underground cavern UX15 at LHC Point~1, including the access shafts PX14 and PX16.
    The area highlighted in orange illustrates the tracking detectors of the \anubis detector on the cavern ceiling and at the bottom of the two shafts.\label{fig:loc_cavern}}
\end{figure}

The \anubis detector will be located in the UX15 experimental cavern at LHC Point~1, directly above the \atlas detector~\cite{Aad:1129811}.
It exploits the air-filled volume between the \atlas detector and the vaulted ceiling of the cavern as a decay volume for neutral LLPs, and reconstructs their decays as displaced vertices (DVs) of charged-particle tracks recorded by tracking detectors mounted on the cavern ceiling and at the bottom of the PX14 and PX16 access shafts~\cite{bib:anubis_orig,ANUBIS:sensitivity2025}.
The layout is shown in Figure~\ref{fig:loc_cavern}.

\subsection{Geometry of ANUBIS and ATLAS}
\label{sec:loc_geometry}

The UX15 cavern is approximately 53~\metre long along the beam axis, up to 31~\metre wide and 35~\metre high.
Its ceiling is a vault with a radius of curvature of 20~\metre.
The beamline through the interaction point (IP) does not coincide with the axis of the vault: it is displaced by 1.7~\metre towards the centre of the LHC ring and by 3.5~\metre downwards relative to the centre of curvature, so that the apex of the ceiling lies approximately 23~\metre above the beamline~\cite{ANUBIS:sensitivity2025}.
\anubis adopts the right-handed coordinate system of \atlas\footnote{%
The origin is the nominal IP in the centre of the \atlas detector and the $z$-axis points along the beam pipe.
The $x$-axis points from the IP to the centre of the LHC ring and the $y$-axis points upwards.
Polar coordinates $(r,\phi)$ are used in the transverse plane, $\phi$ being the azimuthal angle around the $z$-axis, and the pseudorapidity is $\eta\equiv-\ln\tan(\theta/2)$.}.

A tracking detector of \anubis consists of two tracking layers separated by $\sim1~\metre$, each made of a multiplet (doublet or triplet) of RPC layers~\cite{bib:anubis_orig,ANUBIS:sensitivity2025}.
On the ceiling, the tracking detector is built from standardised tracking modules with an edge width of $\order{1~\metre}$.
Each module comprises two RPC multiplets positioned $\sim1~\metre$ apart, which correspond to the upper and lower tracking layer, and the modules are tiled such that they follow the curvature of the vault.
In the nominal geometry, the upper tracking layer is located 0.2~\metre below the ceiling, leaving space for the fixings and services, and the lower layer 1~\metre below it.
This instruments an area of about 1600~m$^2$ per tracking layer.
The modular layout and the performance requirements on the tracking layers are discussed in Section~\ref{sec:performance}.

The openings of the PX14 and PX16 shafts in the ceiling are covered by circular tracking detectors, referred to as shaft plugs, which follow the same two-layer structure at the bottom of each shaft.
The shafts have diameters of 18~\metre (PX14) and 12.6~\metre (PX16).
The shaft plugs ensure that the coverage of the ceiling is hermetic, and are designed to be retractable for the Year-End Technical Stops (YETS) and Long Shutdowns (LS).
A plug at the bottom of PX14 already exists and is removed and reinstalled at the beginning and end of each YETS and LS; instrumenting it is therefore a matter of replacing the steel girders that currently make up the plug, which rest on steel beam stubs at the bottom of the shaft, by two tracking layers.
The mechanical challenges associated with the ceiling structure and the retractable shaft plugs are summarised in Section~\ref{sec:engineering}.
This ceiling configuration supersedes the shaft configuration of the original proposal, with four tracking detectors suspended inside the PX14 shaft~\cite{bib:anubis_orig}, which was dropped due to practical constraints.\footnote{%
All four tracking detectors would have had to be extracted for crane access to UX15 during every YETS, with insufficient space for their disassembly and storage at Point~1~\cite{bib:ludo}.
In addition, the shaft configuration subtends a much smaller solid angle as seen from the IP~\cite{ANUBIS:sensitivity2025}.}

\subsection{Fiducial volume and continuity with ATLAS}
\label{sec:loc_fiducial}

\begin{figure*}[tb]
    \centering
    \begin{subfigure}[b]{0.49\linewidth}
        \centering
        \includegraphics[height=0.67\linewidth]{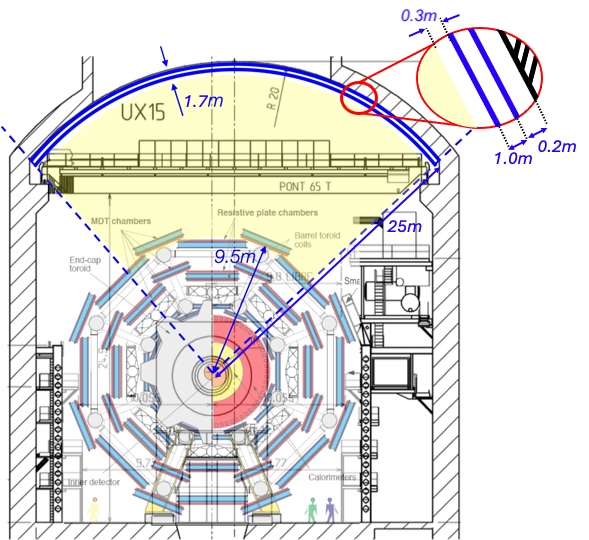}
        \caption{}
        \label{fig:loc_fiducial_XY}
    \end{subfigure}
    \begin{subfigure}[b]{0.49\linewidth}
        \centering
        \includegraphics[height=0.7\linewidth]{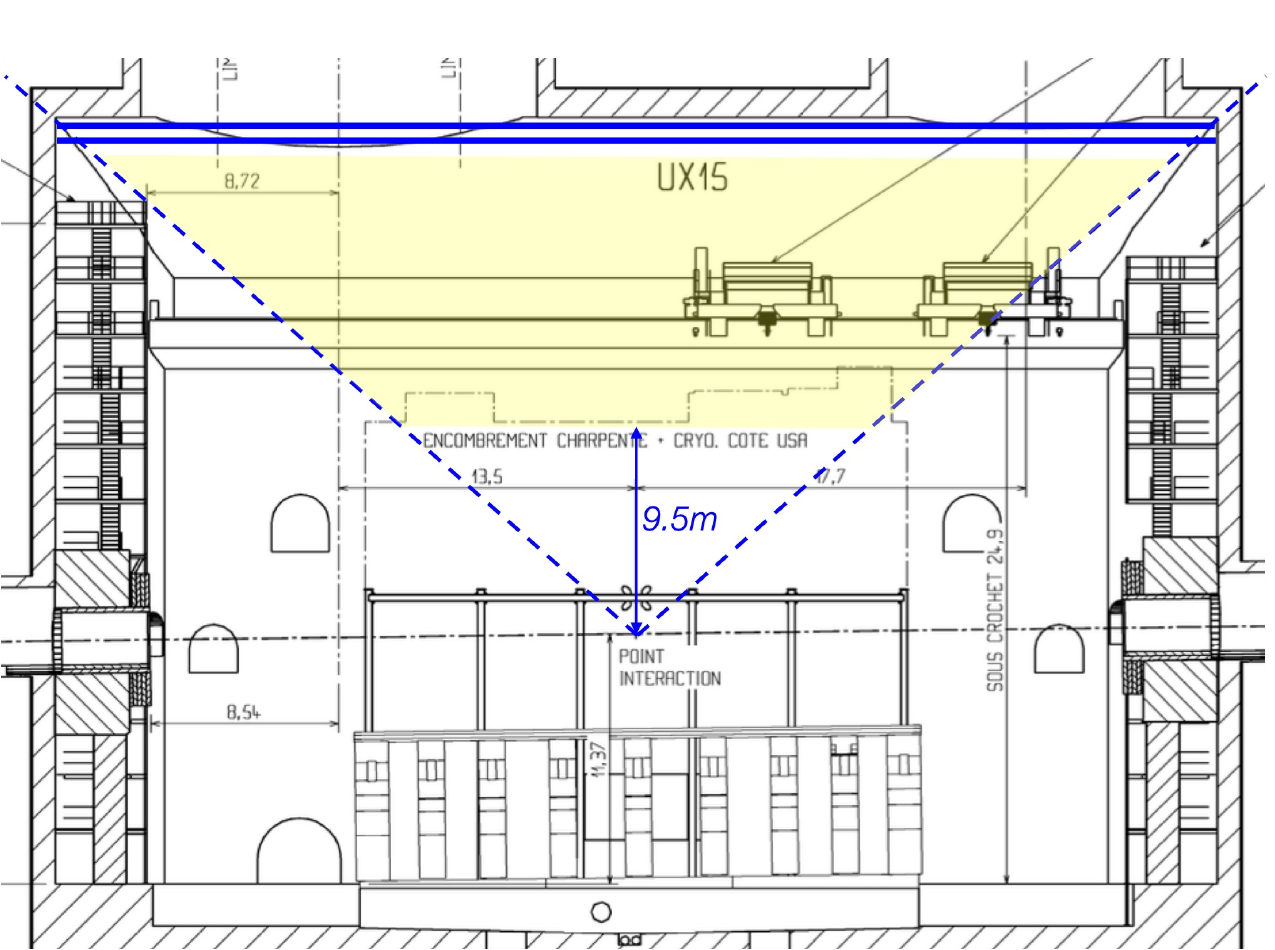}
        \caption{}
        \label{fig:loc_fiducial_ZY}
    \end{subfigure}
    \caption{The two tracking layers of the \anubis detector in the ceiling configuration, shown as thick blue lines inside the UX15 cavern alongside the \atlas detector, in (a)~the $(x,y)$ plane and (b)~the $(y,z)$ plane.
    The fiducial volume is indicated as a shaded yellow region and the acceptance boundaries in $\eta$ and $\phi$ are shown as dashed blue lines~\cite{Bauer:2026ivt}.
    The drawings of the \atlas detector and the cavern are courtesy of \atlas.}
    \label{fig:loc_fiducial}
\end{figure*}

The fiducial volume in which \anubis reconstructs DVs from LLP decays is shown in Figure~\ref{fig:loc_fiducial}.
Its upper boundary lies 0.3~\metre below the lower tracking layer.
This margin ensures a reliable rejection of DVs originating from hadronic interactions in the material of the tracking detector itself.
Towards the beamline, the guiding principle is that a generous lower boundary accounts for the fiducial volume lost due to the proximity of DV candidates to dense material structures.
The density of material decreases rapidly beyond the barrel of the \atlas muon spectrometer at 4.5~\metre from the beamline, and 5~\metre are conservatively added to exclude dense regions such as support structures and muon chambers, including a safety margin for the finite vertex resolution~\cite{ANUBIS:sensitivity2025}.
For the purpose of sensitivity studies, the fiducial volume is therefore assumed to start 9.5~\metre from the beamline.
In addition, the decay vertex must lie inside the cavern~\cite{Bauer:2026ivt}.
For an LLP candidate to be accepted, at least two charged tracks from its decay are required to cross both tracking layers.

A defining feature of this location is the proximity and complementary coverage of the \anubis and \atlas decay volumes.
LLPs decaying close to the IP are reconstructed by the \atlas inner detector, and those decaying further out can be identified using the calorimeters and the muon spectrometer, with the latter providing vertex reconstruction out to approximately 7.5~\metre~\cite{ATLAS:2018tup}.
\anubis complements the \atlas muon spectrometer from 9.5~\metre up to the cavern ceiling, so that \atlas and \anubis together form a single, continuous tracking and vertexing volume from the IP up to approximately 23~\metre above the beamline.
This maximises the combined sensitivity to LLPs with decay lengths ranging from a few \mum to tens of metres from the IP~\cite{ANUBIS:sensitivity2025}.

The proximity to \atlas brings two further advantages.
First, the \atlas detector, in particular its calorimeters, solenoid magnet and support structures, acts as a passive shield against SM particles produced in the collisions, while the \atlas event record of the same bunch crossing provides an active veto, as discussed in Section~\ref{sec:backgrounds}.
Second, \anubis will be integrated with the \atlas trigger and data-acquisition systems, so that a DV in \anubis can trigger \atlas and vice versa, giving access to the full $pp$ collision for every LLP candidate (Sections~\ref{sec:trigger} and~\ref{sec:interfacing}).
The environment just below the ceiling of UX15 has already been explored with the \proanubis demonstrator, which recorded 104~\invfb of $pp$ collision data in 2024 and 73~\invfb in 2025 from level~12 of the HO gantry on the A side of the cavern, close to the ceiling~\cite{ANUBIS:2025mme,ANUBIS:2025inn}, as described in Section~\ref{sec:proanubis}.

%% file: sections/physics.tex
Long-lived particles naturally occur in many scenarios beyond the SM (BSM)~\cite{Antel:2023hkf,Alimena:2019zri,Lee:2018pag,Curtin:2018mvb}.
This is especially the case in models that provide a dark matter (DM) candidate, where the feeble interactions required between the dark and the visible sector, \eg in freeze-in or asymmetric DM scenarios, translate into long lifetimes of the dark-sector states or of their partners~\cite{Hall:2010jx,Cheung:2010gk,Zurek:2013wia}; in seesaw models~\cite{Minkowski:1977sc,Yanagida:1979as,Mohapatra:1979ia,Gell-Mann:1979vob,Schechter:1980gr,Mohapatra:1986bd} that explain the non-vanishing neutrino masses established by observations of neutrino oscillations~\cite{Canetti:2010aw}; and in models motivated by baryogenesis, where the out-of-equilibrium condition is naturally met by slowly decaying particles~\cite{Barry:2013nva,Brdar_2019}.

Long lifetimes arise from a few basic mechanisms generically present in quantum field theories and hence familiar from the SM~\cite{Alimena:2019zri,Lee:2018pag}:
small couplings, such as feeble portal couplings to a hidden sector or couplings protected by approximate symmetries;
decays mediated by heavy, off-shell states, as for the muon in the SM;
and a small available phase space, \eg in compressed mass spectra, as for the neutron.
Since the decay width depends strongly on the masses and couplings involved, \eg $\Gamma\propto g^2m^5/M^4$ for a decay of a particle of mass $m$ mediated by a heavy state of mass $M$, lifetimes spanning many orders of magnitude arise naturally.
A realistic assessment of the discovery potential of the \anubis detector therefore starts from a map of the LLP landscape, identifying which classes of scenarios a transverse detector at the High-Luminosity LHC (HL-LHC) is uniquely placed to probe, and which are better covered by forward, beam-dump or general-purpose experiments.

This section first summarises the portal framework of the Physics Beyond Colliders (PBC) initiative (Section~\ref{sec:phys_portals}), classifies LLP scenarios by their production mechanism (Section~\ref{sec:phys_landscape}), and identifies the strengths of the \anubis detector (Section~\ref{sec:phys_strengths}).
After summarising the common analysis framework (Section~\ref{sec:phys_framework}), the projected sensitivities to the benchmark models studied so far are presented (Sections~\ref{sec:phys_higgs}--\ref{sec:phys_alp}), followed by an outlook on further scenarios (Section~\ref{sec:phys_further}) and a comparison with other proposals (Section~\ref{sec:phys_comparison}).

\subsection{The portal framework}
\label{sec:phys_portals}

The PBC initiative classifies models of feebly interacting particles according to the `portals' through which a hidden or dark sector (DS) couples to the SM~\cite{Beacham:2019nyx,Alemany:2019vsk}.
The sensitivity studies for the \anubis detector are aligned with this framework and with the benchmark cases (BCs) introduced by the PBC initiative~\cite{Beacham:2019nyx,PBC:2025sny}.
The DS is assumed to be neutral under the SM gauge group, and its interactions with the SM are described by
\begin{equation}
\mathcal{L} = \mathcal{L}_\text{SM} + \mathcal{L}_\text{DS} + \mathcal{L}_\text{portal}\,,\qquad
\mathcal{L}_\text{portal} = \sum \mathcal{O}_\text{SM}\times\mathcal{O}_\text{DS}\,,
\label{eq:phys_portal}
\end{equation}
where $\mathcal{O}_\text{SM}$ and $\mathcal{O}_\text{DS}$ are operators composed of SM and DS fields, respectively.
The phenomenologically most important interactions are those with the lowest-dimension operators that respect the SM gauge symmetries.
They define the four portals listed in Table~\ref{tab:phys_portals}~\cite{Beacham:2019nyx}:
\begin{itemize}
\item {\em Vector portal.}
A new massive $U(1)$ gauge boson, the dark photon, $A'$, mixes kinetically with the SM hypercharge boson with a strength $\epsilon$, and thereby acquires suppressed couplings to the electromagnetic current (BC1--BC3).
\item {\em Scalar portal.}
A dark scalar, $S$, couples to the Higgs bilinear $H^\dagger H$.
The term linear in $S$ induces a mixing with the Higgs boson by an angle $\theta\simeq\mu v/(m_h^2-m_S^2)$, so that $S$ inherits the Higgs-boson couplings to SM particles suppressed by $\sin\theta$, while the quartic term $\lambda$ induces the exotic decay $h\to SS$ (BC4, BC5).
\item {\em Neutrino portal.}
Sterile neutrinos or heavy neutral leptons (HNLs), $N$, couple to the SM lepton doublets $L$ through a Yukawa interaction.
After electroweak symmetry breaking, they mix with the active neutrinos, and can explain the neutrino masses through a seesaw mechanism and the baryon asymmetry of the Universe through leptogenesis~\cite{Bondarenko:2018ptm,Gorbunov:2007ak,Brdar_2019} (BC6--BC8).
\item {\em Axion portal.}
Axion-like particles (ALPs), $a$, are pseudo-Nambu--Goldstone bosons of a spontaneously broken global symmetry, whose couplings to photons, gluons and fermions arise from dimension-five operators suppressed by a high scale $f_a$~\cite{Bauer:2017ris} (BC9--BC11).
\end{itemize}
The vector, scalar and neutrino portals are renormalisable.
In the absence of symmetries forbidding them, \eg a $Z_2$ symmetry, the kinetic mixing and the $S^2H^\dagger H$ operator are generated at loop level, so that small couplings arise naturally.
The axion portal is an effective, non-renormalisable interaction; for the PBC benchmarks, couplings induced at loop level are evaluated at a cut-off scale of $\Lambda=1~\TeV$~\cite{Beacham:2019nyx}.

\begin{table}[tb]
  \centering
  \caption{The four portals between the SM and a dark sector defined by the PBC initiative, and the corresponding lowest-dimension couplings~\cite{Beacham:2019nyx}.
  $F'_{\mu\nu}$, $B_{\mu\nu}$, $F_{\mu\nu}$ and $G_{i,\mu\nu}$ denote the field-strength tensors of the dark photon, the hypercharge boson, the photon and the gluons, respectively, $\tilde F_{\mu\nu}$ and $\tilde G_{i,\mu\nu}$ the dual tensors, $H$ the Higgs doublet, $L$ a lepton doublet, and $\psi$ an SM fermion.}
  \label{tab:phys_portals}
  \renewcommand{\arraystretch}{1.25}
  \begin{tabular}{ll}
    \toprule
    Portal & Coupling \\
    \midrule
    Vector: dark photon, $A'_\mu$ & $-\dfrac{\epsilon}{2\cos\theta_W}F'_{\mu\nu}B^{\mu\nu}$ \\[1.2ex]
    Scalar: dark Higgs, $S$ & $(\mu S+\lambda S^2)H^\dagger H$ \\[0.6ex]
    Neutrino: HNL, $N$ & $y_N L H N$ \\[0.6ex]
    Axion: ALP, $a$ & $\dfrac{a}{f_a}F_{\mu\nu}\tilde F^{\mu\nu}$,\quad $\dfrac{a}{f_a}G_{i,\mu\nu}\tilde G_i^{\mu\nu}$, \\[1.6ex]
     & $\dfrac{\partial_\mu a}{f_a}\bar\psi\gamma^\mu\gamma^5\psi$ \\[0.8ex]
    \bottomrule
  \end{tabular}
\end{table}

The BCs defined for each portal are summarised in Table~\ref{tab:phys_bc}.
The minimal dark photon (BC1) decays visibly into SM particles, while BC2 and BC3 describe light DM coupled through a dark photon and millicharged particles, respectively.
In the scalar portal, BC4 assumes $\lambda=0$, so that the scalar is produced predominantly in $B$-meson decays, while BC5 assumes a sizeable $\lambda$ corresponding to $\br(h\to SS)=10^{-2}$, which opens up the production in decays of the Higgs boson for masses up to $m_h/2$.
BC6--BC8 describe a single Majorana HNL mixing exclusively with the electron, muon or tau neutrino, respectively, and BC9--BC11 describe ALPs that couple predominantly to photons, fermions or gluons.

Of these, the scalar, neutrino and axion portals have been studied for the \anubis detector (Sections~\ref{sec:phys_higgs}--\ref{sec:phys_alp}), including two well-motivated ALP scenarios that go beyond the PBC benchmarks.
The vector portal has not yet been studied, since the vector-portal benchmarks fall largely into the domain of other experiments.
Long-lived dark photons in BC1 are light and produced mainly in meson decays, while BC2 and BC3 yield no displaced vertices; these benchmarks are best covered by the complementary forward and beam-dump experiments~\cite{FASER:2018eoc,SHiP:2021nfo}, whereas heavier dark photons from $h\to A'A'$ decays~\cite{Curtin:2014cca} are accessible to the \anubis detector.

\begin{table}[tb]
  \centering
  \caption{Summary of the PBC benchmark cases~\cite{Beacham:2019nyx}, their free parameters, and the corresponding \anubis sensitivity studies, where available.
  For the HNLs, the notation $|V_{1\ell}|^2$ of Section~\ref{sec:phys_hnl} is used for the mixing with the active neutrinos.
  Further ALP scenarios beyond the PBC benchmarks are discussed in Section~\ref{sec:phys_alp}.}
  \label{tab:phys_bc}
  \begin{tabular}{llll}
    \toprule
    BC & Model & Parameters & Study \\
    \midrule
    \multicolumn{4}{l}{\em Vector portal} \\
    BC1 & minimal dark photon & $m_{A'}$, $\epsilon$ & -- \\
    BC2 & light DM via $A'$ & $m_{A'}$, $\epsilon$, $m_\chi$, $\alpha_D$ & -- \\
    BC3 & millicharged particles & $m_\chi$, $Q_\chi/e$ & -- \\
    \midrule
    \multicolumn{4}{l}{\em Scalar portal} \\
    BC4 & Higgs-mixed scalar & $m_S$, $\theta$ & -- \\
    BC5 & BC4 with $h\to SS$ & $m_S$, $\theta$, $\lambda$ & \cite{ANUBIS:sensitivity2025} \\
    \midrule
    \multicolumn{4}{l}{\em Neutrino portal} \\
    BC6 & HNL, $e$ mixing & $m_{N_1}$, $|V_{1e}|^2$ & \cite{Bauer:2026ivt} \\
    BC7 & HNL, $\mu$ mixing & $m_{N_1}$, $|V_{1\mu}|^2$ & \cite{Bauer:2026ivt} \\
    BC8 & HNL, $\tau$ mixing & $m_{N_1}$, $|V_{1\tau}|^2$ & \cite{ANUBIS:tauHNL2026} \\
    \midrule
    \multicolumn{4}{l}{\em Axion portal} \\
    BC9 & ALP, photon coupling & $m_a$, $g_{a\gamma\gamma}$ & -- \\
    BC10 & ALP, fermion coupling & $m_a$, $f_\ell^{-1}=f_q^{-1}$ & \cite{ANUBIS:ALP2026} \\
    BC11 & ALP, gluon coupling & $m_a$, $f_G^{-1}$ & \cite{ANUBIS:ALP2026} \\
    \bottomrule
  \end{tabular}
\end{table}

\subsection{Classification by production mechanism}
\label{sec:phys_landscape}

For the experimental reach, the decisive property is less the portal than the production mechanism, which sets the partonic centre-of-mass energy $\sqrt{\hat s}$, the boost and angular distribution of the LLP, and the activity accompanying it in the event.
From this perspective, LLP scenarios fall into four broad types, summarised in Table~\ref{tab:phys_categories}:
\begin{enumerate}[label=(\roman*)]
\item {\em LLPs from decays of electroweak-scale SM bosons.}
The Higgs boson, with its narrow width, is a uniquely sensitive window to hidden sectors: even a small coupling can induce sizeable exotic decays into LLPs~\cite{Curtin:2013fra}.
Such $h\to\text{LLP}$ decays occur in many models, \eg in neutral-naturalness models such as the twin Higgs~\cite{Chacko:2005pe}, folded supersymmetry~\cite{Burdman:2006tz} and quirky little Higgs models~\cite{Cai:2008au}, in hidden valleys~\cite{Strassler:2006im,Han:2007ae}, in singlet-scalar extensions of the Higgs sector~\cite{Curtin:2013fra}, in models with dark photons from a dark Higgs sector, $h\to A'A'$~\cite{Curtin:2014cca}, and in ALP models with $h\to Za$ and $h\to aa$ decays~\cite{Bauer:2017ris}.
These models share many common kinematic features: $\sqrt{\hat s}\approx m_h$, LLP energies of order $m_h/2$, moderate boosts, and a wide angular distribution at central rapidities.
The model dependence enters mainly through the LLP mass, lifetime and decay modes, so that this class can be benchmarked by a concrete scenario such as $h\to ss$ in neutral naturalness, which corresponds to BC5.
These models cannot be probed at forward and fixed-target experiments due to their limited kinematic reach.
\item {\em LLPs produced in association with prompt SM objects.}
Many models produce the LLP together with a charged lepton, a $W^\pm$ or $Z$ boson, or a photon, as in charged-current HNL production, $W^{\pm(*)}\to\ell N$, or for ALPs radiated off a weak boson, $pp\to W^\pm a$ and $Za$.
These prompt objects represent a distinct experimental signature.
They can hence be used for triggering the \atlas experiment and for suppressing the backgrounds to negligible levels.
\item {\em LLPs from decays of heavy BSM states.}
Pair-produced gluinos, squarks, neutralinos or heavy hidden-sector mediators at the TeV scale can decay to LLPs or be long-lived themselves, and confining dark sectors can produce dark showers.
These scenarios feature not only the largest $\sqrt{\hat s}$, but also the largest LLP boosts and the highest LLP decay product multiplicities, typically resulting in very distinct experimental signatures.
Since they tend to produce experimentally more accessible signatures than LLP models at the electroweak scale, they are not targeted specifically in the optimisation of the \anubis detector, although they will be an integral part of its future physics programme.
\item {\em Light LLPs from hadron decays.}
Dark photons, light scalars, ALPs and HNLs with masses of $\order{\GeV}$ are copiously produced in decays of $K$, $D$ and $B$ mesons, or in bremsstrahlung processes.
Here $\sqrt{\hat s}=\order{\GeV}$ and the LLPs are strongly boosted in the forward direction, which is the domain of forward and beam-dump experiments such as FASER and SHiP~\cite{FASER:2018eoc,SHiP:2021nfo}.
\end{enumerate}
Types~(i), (ii), and~(iii) populate the region of LLP parameter space characterised by decay lengths of $\gtrsim10~\metre$ and $\sqrt{\hat s}$ at or above the electroweak scale, which is not covered by any existing or approved experiment, cf.~Figure~\ref{fig:LLP_Complimentarity}.
This region is the primary target of the \anubis detector.

The sensitivity of the \anubis detector to type~(iv) scenarios is mostly complementary to that of other experiments.
In these scenarios the LLPs produced in hadron decays are typically accompanied by hadronic activity.
Such LLP candidates are currently rejected by the isolation requirements (Section~\ref{sec:phys_framework}) that are tied to the data-driven background estimate of Section~\ref{sec:backgrounds}, which rescales the background observed in an \atlas search that applied the same isolation.
\anubis is expected to probe type~(iv) scenarios through dedicated selections considering information from both the \anubis and \atlas detectors, cf.~Section~\ref{sec:phys_further}.
Hence, all sensitivity estimates for such models presented in this Letter of Intent (LoI) are lower-bound estimates.

\begin{table*}[tb]
  \centering
  \caption{Classification of LLP scenarios by production mechanism, with representative models and PBC benchmark cases (BC)~\cite{Beacham:2019nyx}, the typical partonic centre-of-mass energy, the main experimental handle at the \anubis detector, and the \anubis sensitivity studies to date.
  Studies that are relevant for more than one type are listed under each of them.}
  \label{tab:phys_categories}
  \begin{tabular}{lllll}
    \toprule
    Type & Representative models & $\sqrt{\hat s}$ & Handle at the \anubis detector & \anubis studies \\
    \midrule
    \tabcell[0.17\textwidth]{(i) Decays of electroweak-scale SM bosons}
      & \tabcell[0.27\textwidth]{exotic Higgs decays: $h\to ss$ in neutral naturalness (BC5), hidden valleys, $h\to A'A'$, $h\to Za$}
      & $\approx m_h$
      & \tabcell[0.2\textwidth]{\met, isolation from jets and tracks; 2DVs}
      & \tabcell[0.17\textwidth]{BC5~\cite{ANUBIS:sensitivity2025}; BC10h~\cite{ANUBIS:ALP2026}} \\
    \tabcell[0.17\textwidth]{(ii) In association with prompt objects}
      & \tabcell[0.27\textwidth]{charged-current HNLs; ALPs in $W^\pm a$, $Za$; LLPs with prompt photons or leptons}
      & $\gtrsim m_W$
      & \tabcell[0.2\textwidth]{\atlas trigger on the prompt object; background suppression}
      & \tabcell[0.17\textwidth]{BC6--8~\cite{Bauer:2026ivt,ANUBIS:tauHNL2026}; photophobic ALPs, BC10, BC11~\cite{ANUBIS:ALP2026}} \\
    \tabcell[0.17\textwidth]{(iii) Decays of heavy BSM states}
      & \tabcell[0.27\textwidth]{$R$-parity-violating, gauge-mediated and split supersymmetry; hidden valleys and dark showers; heavy dark-sector mediators}
      & $\gtrsim1~\TeV$
      & \tabcell[0.2\textwidth]{large $\sqrt{\hat s}$, LLP boosts and multiplicities}
      & \tabcell[0.17\textwidth]{not targeted specifically} \\
    \tabcell[0.17\textwidth]{(iv) Light LLPs from hadron decays}
      & \tabcell[0.27\textwidth]{dark photons (BC1--3), light scalars (BC4), GeV-scale ALPs (BC9--11), HNLs (BC6--8) from $D$/$B$ decays}
      & $\order{\GeV}$
      & \tabcell[0.2\textwidth]{production with prompt EW bosons or leptons; LLPs in jets require dedicated selections}
      & \tabcell[0.17\textwidth]{BC6--8~\cite{Bauer:2026ivt,ANUBIS:tauHNL2026}; BC10h, photophobic ALPs~\cite{ANUBIS:ALP2026}; first estimates~\cite{Wang:2025esc}} \\
    \bottomrule
  \end{tabular}
\end{table*}

\subsection{Strengths and uniqueness of the ANUBIS detector}
\label{sec:phys_strengths}

Several features make the \anubis detector particularly well suited to the scenarios of types~(i) and~(ii):
\begin{itemize}
\item {\em Proximity to the IP and large solid angle.}
With its tracking detectors located about 23~\metre above the beam pipe and covering most of the cavern ceiling, the \anubis detector has a large geometric acceptance for LLPs emitted at central rapidities, which are characteristic of electroweak-scale and heavier mediators.
A convenient figure of merit is the effective path length,
\begin{equation}
\leff \equiv \int_V \mathrm{d}\Omega\,\mathrm{d}l = \int_V \frac{\mathrm{d}^3r}{|\vec r\,|^2}\,,
\label{eq:phys_leff}
\end{equation}
where the integral runs over the decay volume $V$, and $l=|\vec r\,|$ is the distance from the IP along the LLP flight direction.
For an isotropic flux of LLPs with a decay length $\lambda$ that is large compared to the distance to the detector, the probability of a decay inside $V$ is approximately $\leff/(4\pi\lambda)$.
For the \anubis fiducial volume, $\leff=32.3~\metre$, compared with $\leff=22.1~\metre$ for the volume of the \atlas muon spectrometer~\cite{ANUBIS:sensitivity2025}.
\item {\em An air-filled decay volume.}
Another major advantage of \anubis is that its decay volume is mostly filled with air, which suppresses the backgrounds from hadronic interactions of SM particles with material.
With a nuclear interaction length of $\Lambda_{I,\text{air}}=7.5\times10^4~\text{cm}$~\cite{PhysRevD.98.030001}, such interactions are $\order{10^4}$ times less likely than in the \atlas tile calorimeter~\cite{satterthwaite_2022_292b9-eck73}.
\item {\em Reach in $\sqrt{\hat s}$ and in lifetime.}
As a transverse detector, the \anubis detector is sensitive to LLPs produced at the electroweak scale and above, with $c\tau$ from $\order{0.1~\metre}$ up to $\order{10^5~\metre}$, closing the gap between the general-purpose detectors and the Big Bang nucleosynthesis limit for some scenarios~\cite{ANUBIS:sensitivity2025}.
\item {\em A continuous decay volume with the \atlas detector.}
Together with the \atlas detector, the \anubis detector covers decay lengths from a few \mum to tens of metres without a gap (Section~\ref{sec:location}).
It effectively extends the \atlas muon system, providing additional measurements at large distances from the IP that allow for an improved muon reconstruction.
For scenarios with two LLPs per event, the two decays are in general not split between the two detectors; instead, the combined large decay volume of the \atlas and \anubis detectors maximises the probability to reconstruct both LLP decays, providing an effectively background-free two-vertex (2DVs) signature.
\item {\em The full \atlas event.}
The information recorded by the \atlas detector in the same bunch crossing provides \met, isolation and associated prompt objects to reject backgrounds and to characterise a signal, \eg to determine the LLP production mode or the mass of its parent.
Two-way triggering between the \atlas and \anubis detectors gives access to signatures that neither detector could trigger on alone.
\item {\em A mature and cost-effective technology} that has been validated in situ with the \proanubis demonstrator (Section~\ref{sec:proanubis}).
\item {\em A stepping stone towards the FCC.}
The \anubis detector is a proof of concept for the transverse LLP detectors LAYCAST~\cite{laycast} and HECATE~\cite{hecate} proposed for future $e^+e^-$ colliders, and hence an important stepping stone towards the FCC.
Moreover, the \anubis detector can be upcycled in case an FCC detector is deployed in the \atlas cavern.
\end{itemize}
These strengths are mutual: the \anubis detector extends the acceptance of the \atlas experiment for long-lived signatures by orders of magnitude in $c\tau$, while the \atlas detector provides the trigger, the event context and the shielding without which a transverse detector this close to an IP could not operate.

\subsection{Analysis framework and common assumptions}
\label{sec:phys_framework}

All projections in this section are obtained for $\mathcal{L}=3~\invab$ of $pp$ collisions at $\sqrt s=14~\TeV$ at the HL-LHC, using \setanubis, a modular and configuration-driven framework for \anubis sensitivity studies that covers event generation, LLP decays, detector geometry, selection and sensitivity extraction~\cite{erner2026setanubismodularpipelineanubis,Reymermier:SETANUBISv100}.
The BC5 study~\cite{ANUBIS:sensitivity2025} and the electron- and muon-coupled HNL study~\cite{Bauer:2026ivt} used pre-release versions; the tau-coupled HNL and ALP studies used the released \setanubis~v1.0.0~\cite{ANUBIS:tauHNL2026,ANUBIS:ALP2026}, which was confirmed to be fully consistent with the pre-release version in terms of physics results.

A common selection is applied in all studies~\cite{ANUBIS:sensitivity2025,Bauer:2026ivt}.
The LLP must decay inside the fiducial volume of Section~\ref{sec:loc_fiducial}, and at least two charged decay products must each cross both tracking layers.
The background estimate of Section~\ref{sec:backgrounds} is data-driven: it rescales the background observed in the \atlas search for LLPs decaying in the muon spectrometer~\cite{ATLAS:2018tup} to the \anubis detector.
This rescaling is only valid if the event selection mirrors that of the \atlas search, which motivates the following requirements: $\met>30~\GeV$, and isolation of the LLP by $\Delta R>0.5$ from jets with $\pt>15~\GeV$ and from charged particles with $\pt>5~\GeV$.
A vertex-level signal efficiency of $\eps_{\rm sig}=50\%$ is conservatively assumed, below the 62--74\% achieved in the \atlas muon spectrometer~\cite{ATLAS:2018tup}.
The number of expected signal events is
\begin{equation}
N_\text{sig} = \mathcal{L}\,\sigma_\text{LLP}\,\br_\text{LLP}\,\acc\,\eps_\text{sig}\,,
\label{eq:phys_nsig}
\end{equation}
where $\sigma_\text{LLP}$ is the production cross-section, $\br_\text{LLP}$ the branching fraction into the considered final states, and \acc the geometric and kinematic acceptance.

For signatures without an associated prompt object, the backgrounds of Section~\ref{sec:backgrounds} apply.
The exclusion sensitivity at 95\% confidence level (CL) with the CLs method then requires $N_\text{sig}\geq36$ for the conservative background estimate of Eq.~\eqref{eq:cons_bkg_est_value} and $N_\text{sig}\geq17$ for the nominal estimate of Eq.~\eqref{eq:nom_bkg_est_value}; a $5\sigma$ discovery requires 102 and 45 events, respectively~\cite{ANUBIS:sensitivity2025}.
For background-free signatures, such as 2DVs or LLPs accompanied by a prompt lepton or weak boson, $N_\text{sig}\geq4$ is used, following the PBC conventions~\cite{PBC:2025sny}.
The conservative and background-free projections bracket the expected sensitivity of the \anubis detector.

\subsection{Scalar portal: exotic Higgs decays to long-lived scalars}
\label{sec:phys_higgs}

The flagship benchmark is the exotic decay of the 125~GeV Higgs boson into a pair of long-lived scalars, $h\to ss$, with $s\to b\bar b$, which corresponds to BC5 of the PBC initiative~\cite{Beacham:2019nyx} and arises, \eg, in neutral-naturalness models~\cite{Curtin:2013fra,Burdman:2006tz}.
As discussed in Section~\ref{sec:phys_landscape}, it is representative of the broad class of $h\to\text{LLP}$ scenarios of type~(i): it features an electroweak-scale mediator, displaced hadronic jets typical of a broad class of models~\cite{Alimena:2019zri}, and two LLPs per event.
The Higgs boson is produced in gluon--gluon fusion with $\sigma_h=54.7~\pb$~\cite{ANUBIS:sensitivity2025}.

\begin{figure*}[tb]
\centering
\begin{subfigure}[b]{0.42\textwidth}
\includegraphics[width=\textwidth]{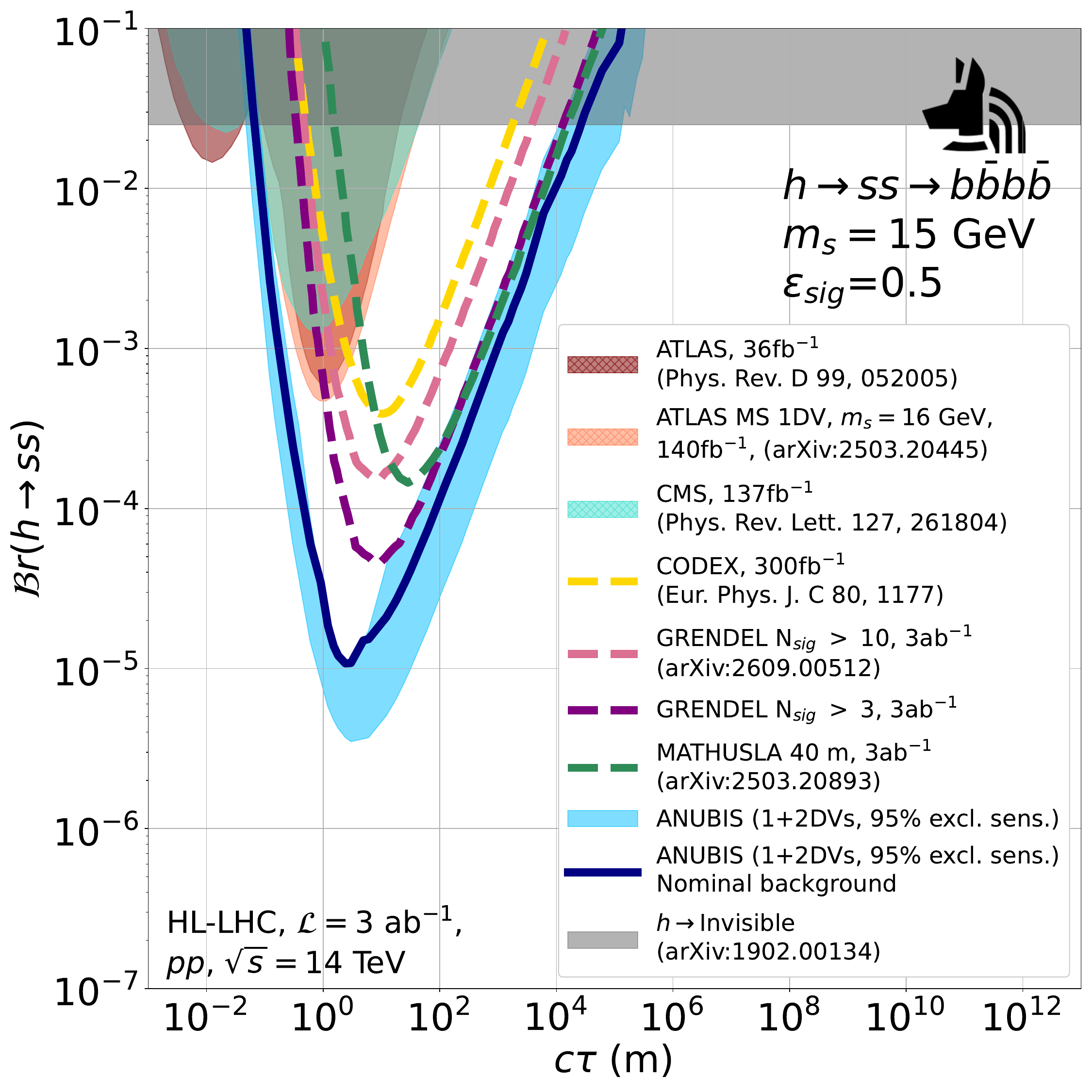}
\caption{}
\end{subfigure}
\qquad
\begin{subfigure}[b]{0.42\textwidth}
\includegraphics[width=\textwidth]{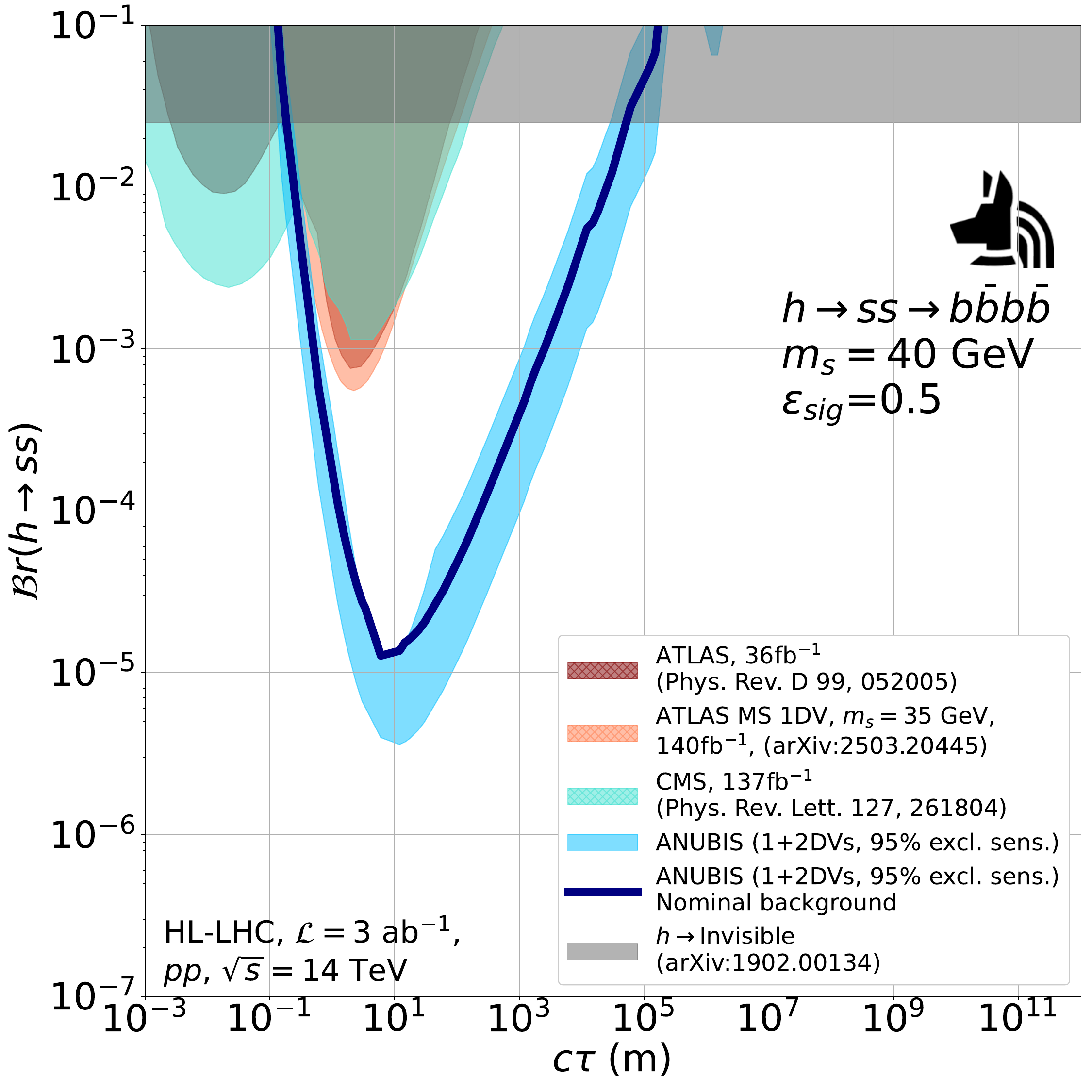}
\caption{}
\end{subfigure}
\caption{
\label{fig:phys_bc5}
The 95\% CL exclusion sensitivity of the \anubis detector to an exotic branching ratio of the 125~GeV Higgs boson into a pair of long-lived scalars $s$ with masses of (a)~15~GeV and (b)~40~GeV using 3~\invab of $pp$ collisions at the HL-LHC~\cite{ANUBIS:sensitivity2025}.
The combined 1DV and 2DVs projections for the \anubis detector are shown as a band, the lower (upper) edge being the background-free (conservative) limit; the nominal background scenario for the 1DV selection is shown as a solid line.
For comparison, the projections from \codex~\cite{CODEX-b:2019jve}, GRENDEL~\cite{citron2026}, and \mathusla{}40~\cite{MATHUSLA:2025zyt} at the HL-LHC, the projected HL-LHC sensitivity from $h\to\text{invisible}$ decays~\cite{Cepeda:2019klc}, and the latest results from the \atlas~\cite{ATLAS:2018tup,ATLAS:2025pak} and \cms~\cite{CMS:2021juv} experiments are shown.
The GRENDEL projections assume an active volume reflecting the rounded cross-section of the drainage gallery and a detection efficiency of 100\%. 
}
\end{figure*}

The resulting sensitivity to $\br(h\to ss)$ is shown in Figure~\ref{fig:phys_bc5} for $m_s=15$ and 40~GeV, combining the single displaced vertex (1DV) and 2DVs selections, together with projections for \codex~\cite{CODEX-b:2019jve} and \mathusla{}40~\cite{MATHUSLA:2025zyt}, the latest results of the \atlas~\cite{ATLAS:2018tup,ATLAS:2025pak} and \cms~\cite{CMS:2021juv} experiments, and the HL-LHC projection for $h\to$~invisible~\cite{Cepeda:2019klc}.
The \anubis detector can probe $\br(h\to ss)$ down to $\order{10^{-5}}$ and $\order{10^{-6}}$ in the nominal and background-free projections, respectively, \eg $1.1\times10^{-5}$ ($3.5\times10^{-6}$) for $m_s=15~\GeV$ and $c\tau=3~\metre$~\cite{ANUBIS:sensitivity2025}.
At its peak sensitivity, this is an improvement by two to three orders of magnitude on the latest \atlas and \cms results.
For $\br(h\to ss)=10^{-3}$ and $m_s=15~\GeV$, the \anubis detector covers $0.17<c\tau<960~\metre$ ($0.11<c\tau<4000~\metre$) in the nominal (background-free) projections, extending the $c\tau$ reach of the \atlas and \cms experiments by three (four) orders of magnitude.
For $\br(h\to ss)=10^{-2}$, the $c\tau$ reach of $10^4~\metre$ ($4\times10^4~\metre$) approaches the limit from Big Bang nucleosynthesis~\cite{Fradette:2017sdd} to within about three orders of magnitude.

\begin{figure}[tb]
\centering
\includegraphics[width=0.95\linewidth]{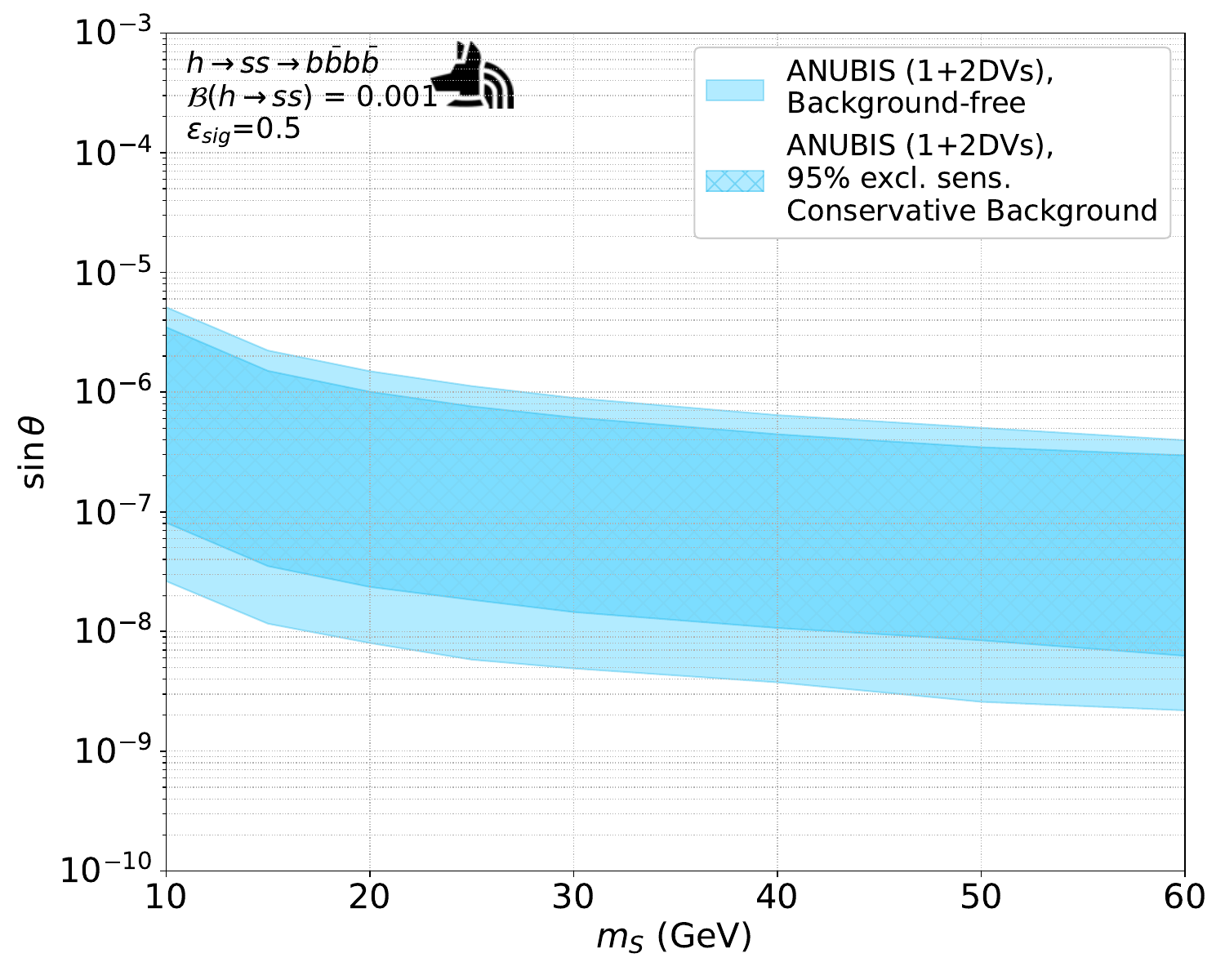}
\caption{
\label{fig:phys_bc5_sintheta}
The combined 1DV and 2DVs 95\% CL exclusion $\sin\theta$ reach of the \anubis detector for $h\to ss$, $s\to b\bar b$, using 3~\invab at the HL-LHC, for $\br(h\to ss)=10^{-3}$, \ie one order of magnitude below the value of the BC5 benchmark~\cite{Beacham:2019nyx,ANUBIS:sensitivity2025}.
Conservative (hatched) and background-free (filled) projections bracket the expected sensitivity.}
\end{figure}

Among the transverse experiments, the nominal \anubis projections surpass those of \codex by more than one order of magnitude in $\br(h\to ss)$ across the whole $c\tau$ range for $m_s=15~\GeV$.
They also surpass those of \mathusla{}40 across the whole $c\tau$ range, by more than two orders of magnitude for $c\tau\lesssim3~\metre$~\cite{ANUBIS:sensitivity2025}.
Translated into the BC5 parameter space, the mixing angle $\theta$ between $s$ and the Higgs boson is probed over a wide range of $m_s$~\cite{ANUBIS:sensitivity2025}.
To better represent the reach of the \anubis detector, Figure~\ref{fig:phys_bc5_sintheta} shows the $\sin\theta$ reach for $\br(h\to ss)=10^{-3}$, \ie one order of magnitude below the BC5 value of $10^{-2}$; despite this reduction, the reach in $\sin\theta$ degrades by only about half an order of magnitude~\cite{ANUBIS:sensitivity2025}.
Owing to the common kinematic features of $h\to\text{LLP}$ decays, comparable sensitivities to the exotic branching ratio are expected for other scenarios of type~(i) with similar LLP masses, lifetimes and decay multiplicities.

\subsection{Neutrino portal: heavy neutral leptons}
\label{sec:phys_hnl}

Heavy neutral leptons $N_1$ provide a minimal extension of the neutrino sector that can generate neutrino masses through a seesaw mechanism; for small mixing $|V_{1\ell}|^2$ with the active neutrinos they are naturally long-lived~\cite{Bondarenko:2018ptm,Antel:2023hkf}.
The PBC benchmarks BC6, BC7 and BC8 consider a single Majorana HNL coupled exclusively to $e$, $\mu$ or $\tau$, respectively~\cite{Beacham:2019nyx}.
HNLs can be produced both in hadron decays and in electroweak processes.
The \anubis detector targets the latter, \ie charged- and neutral-current Drell--Yan production, $W^{\pm(*)}\to\ell N_1$ and $Z^{(*)}\to\nu N_1$, and $W^\pm\gamma$ fusion~\cite{Bauer:2026ivt}, which correspond to type~(ii) and extend to masses well above those accessible in $B$-meson decays.
These processes are accompanied by a prompt lepton or by \met but not by nearby hadronic activity, and the search is taken to be background-free.

\begin{figure*}[tb]
\centering
\begin{subfigure}[b]{0.48\textwidth}
\includegraphics[width=\textwidth]{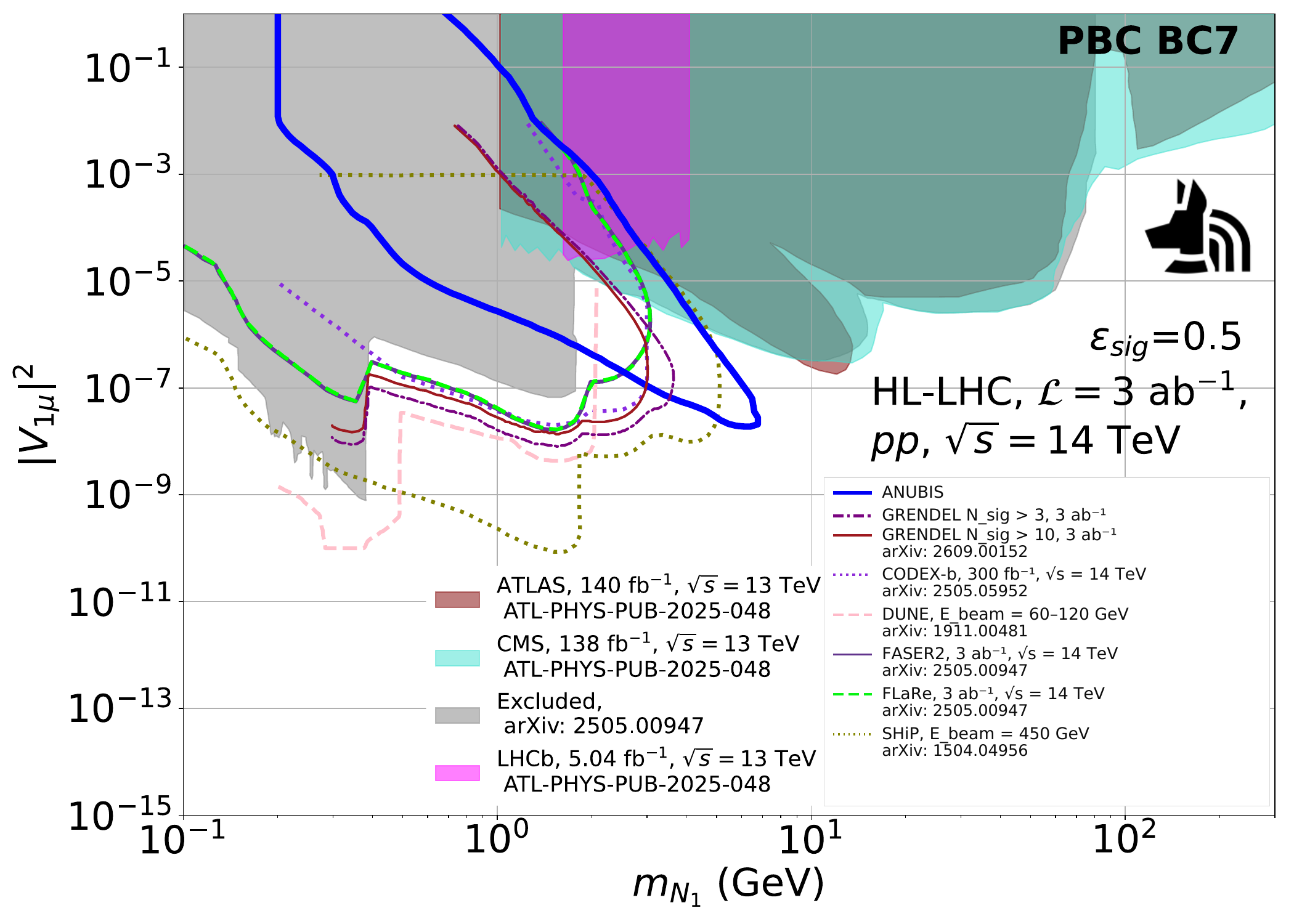}
\caption{}
\end{subfigure}
\hfill
\begin{subfigure}[b]{0.48\textwidth}
\includegraphics[width=\textwidth]{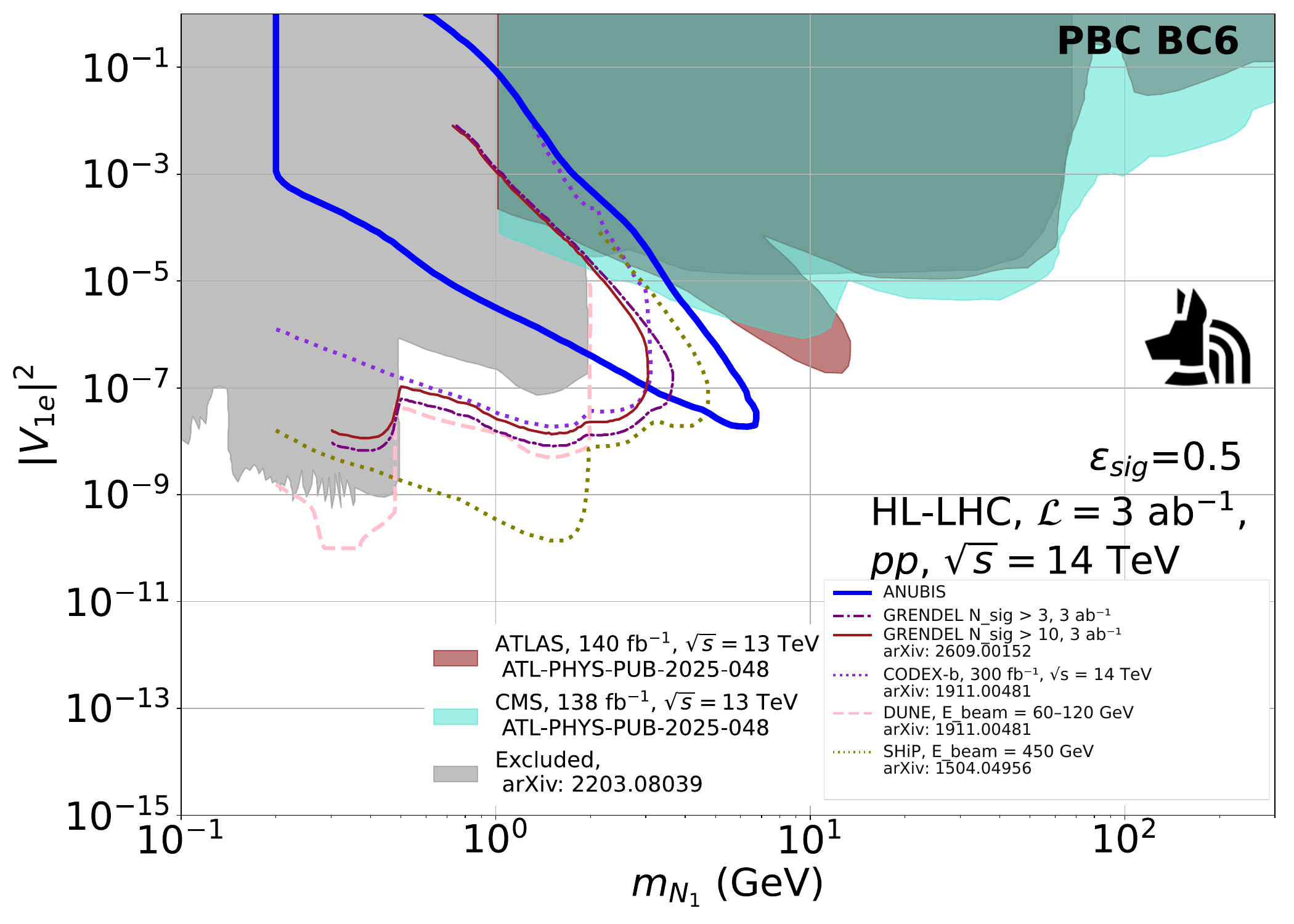}
\caption{}
\end{subfigure}
\caption{
\label{fig:phys_hnl_emu}
Projected 95\% CL exclusion sensitivity of the \anubis detector to a minimal Majorana HNL with exclusive (a)~$|V_{1e}|$ and (b)~$|V_{1\mu}|$ couplings, corresponding to the BC6 and BC7 PBC benchmarks, using 3~\invab at the HL-LHC~\cite{Bauer:2026ivt}.
Also shown are projected sensitivities and excluded regions from previous searches~\cite{ATL-PHYS-PUB-2025-048}, exclusion sensitivity projections summarised in Ref.~\cite{PBC:2025sny} as well as from the \codex~\cite{CODEX-b:2019jve} and GRENDEL experiments~\cite{citron2026}.
The GRENDEL projections assume an active volume reflecting the rounded cross-section of the drainage gallery and a detection efficiency of 100\%. 
}
\end{figure*}

The projected exclusion sensitivities for BC6 and BC7 are shown in Figure~\ref{fig:phys_hnl_emu}.
The \anubis detector extends the reach of existing searches for $3\lesssim m_{N_1}\lesssim8~\GeV$ and $10^{-8}\lesssim|V_{1\ell}|^2\lesssim10^{-5}$, reaching a maximum sensitivity of $|V_{1e}|^2=1.8\times10^{-8}$ at $m_{N_1}=6.4~\GeV$ and $|V_{1\mu}|^2=1.9\times10^{-8}$ at $m_{N_1}=6.3~\GeV$~\cite{Bauer:2026ivt}.
Part of this region may also be reached by \codex~\cite{CODEX-b:2019jve} and SHiP~\cite{SHiP:2021nfo}, but the \anubis detector provides unique sensitivity for $m_{N_1}\gtrsim5~\GeV$.
It is complementary to SHiP and FASER2, which dominate for $m_{N_1}<1.6~\GeV$ through HNL production in hadron decays.

\begin{figure}[tb]
\centering
\includegraphics[width=0.95\linewidth]{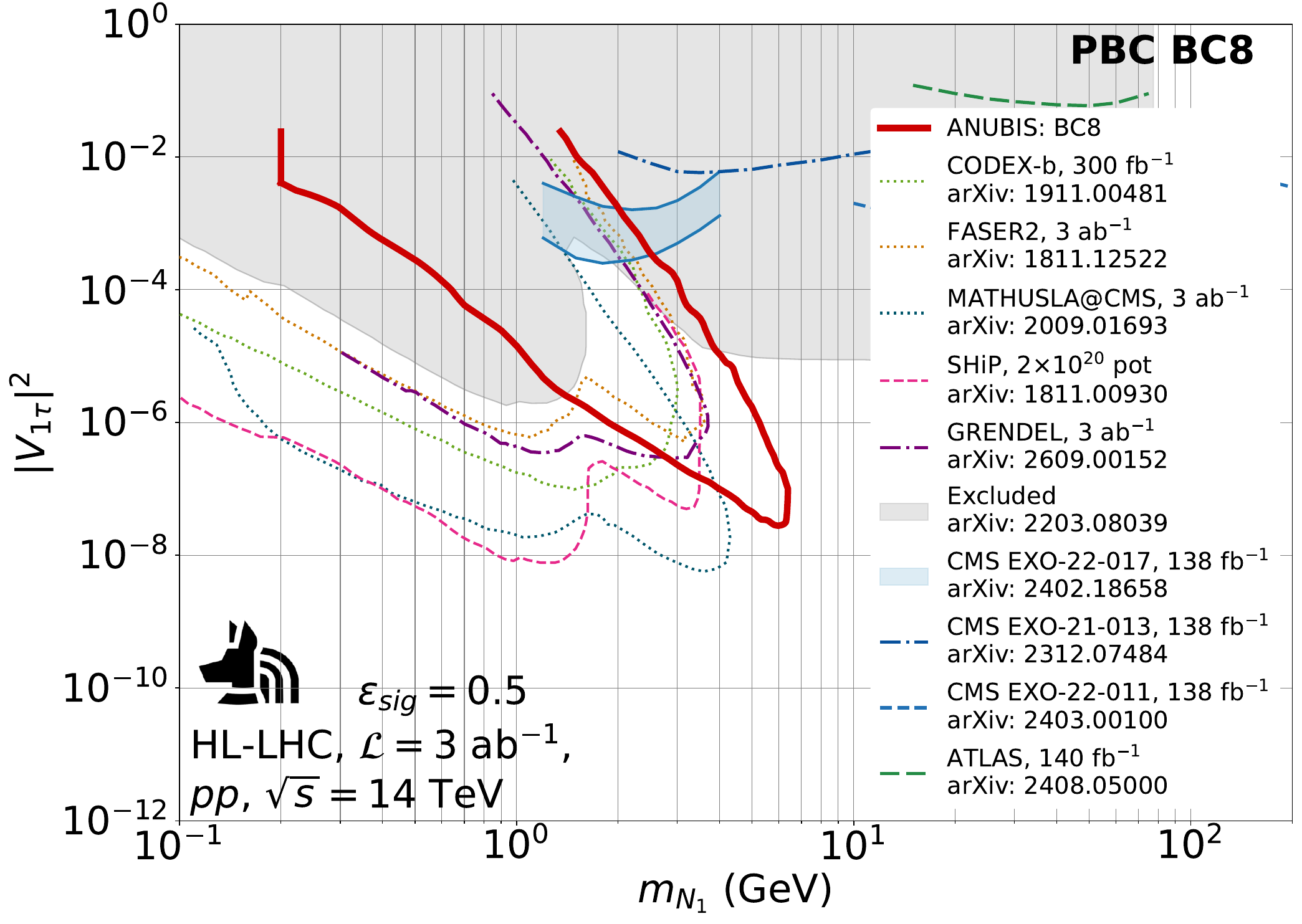}
\caption{
\label{fig:phys_hnl_tau}
Projected 95\% CL exclusion sensitivity of the \anubis detector to a minimal Majorana HNL with exclusive $|V_{1\tau}|$ couplings, corresponding to the BC8 PBC benchmark, using 3~\invab at the HL-LHC~\cite{ANUBIS:tauHNL2026}.
Also shown are projected sensitivities and excluded regions from previous searches~\cite{ATL-PHYS-PUB-2025-048}, exclusion sensitivity projections summarised in Ref.~\cite{PBC:2025sny} as well as from the \codex~\cite{CODEX-b:2019jve} and GRENDEL experiments~\cite{citron2026}.
The GRENDEL projections assume an active volume reflecting the rounded cross-section of the drainage gallery and a detection efficiency of 100\%.}
\end{figure}

The tau-coupled BC8 scenario cannot be obtained by rescaling the electron- and muon-coupled results, since the $\tau$ lepton is itself unstable and charged-current HNL decays become available only above the $\tau$ production threshold~\cite{ANUBIS:tauHNL2026}.
As shown in Figure~\ref{fig:phys_hnl_tau}, the \anubis detector extends the reach of existing searches over $2\lesssim m_{N_1}\lesssim7~\GeV$ and $10^{-8}\lesssim|V_{1\tau}|^2\lesssim10^{-5}$.
For $m_{N_1}\gtrsim4~\GeV$, it is the only proposal considered with sensitivity to small couplings, reaching $|V_{1\tau}|^2=2.5\times10^{-8}$ at $m_{N_1}=6.2~\GeV$~\cite{ANUBIS:tauHNL2026}.
Since the existing limits on $|V_{1\tau}|^2$ are considerably weaker than those on $|V_{1e}|^2$ and $|V_{1\mu}|^2$, this region is particularly well motivated for a dedicated search.
For all three flavours, the sensitivity at lower masses, where HNLs are produced predominantly in $D$- and $B$-meson decays, is currently limited by the isolation requirements, and the projections in this region are therefore lower bounds on the full sensitivity of the \anubis detector, cf.~Section~\ref{sec:phys_further}.

\subsection{Axion portal: axion-like particles}
\label{sec:phys_alp}

ALPs are pseudo-Nambu--Goldstone bosons whose couplings are suppressed by a high scale and can therefore be long-lived on detector scales~\cite{Bauer:2017ris}.
The ALP scenarios studied for the \anubis detector exploit production modes at the electroweak scale: in association with a weak boson $V=W^\pm,Z$, \ie $pp\to Va$, in which the leptonic decay of the boson provides a trigger for the \atlas experiment and renders the search effectively background-free, and in exotic Higgs-boson decays~\cite{ANUBIS:ALP2026}.

\begin{figure}[tb]
\centering
\includegraphics[width=0.95\linewidth]{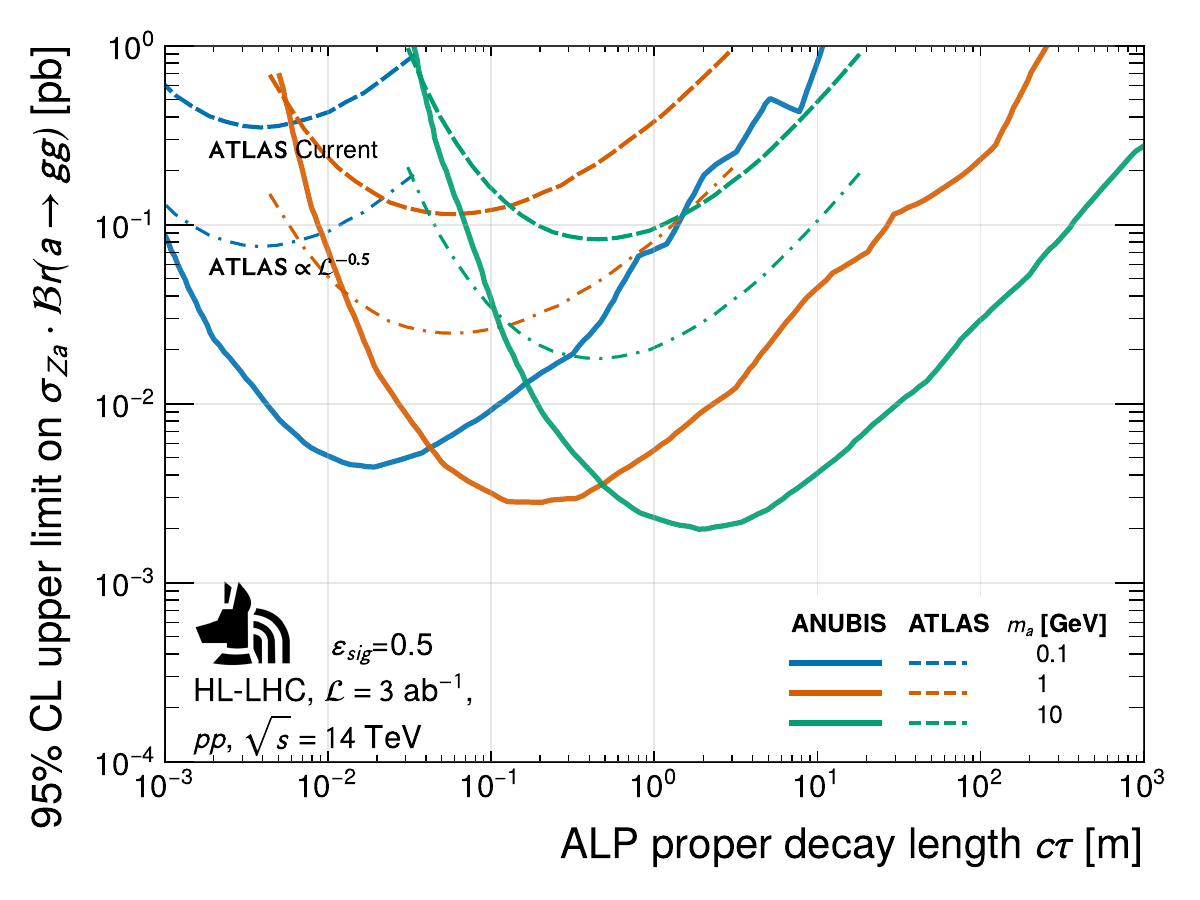}
\caption{
\label{fig:phys_alp_photophobic}
Projected 95\% CL exclusion sensitivity of the \anubis detector to $\sigma_{Za}\cdot\br(a\to gg)$ as a function of $c\tau$ for the photophobic ALP model with $Za$ production, using 3~\invab at the HL-LHC~\cite{ANUBIS:ALP2026}.
The existing \atlas limits~\cite{ATLAS:2025pak} are shown for reference, together with their extrapolation to the HL-LHC luminosity.}
\end{figure}

In the {\em photophobic} model, the ALP couples to gluons with suppressed photon couplings, a benchmark widely adopted by the \atlas Collaboration~\cite{Craig:2018kne,ATLAS:2025pak}.
The \anubis detector reaches a peak sensitivity to the gluon coupling of $C_{\tilde G}=1.4\times10^{-10}$ at $m_a=71.4~\GeV$.
In terms of the production cross-section times branching ratio, the sensitivity peaks at $\sigma_{Va}\cdot\br(a\to gg)=\order{10^{-3}}~\pb$ for all ALP masses studied, \eg $6.2\times10^{-4}~\pb$ at $c\tau=1.95~\metre$ for $m_a=10~\GeV$ in $W^\pm a$ production~\cite{ANUBIS:ALP2026}.
As shown in Figure~\ref{fig:phys_alp_photophobic}, this improves on the current \atlas limits~\cite{ATLAS:2025pak} by two orders of magnitude, and on their luminosity-scaled HL-LHC extrapolation by one, while extending the $c\tau$ coverage by about three orders of magnitude.

\begin{figure}[tb]
\centering
\includegraphics[width=0.95\linewidth]{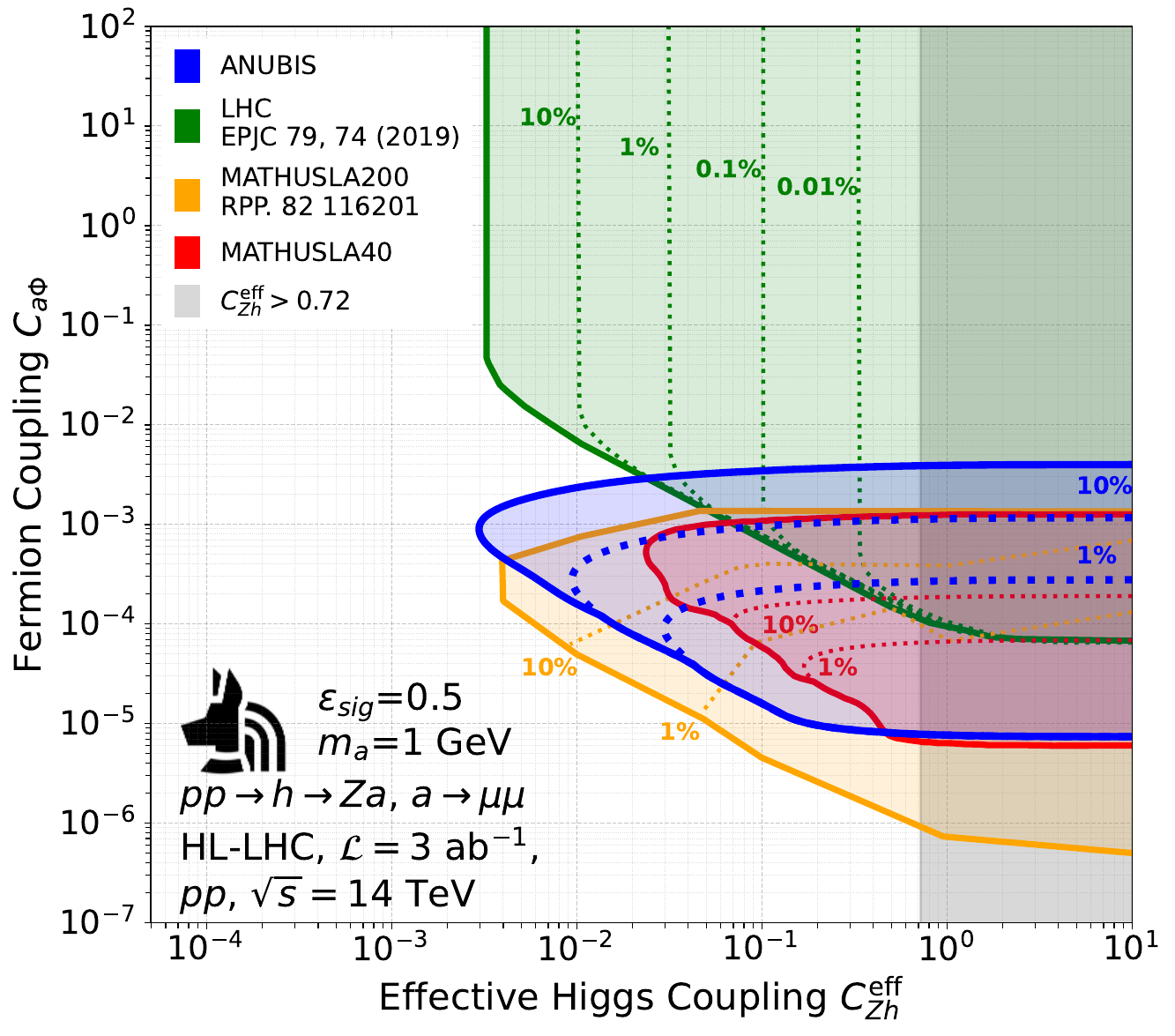}
\caption{
\label{fig:phys_alp_bc10h}
Projected 95\% CL exclusion sensitivity of the \anubis detector in the BC10h model as a function of $C_{Zh}^{\rm eff}$ and $C_{a\Phi}$ for $m_a=1~\GeV$ and $a\to\mu^+\mu^-$, using 3~\invab at the HL-LHC~\cite{ANUBIS:ALP2026}.
The \anubis contour is compared with the projected limits for the main LHC detectors~\cite{Bauer:2018uxu}, legacy \mathusla{}200 projections~\cite{Curtin:2018mvb}, and an approximate \mathusla{}40 contour.
Solid contours assume a unit muonic branching fraction; decreasing branching fractions are indicated by dotted curves.}
\end{figure}

For the fermion-coupled BC10 benchmark, the $pp\to Va$ production rate yields at most $\order{1}$ selected event with 3~\invab, and no sensitivity was found for the gluon-coupled BC11 benchmark in $Va$ final states~\cite{ANUBIS:ALP2026}.
The situation changes in the `BC10h' extension, in which the large Higgs-boson sample at the HL-LHC is exploited through the exotic decay $h\to Za$, governed by an effective Higgs coefficient $C_{Zh}^{\rm eff}$, while the ALP lifetime is controlled by its fermion coupling $C_{a\Phi}$~\cite{Bauer:2017ris}.
For $C_{Zh}^{\rm eff}=0.72$, corresponding to the legacy global constraint $\br(h\to\text{BSM})<0.34$, the \anubis detector reaches $C_{a\Phi}=4.4\times10^{-8}$ for $16\lesssim m_a\lesssim23~\GeV$.
Figure~\ref{fig:phys_alp_bc10h} compares the reach in the $(C_{Zh}^{\rm eff},C_{a\Phi})$ plane for $m_a=1~\GeV$ and $a\to\mu^+\mu^-$ with the projections for the main LHC detectors~\cite{Bauer:2018uxu} and for \mathusla{}~\cite{Curtin:2018mvb}.
The \anubis detector covers a significant region that cannot be probed by the main LHC experiments, reaching fermion couplings about forty times smaller, and the approximate \mathusla{}40 contour, obtained with the same \setanubis samples, is entirely enclosed in that of the \anubis detector~\cite{ANUBIS:ALP2026}.
This example illustrates how GeV-scale LLPs, which are otherwise the domain of type~(iv) experiments, can be probed by the \anubis detector when they are produced at the electroweak scale.

\subsection{Further scenarios and improvements}
\label{sec:phys_further}

The studies above cover benchmarks of types~(i) and~(ii), and partly of type~(iv), with deliberately conservative assumptions.
Several directions are expected to extend the discovery potential of the \anubis detector further:
\begin{itemize}
\item {\em Hadronic production modes.}
The isolation requirements, which are tied to the data-driven background estimate, reject LLPs produced inside or near jets, such as HNLs or ALPs from $D$- and $B$-meson decays.
Dedicated selections without these requirements, exploiting multivariate techniques and the detailed event information from the \atlas and \anubis detectors, are expected to recover a significant fraction of these modes, which would unlock sensitivity to, \eg, HNLs with $m_{N_1}<3~\GeV$~\cite{Bauer:2026ivt}; first estimates without background treatment indicate a substantial gain~\cite{Wang:2025esc}.
Until \anubis is realised, such selections will require background estimation techniques based on Monte Carlo (MC) simulations, for which the \proanubis data provide a first testing ground (Section~\ref{sec:proanubis}).
\item {\em The vector portal and the remaining PBC benchmarks}, in particular dark photons produced in exotic Higgs-boson decays~\cite{Curtin:2014cca}, BC4, and the ALP benchmarks with alternative production modes, which can be evaluated with \setanubis in the same framework.
\item {\em Heavy BSM states and dark showers} (type~(iii)).
Although these scenarios are not targeted specifically, the large LLP boosts and multiplicities, and the possibility of several DVs per event, play to the strengths of the \anubis detector and of its integration with the \atlas experiment, and can extend the $c\tau$ reach of the \atlas searches.
\item {\em Analysis improvements}, \eg exploiting the higher track multiplicity of $\tau$ and $b$-hadron decays, which justifies a higher signal efficiency than the conservative $\eps_{\rm sig}=50\%$, and refining the background estimate with \proanubis data (Section~\ref{sec:proanubis}).
\end{itemize}

\subsection{Comparison and complementarity with other proposals}
\label{sec:phys_comparison}

Table~\ref{tab:comparison} summarises the dedicated LLP experiments that are operating, approved, or proposed for the HL-LHC era.
They fall into two classes.
Forward and beam-dump experiments, such as FASER~\cite{FASER:2018eoc} and SHiP~\cite{SHiP:2021nfo,SHiP:ECN3}, probe light LLPs from hadron decays at $\sqrt{\hat s}=\order{\GeV}$, \ie type~(iv) of Section~\ref{sec:phys_landscape}, with excellent sensitivity at very small couplings.
Transverse experiments, \ie the \anubis detector, \codex~\cite{CODEX-b:2019jve}, \mathusla~\cite{Curtin:2018mvb,MATHUSLA:2025zyt} and GRENDEL~\cite{citron2026}, target LLPs produced at the electroweak scale and above, \ie types~(i), (ii), and~(iii).
The two classes are therefore complementary, as illustrated for HNLs in Figures~\ref{fig:phys_hnl_emu} and~\ref{fig:phys_hnl_tau}: SHiP and FASER2 dominate for $m_{N_1}<1.6~\GeV$, while the \anubis detector provides the leading reach at higher masses.

\begin{table*}[tb]
  \centering
  \caption{Overview of the \anubis detector and comparable dedicated LLP experiments at CERN.
  The distances are approximate.
  }
  \label{tab:comparison}
  \begin{tabular}{lllll}
    \toprule
    Experiment & Location & Type & Distance from IP & Status \\
    \midrule
    \anubis{}~\cite{ANUBIS:sensitivity2025}
      & \tabcell[0.3\textwidth]{ceiling of the \atlas cavern UX15 and PX14/PX16 shaft plugs, LHC Point~1}
      & transverse & 10--23~\metre
      & proposal (this LoI) \\
    \codex~\cite{CODEX-b:2019jve}
      & \tabcell[0.3\textwidth]{LHCb cavern behind the shielding wall, LHC Point~8}
      & transverse & $\sim25~\metre$
      & \tabcell[0.17\textwidth]{expression of interest} \\
    GRENDEL~\cite{citron2026}
      & \tabcell[0.3\textwidth]{PX56 drainage gallery above CMS, LHC Point~5}
      & transverse & 24--55~\metre
      & proposal \\
    \mathusla{}40~\cite{MATHUSLA:2025zyt}
      & \tabcell[0.3\textwidth]{surface above CMS, LHC Point~5}
      & transverse & $\order{100~\metre}$
      & \tabcell[0.17\textwidth]{conceptual design report} \\
    \midrule
    FASER~\cite{FASER:2018eoc}
      & \tabcell[0.3\textwidth]{TI12 tunnel along the beam axis, downstream of \atlas}
      & forward & $\sim480~\metre$
      & operating \\
    SHiP~\cite{SHiP:2021nfo,SHiP:ECN3}
      & \tabcell[0.3\textwidth]{ECN3 beam-dump facility at the SPS}
      & beam dump & --
      & approved \\
    \bottomrule
  \end{tabular}
\end{table*}

Among the transverse experiments, the \anubis detector combines a short distance to the IP, and hence a large solid angle of 2.9~sr, with the full integrated luminosity of 3~\invab delivered to the \atlas experiment and with the event information of the \atlas detector.
\codex is also close to its IP but has a smaller decay volume and relies on the lower integrated luminosity delivered to LHCb~\cite{CODEX-b:2019jve}.
GRENDEL benefits from the shielding by at least 7~\metre of rock and from the large integrated luminosity delivered to CMS, but is limited by the small and irregularly shaped cross section of the drainage gallery, which requires a high spatial resolution to reconstruct LLP decays~\cite{citron2026}.
\mathusla, located at the surface, benefits from the rock overburden as passive shielding relative to the IP of CMS, but has no shielding against cosmic rays and subtends a moderate solid angle.

As a result, the \anubis detector provides the best projected sensitivity among the transverse proposals for the BC5 benchmark over most of the $c\tau$ range (Figure~\ref{fig:phys_bc5}), and uniquely covers HNLs with $m_{N_1}\gtrsim5~\GeV$ and small couplings for all three lepton flavours (Figures~\ref{fig:phys_hnl_emu} and~\ref{fig:phys_hnl_tau}).
In the BC10h plane, the original \mathusla{}200 design reaches smaller effective Higgs couplings for part of the long-lifetime band because of its large surface decay volume~\cite{Curtin:2018mvb,ANUBIS:ALP2026}, which illustrates that the transverse proposals are themselves partly complementary in $c\tau$.
A distinctive advantage of the \anubis detector is the continuous decay volume with the \atlas detector, which provides seamless coverage from prompt decays to $c\tau\approx10^5~\metre$ and allows a discovery to be characterised with the full event.

In summary, the projections for the scalar, neutrino and axion portals, \ie for exotic Higgs-boson decays, HNLs of all three flavours, and ALPs, consistently show that the \anubis detector would extend the reach of the HL-LHC by up to two to three orders of magnitude in coupling or branching ratio, and by up to three to four orders of magnitude in $c\tau$, in a region of parameter space that cannot be covered by forward, beam-dump or general-purpose experiments.
This makes the \anubis detector a strong and complementary component of the experimental programme for LLP searches at the HL-LHC.

%% file: sections/backgrounds.tex
The signal primarily targeted by the \anubis detector consists of DVs produced in the fiducial volume between the \atlas detector and the \anubis tracking layers (Section~\ref{sec:loc_fiducial}).
The following background processes can also produce DVs in the \anubis detector:

\begin{itemize}
    \item cosmic rays;
    \item beam-induced background processes;
    \item decays of quasi-thermal neutrons;
    \item random coincidences of unrelated tracks;
    \item decays and material interactions of SM particles with macroscopic lifetimes, referred to as SM LLPs.
\end{itemize}

Most of these processes can be vetoed or strongly reduced thanks to the capabilities of the \anubis detector and its integration with the \atlas experiment, so that the remaining background is dominated by material interactions of SM LLPs and can be estimated with data-driven methods (Section~\ref{subsec:cons_bkg_est}).
Searches for two or more DVs in the same event are expected to be essentially background-free.

\subsection{Non-collision backgrounds}
\label{subsec:noncol_bkg}

Downward-going cosmic-ray tracks can be rejected thanks to the time-of-flight measurement between the tracking layers.
For two tracking layers about 1~\metre apart and a time resolution of $\delta t = 0.5~\unit{ns}$ (Table~\ref{tab:requirements}), downward-going tracks are separated from the signal acceptance window by about $7\sigma$, as outlined in Section~\ref{sec:performance}; a more detailed estimate, using all hits of each track, is given in Section~\ref{subsec:trig_offline}.
Upward-going tracks from cosmic-ray interactions in the air below the detector are strongly suppressed by the low interaction probability in air, by the need to produce at least two upward-going tracks to form a vertex, and by the detection of the incoming downward-going particle that initiates the interaction~\cite{bib:anubis_orig} (Section~\ref{subsec:trig_cosmicveto}).

Beam-induced backgrounds, such as beam--collimator or beam--gas interactions, are negligible given the position of the \anubis detector, whose fiducial volume starts 9.5~\metre from the beamline.
Moreover, these interactions have a characteristic topology, with large boosts along the beamline, which provides strong discrimination~\cite{bib:anubis_orig}.

Backgrounds from the decays of quasi-thermal neutrons~\cite{Hedberg:806274} are negligible, since their decay products have momenta of at most about 1~\MeV, which is insufficient to produce signals in both tracking layers~\cite{bib:anubis_orig}.

\subsection{Instrumental backgrounds}
\label{subsec:instrum_bkg}

The main instrumental background is due to random coincidences of unrelated particle tracks~\cite{bib:anubis_orig}.
Its contribution scales with the density of tracks in space and time, which decreases rapidly with the distance from the beamline.
For the proposed location of the \anubis detector, more than 20~\metre from the beamline, this contribution is therefore negligible.
Random coincidences of unrelated hits at the trigger level are discussed in Section~\ref{subsec:trig_rates}.

\subsection{Collision backgrounds}
\label{subsec:col_bkg}

The dominant background for the \anubis detector arises from $pp$ collisions, in particular from electrically neutral SM LLPs.
The only two such particles with a sufficiently long lifetime to contribute are the long-lived neutral kaon, $K_L$, and the neutron, $n$, with proper lifetimes of $c \tau = 15.3~\metre$ and $2.6 \times 10^{11}~\metre$, respectively.
These SM LLPs can contribute to the background either by decaying into charged particles, or by producing hadrons in material interactions~\cite{bib:anubis_orig, satterthwaite_2022_292b9-eck73}.

Given the long lifetime of the neutron, only $K_L$ contribute significantly to the background through decays into charged particles.
Apart from a few rare decay modes, $K_L$ decays produce at most two charged tracks.
Moreover, $K_L$ are usually strongly boosted, with a Lorentz factor of $\gamma \gtrsim 20$, which results in vertices with small opening angles of $\mathcal{O}\left(1^\circ \right)$.
BSM LLPs tend to be more massive, which results in larger opening angles and provides a useful handle to reduce this background.
Furthermore, most of the $K_L$ escaping the \atlas detector do not decay in the \anubis fiducial volume because of their large $\gamma$~\cite{bib:anubis_orig, satterthwaite_2022_292b9-eck73}.

Material interactions of SM LLPs are the dominant background contribution.
They produce vertices with wide opening angles, which mimic the expected signal from BSM LLPs.
The rate of material interactions is directly proportional to the density of the material traversed by the SM LLPs.
The large nuclear interaction length of air, $\Lambda_{I,\text{air}} = 7.5 \times 10^4~\unit{cm}$~\cite{PhysRevD.98.030001}, significantly reduces the number of material interactions, which are $\mathcal{O} \left( 10^4\right)$ times less likely than in the \atlas tile calorimeter~\cite{satterthwaite_2022_292b9-eck73}.

The main method to reduce the collision backgrounds is to use the \atlas detector as both a passive and an active veto.
Given its high material density, the \atlas detector acting as a passive veto reduces the background from SM LLPs by $\mathcal{O} \left( 10^5 \right)$.
Its material provides a total shielding equivalent to $11.2~\Lambda_I$ at $\eta = 0$ and up to $13.1~\Lambda_I$ at $\left| \eta \right| = 0.7$, which reduces the SM LLP flux to $1.4 \times 10^{-5}$ and $2.0 \times 10^{-6}$ of its original value, respectively~\cite{satterthwaite_2022_292b9-eck73}.

However, given the total inelastic cross section of about 80~mb~\cite{bib:xsectotem, bib:xseclhcb, bib:xsecalfa}, the \anubis detector is still exposed to a significant flux of SM LLPs at the HL-LHC.
Thanks to the synchronisation of the \anubis and \atlas events, \atlas event-level information can be used for kinematic and topological selections that further reduce these backgrounds.
The chosen selections are inspired by the \atlas search for LLPs with decay lengths of $\mathcal{O} \left( 3~\metre \right)$ using 36~\invfb of $pp$ collisions at $\sqrt{s} = 13~\TeV$~\cite{ATLAS:2018tup}.
First, a missing transverse momentum of $E_T^{\text{miss}} > 30~\GeV$ is required, since an LLP decaying in the \anubis fiducial volume would contribute to the transverse momentum imbalance measured by the \atlas detector.
Second, since hadrons such as $n$ and $K_L$ are produced in jets, the SM LLP contribution is reduced by requiring the vertices in the \anubis detector to be geometrically isolated from jets and from prompt tracks with high transverse momentum.
In particular, an isolation of $\Delta R \equiv \sqrt{ \left( \Delta \phi \right) ^2 + \left( \Delta \eta \right) ^2} \geq 0.5$ is required between the \anubis vertices and \atlas jets with transverse momentum $p_T > 15~\GeV$, and prompt tracks with $p_T > 5~\GeV$.
The isolation from high-$p_T$ tracks ensures that SM LLPs originating from jets with few high-$p_T$ charged particles are also vetoed~\cite{bib:anubis_orig}.

\subsection{Data-driven background estimates}
\label{subsec:cons_bkg_est}

Since material interactions, the dominant source of background, are difficult to model with MC simulations, a data-driven approach is used to estimate the background~\cite{satterthwaite_2022_292b9-eck73}.

Given the similarity of the targeted signal, backgrounds, analysis strategy and technology, the \atlas search that inspired the active-veto selections~\cite{ATLAS:2018tup} can be regarded as an analogous scenario to the \anubis detector~\cite{bib:anubis_orig}.
Its background estimate for the $E_T^{\text{miss}}$ + DV channel is therefore scaled to obtain a conservative data-driven background estimate for the \anubis detector.
In the original analysis, $N_\text{obs} = 224$ events were observed in the barrel, which was compatible with the background estimate of $N_{\text{bgr,ATLAS}}= 243 \pm 38 \text{ (stat) } \pm 29 \text{ (syst)}$, obtained with the ABCD method, which uses uncorrelated sidebands.

The rescaling of the \atlas search to the \anubis conditions is performed as

\begin{equation}
    \begin{split}
    N_{\text{bgr,ANUBIS}} = {} & \dfrac{\mathcal{L}}{\mathcal{L}^\prime} \times \dfrac{\Lambda_{I,\text{MS}}}{\Lambda_{I,\text{air}}} \times \dfrac{\varepsilon}{\varepsilon^\prime} \\
    & \times \dfrac{L_{\text{eff}}}{L_{\text{eff}}^\prime} \times f_{\text{bgr,MS}} \times N_{\text{bgr,ATLAS}},
    \end{split}
    \label{eq:cons_bkg_est}
\end{equation}

\noindent where $\mathcal{L}$ is the integrated luminosity, $\varepsilon$ is the reconstruction efficiency, $L_{\text{eff}}$ is the effective path length and $f_{\text{bgr,MS}}$ is the fraction of \atlas background vertices in the muon spectrometer (MS); primed quantities refer to the \atlas search and unprimed ones to the \anubis detector~\cite{satterthwaite_2022_292b9-eck73}.

The difference in volume, $V$, between the \atlas MS and the \anubis fiducial volume is accounted for in the effective path length, $L_{\text{eff}} \equiv \int_V d \Omega\, dl$, where $\Omega$ is the solid angle and $l$ is the path length, measured away from the \atlas interaction point.
The angular flux of SM LLPs, $\tfrac{d^2N}{d\Omega dt}$, is conservatively assumed to be constant, neglecting the decrease of the flux due to material interactions.
The resulting values are $L_{\text{eff}}^\prime = 22.1~\metre$ and $L_{\text{eff}} = 32.3~\metre$~\cite{satterthwaite_2022_292b9-eck73}.

The \atlas search includes background vertices both from the outer part of the calorimeter and from the MS.
Since the reconstruction efficiency in the calorimeter varies significantly with $l$, while it is stable in the MS, only the background vertices in the MS are used for the \anubis background estimate.
The fraction of MS vertices in the \atlas search is $f_{\text{bgr,MS}} = 0.65$.
As a cross-check, the scaling was also performed using only the calorimeter vertices, which resulted in a smaller background expectation.
The MS vertices are therefore used as the more conservative option~\cite{satterthwaite_2022_292b9-eck73}.

The nuclear interaction length of the MS, $\Lambda_{I,\text{MS}}$, is calculated by averaging the densities of iron and aluminium in the MS, conservatively ignoring other materials, which results in $\Lambda_{I,\text{MS}} = 385~\unit{cm}$.
Finally, using the projected HL-LHC integrated luminosity of $\mathcal{L} = 3~\invab$, and conservatively assuming that the \anubis detector is twice as efficient as the \atlas MS at reconstructing background vertices, Eq.~\ref{eq:cons_bkg_est} gives the conservative estimate
\begin{equation}
    N_{\text{bgr,ANUBIS}}^{\text{(cons)}} = 182.4 \pm 12.2.
    \label{eq:cons_bkg_est_value}
\end{equation}

\atlas has recently published a new search for LLPs decaying in the MS with decay lengths of $\mathcal{O} \left( 3~\metre \right)$, using 140~\invfb of $pp$ collision data at $\sqrt{s} = 13~\TeV$~\cite{ATLAS:2025pak}.
This analysis achieves a higher background rejection using machine-learning techniques, with a barrel background estimate of $N_{\text{bgr,ATLAS}}= 241 \pm 18 \text{ (stat) } \pm 6 \text{ (syst)}$, compatible with the observation of $N_\text{obs} = 245$ events.
Rescaling this background estimate with Eq.~\ref{eq:cons_bkg_est} gives a more realistic indication of the background rejection achievable in the \anubis detector with machine-learning techniques.
This estimate is referred to as the nominal background estimate and corresponds to
\begin{equation}
    N_{\text{bgr,ANUBIS}}^{\text{(nom)}} = 51.3 \pm 3.2.
    \label{eq:nom_bkg_est_value}
\end{equation}

%% file: sections/performance.tex
\subsection{Detector layout}

As described in Section~\ref{sec:location}, the \anubis detector consists of two tracking layers separated by $\sim1~\metre$, mounted on the ceiling of the \atlas cavern and at the bottom of the PX14 and PX16 access shafts, and covering an area of about 1600~m$^2$ per tracking layer.

The layout of the detector has been chosen to address several engineering and detector challenges.
For ease of construction and maintenance, the detector follows a modular design.
It is built from standardised tracking modules with an edge width of $\order{1~\metre}$, each of which reproduces the two-layer structure of the \anubis detector with an upper and a lower RPC multiplet separated by $\sim1~\metre$.
Adjacent modules are mounted at alternating heights, following a checkerboard-like pattern, so that neighbouring modules overlap and the coverage of the detector is hermetic.
The mechanical structure that implements this pattern on the cavern vault is described in Section~\ref{sec:engineering}.

The openings of the PX14 and PX16 shafts are covered by shaft plugs rather than by the tracking modules mounted on the vault.
The shaft plugs are modular: they are made from strip-shaped units equipped with the same RPC multiplets, which are suspended into the shaft individually and rest on steel beam stubs protruding from the shaft walls, similar to the established solution for the shaft plugs during LHC operation.
This allows the plugs to be removed quickly during the YETS and long shutdowns, when the shafts are needed for crane access to the \atlas cavern.
A more detailed discussion of these engineering challenges is given in Section~\ref{sec:engineering}.

\subsection{Detector performance}

\begin{table}[htbp]
  \centering
  \caption{Performance requirements for the tracking layers of the \anubis detector.}
  \label{tab:requirements}
  \begin{tabular}{ll}
    \toprule
    Quantity & Requirement \\
    \midrule
    Single-layer efficiency            & $> 98\%$ \\
    Trigger output rate (accidental)   & $< 100~\unit{Hz}$ \\
    Trigger latency                    & $< 10~\unit{\mu s}$ \\
    Readout latency envelope           & $> 10~\unit{\mu s}$ \\
    Time resolution per layer          & $< 0.5~\unit{ns}$  \\
    Angular resolution                 & $< 0.01~\unit{rad}$ \\
    Spatial resolution per layer       & $\lesssim 3~\unit{mm}$ \\
    Channel granularity                & $\lesssim 10~\unit{mm}$ \\
    \bottomrule
  \end{tabular}
\end{table}

The performance requirements on the tracking layers of the \anubis detector that follow from the physics programme are summarised in Table~\ref{tab:requirements} and motivated below.

{\em Efficiency.}
The ability of the detector to record signal decays, as well as to detect and veto backgrounds, depends directly on its hit detection efficiency.
A reconstruction efficiency of $\geq95\%$, defined as the efficiency to record a combined hit in each of the two tracking layers, requires a hit detection efficiency of $>98\%$ per tracking layer~\cite{bib:anubis_orig}.

{\em Trigger rate.}
The \anubis trigger provides an input to the \atlas Level-0 (L0) trigger, whose accept rate at the HL-LHC is about 1~MHz~\cite{ATL-DAQ-PROC-2026-016}.
To keep its contribution negligible and the offline selection of LLP candidates manageable, the accidental rate at the output of the \anubis trigger is required to be below 100~Hz.
The trigger study of Section~\ref{sec:trigger} shows that a six-hit coincidence between the two tracking layers retains about 91--99\% of the signal at an accidental rate below this value (Section~\ref{subsec:trig_outlook}).

{\em Trigger latency.}
The \atlas L0 trigger at the HL-LHC will have a maximum latency envelope of $10~\unit{\mu s}$~\cite{ATL-DAQ-PROC-2026-016}.
For the \anubis detector to provide a trigger to the \atlas experiment, its trigger latency must therefore be below this value; the trigger concept of Section~\ref{sec:trigger} targets a latency of 5~\unit{\mu s}.
Conversely, for the \atlas experiment to trigger the readout of the \anubis detector, the latency envelope of the \anubis readout buffers must exceed that of the \atlas experiment.

{\em Time resolution.}
For particles with $\beta=1$, a time resolution of $\delta t=0.5~\unit{ns}$ corresponds to a resolution of $\delta y_\text{DV}\approx15~\unit{cm}$ on the vertical position of a DV~\cite{bib:anubis_orig}.
This allows the fiducial volume for accepting DVs from LLP candidate decays to be defined to within 30~cm of the lower tracking layer at the $2\sigma$ level.
Moreover, the time resolution is instrumental in eliminating backgrounds from cosmic rays.
Downward-going charged particles with $\beta\approx1$ at normal incidence correspond to a time difference between the upper and lower tracking layer of $\Delta t\approx-3.3~\unit{ns}$.
In contrast, tracks from DVs of LLPs produced in $pp$ collisions are upward-going and correspond to $\Delta t\gtrsim 3.3~\unit{ns}$.
With $\delta t = 0.5~\unit{ns}$ per layer, \ie $\sigma(\Delta t)\approx0.7~\unit{ns}$, and a margin of $2\sigma(\Delta t)$ around the $\beta=1$ edge to retain signal tracks, downward-going tracks are separated from the acceptance time window by $\sim7\sigma$.
A time resolution of $\delta t<0.5~\unit{ns}$ therefore ensures that downward-going cosmic-ray tracks can be rejected with a large margin (Section~\ref{sec:backgrounds}); a more detailed estimate, using all hits of each track, is given in Section~\ref{subsec:trig_offline}.

{\em Angular resolution.}
The angular resolution of the detector, $\delta\alpha$, determines its sensitivity to low LLP masses.
For a fixed mediator mass, lighter LLPs are more strongly boosted, so that their decay products are more collimated.
The required resolution is derived for the most challenging case, \ie the lightest LLP to which the \anubis detector should be sensitive, decaying symmetrically into two bodies.
Assuming a two-body decay is conservative, since decays with a higher multiplicity are easier to identify.
The relevant angle, $\alpha$, is that between the LLP flight direction and the momentum of either decay product, which must be resolved to distinguish a track originating from a DV from a track originating from the IP.
For LLPs as light as the $K_L$, with $m_\text{LLP}\approx0.5~\GeV$, produced in the decay of a mediator at rest with a mass at the electroweak scale, $m_\text{med}\approx100~\GeV$, the required angular resolution is $\delta\alpha\lesssim0.01~\unit{rad}$~\cite{bib:anubis_orig}.

{\em Spatial resolution.}
The spatial resolution required in the plane of a tracking layer, $(x,z)$, follows from the angular resolution.
Since the direction of a track is measured from its combined hits in the two tracking layers, which are separated by $L\approx1~\metre$ (each tracking layer consisting of two RPC singlets about 8~cm apart in the 2~mm gas-gap scenario), an angular resolution of $\delta\alpha=0.01~\unit{rad}$ corresponds to $L\tan(\delta\alpha)\approx L\,\delta\alpha\approx10~\unit{mm}$.
Assuming the same spatial resolution in both tracking layers, the resolution per layer must be $\delta x\approx10~\unit{mm}/\sqrt{2}\approx7~\unit{mm}$.
Including a further safety factor of two, a spatial resolution of $\delta x\lesssim3~\unit{mm}$ per layer is required.

{\em Channel granularity.}
The channel granularity required to resolve two separate tracks is determined by their opening angle, $\omega$.
Since $\alpha$ is the angle between each decay product and the LLP flight direction, and the decay is assumed to be symmetric, $\omega=2\alpha$ and $\delta\omega=\sqrt{2}\,\delta\alpha$.
For a $K_L$ decaying at a distance $h=0.75~\metre$ below the lower tracking layer, the granularity required in this layer is then $\delta x_\text{2trk}\approx h\sqrt{2}\,\delta\alpha\approx10~\unit{mm}$.
With this granularity, two-body decays of heavier LLPs can be resolved closer to the detector: for $m_\text{LLP}\geq1.2~\GeV$, decays can be resolved as close as $h=0.3~\metre$ below the lower tracking layer, which is the boundary of the fiducial volume used in the \anubis sensitivity studies (Section~\ref{sec:loc_fiducial})~\cite{satterthwaite_2022_292b9-eck73,Bauer:2026ivt}, and for $m_\text{LLP}\geq3.6~\GeV$ as close as $h=0.1~\metre$.
These estimates assume that hits in adjacent strips from two different tracks can always be resolved.
While this is not trivial for adjacent hits, staggering the RPC layers within a tracking layer would facilitate this task.

\subsection{Proposed technology}

The \anubis detector is proposed to use the RPC technology developed for the Phase~II upgrade of the \atlas muon spectrometer~\cite{CERN-LHCC-2017-017}, with the gas-gap architecture, gas mixture and front-end electronics optimised as described in Section~\ref{sec:design}.
RPCs offer high efficiency and good time and spatial resolution, as demonstrated with the \proanubis demonstrator (Section~\ref{sec:proanubis}).
Their affordability is also essential to keep the cost of a detector with such a large area under control (Section~\ref{sec:costing}).
The main concern regarding RPCs is the high global warming potential (GWP) of the gas mixture currently used in the \atlas RPC system.
Low-GWP gas mixtures have therefore been studied by the \anubis and \atlas RPC community, as well as by other collaborations working on low-GWP gases for RPCs, such as the RPC EcoGas@GIF++ Collaboration~\cite{Shah:2025luk,RPCEcoGasGIF:2025gqp}, as discussed further in Section~\ref{sec:design}.
The \atlas RPCs have also experienced gas leaks~\cite{Simsek:2024wnt}, an issue that would be exacerbated by the large area of the \anubis detector.
To address this, a new technique to assure the mechanical integrity of RPC gas gaps has been established, which allows for accelerated ageing by applying controlled mechanical stress through thermal cycling, exploiting the differences in the thermal expansion coefficients across the interfaces of the gas gap~\cite{Shah:2025pcy}.

The choice of RPCs also reflects the close connection between the \anubis detector and the \atlas experiment.
RPCs are used in the \atlas barrel muon trigger system, so that the \anubis detector can benefit from the existing expertise and ongoing detector developments.
Conversely, the R\&D on the \anubis RPCs and on low-GWP gas mixtures is carried out within the \atlas RPC community and is of common interest (Section~\ref{sec:int_rnd}).

Plastic scintillators are also being considered as a complementary technology.
Their potential for cost-effective large-area instrumentation, their
use in the \mathusla test stand, and their adoption in the proposed
\mathusla detector design motivate their consideration for
\anubis~\cite{ALIDRA2021164661,MATHUSLA:2025zyt}.
Their spatial and timing performance depends on the segmentation,
geometry and readout configuration; any performance evaluation of hybrid solutions using scintillators in addition to RPCs must
therefore be made for specific detector designs and requires further study.

%% file: sections/proanubis.tex
The \proanubis demonstrator consisted of a single prototype module of \anubis with three tracking layers~\cite{ANUBIS:2025mme}.
The layers had horizontal dimensions of $0.99\times1.78~\metre^2$ and were separated vertically by 0.6~m between the lower two layers and by 0.5~m between the upper two layers.
The tracking layers were made of BIS7S RPC singlets: three in the bottom layer, one in the central layer and two in the top layer.

The demonstrator was installed on level~12 of the HO gantry on the A side of the \atlas cavern in 2023, and was re-commissioned in early 2024 to collect data in 2024 and 2025~\cite{ANUBIS:2025inn}.
It recorded data corresponding to about 104~\invfb of LHC collisions in 2024 and 73~\invfb in 2025, using a custom hardware trigger that required coincident hits in four individual RPC planes of the bottom and top tracking layers.
The large sample of tracks from collision muons, cosmic-ray muons and beam hadrons has been used to validate the performance of the RPCs and to confirm that the RPC technology can achieve the efficiency and time-resolution targets of Table~\ref{tab:requirements}~\cite{ANUBIS:2026jxg}.

\begin{figure*}[ht]
\centering
\subfloat[]{\includegraphics[width=0.5\textwidth]{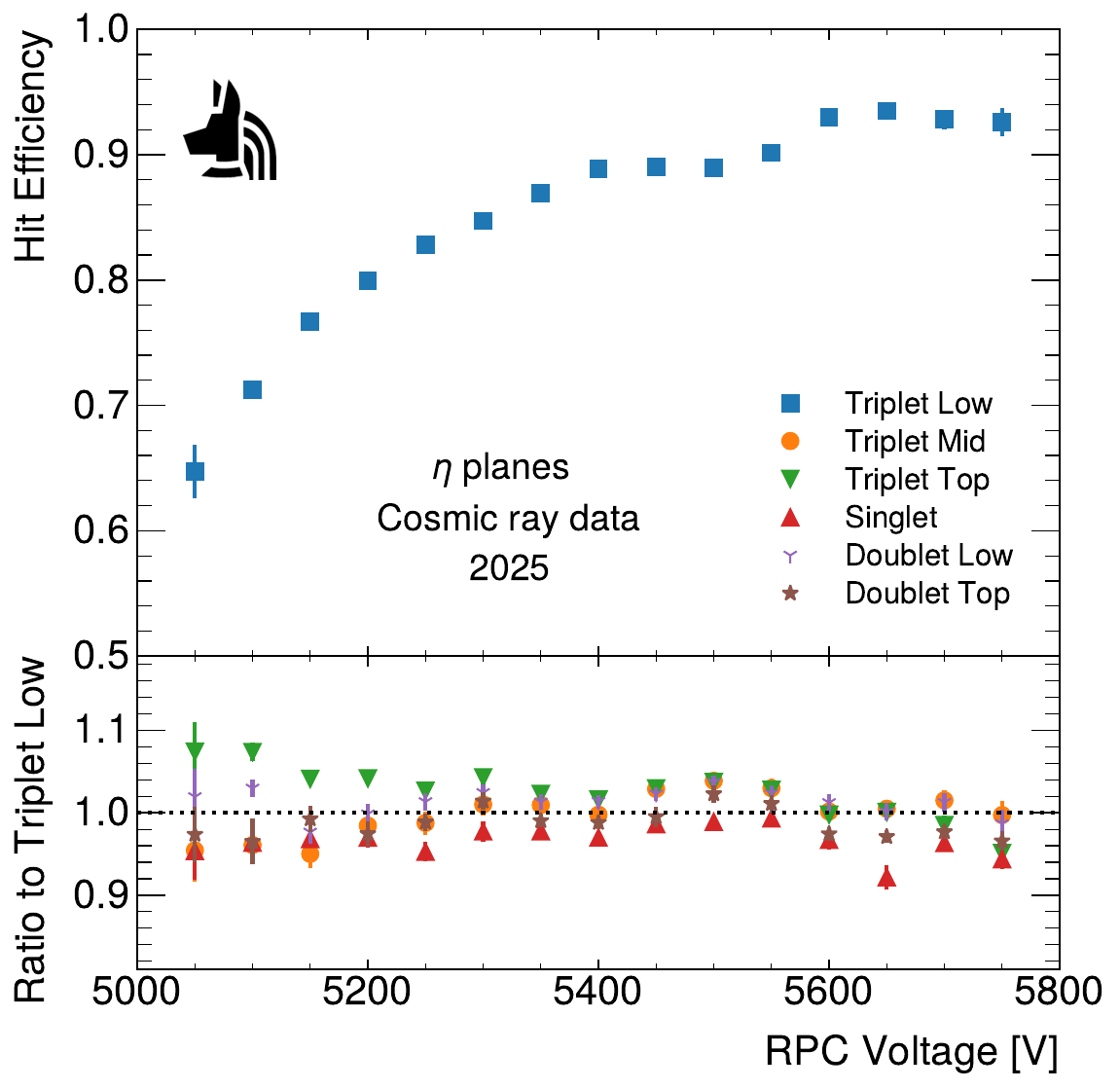}}
\subfloat[]{\includegraphics[width=0.5\textwidth]{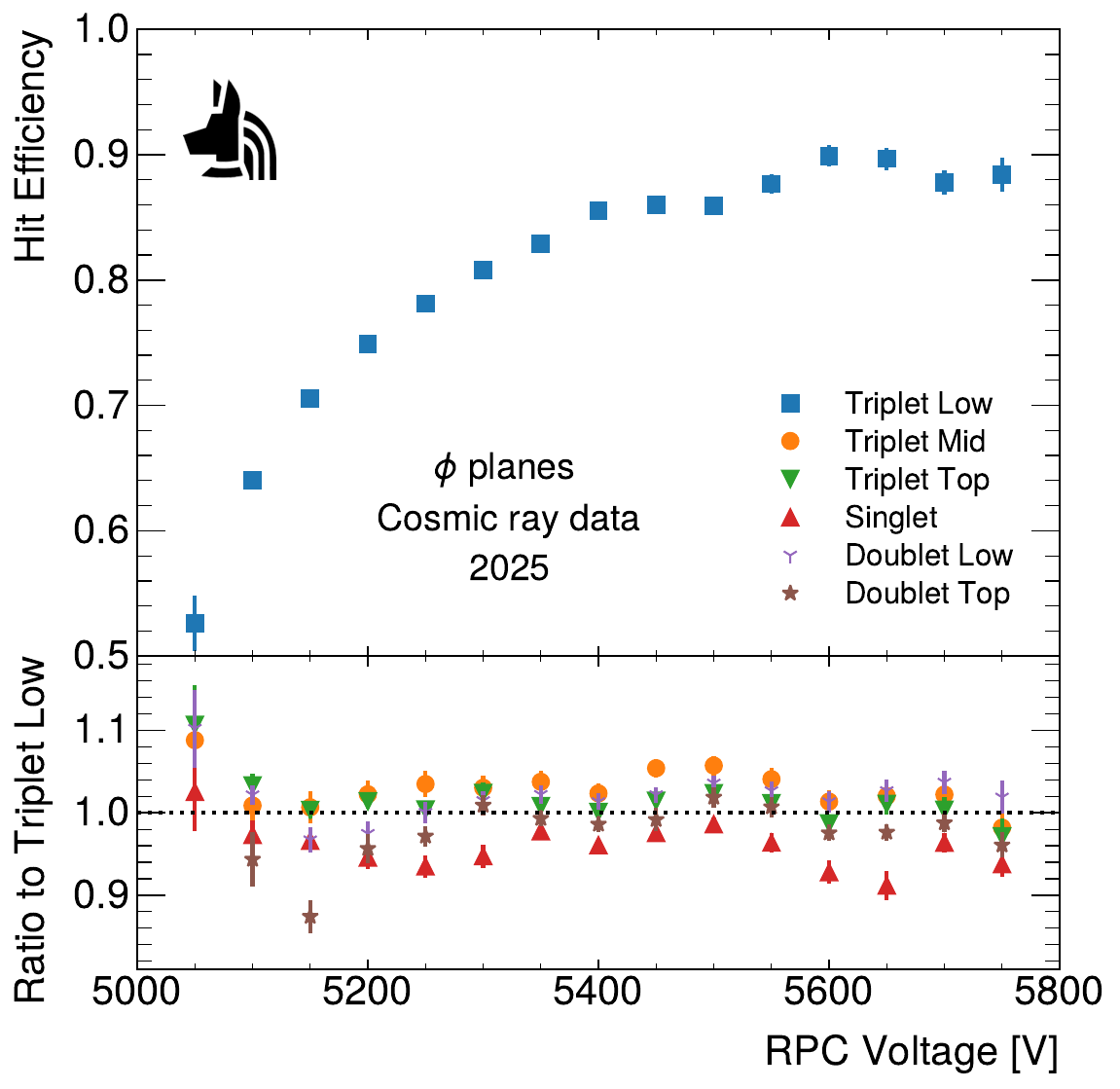}}
\caption{The RPC efficiency measured in cosmic-ray data as a function of the applied voltage.
The efficiencies of the $\eta$ planes are shown in panel~(a), and the $\phi$ planes in panel~(b).} 
\label{fig:stripEff}
\end{figure*}

The efficiency of each RPC was measured as a function of the applied high voltage by excluding the RPC under test from the track fit, requiring hits above threshold in the remaining RPCs, projecting the track onto the excluded RPC and measuring the fraction of tracks with a cluster within 4~cm of the expected position.
All RPCs reach efficiencies above 85\%, as shown in Figure~\ref{fig:stripEff}, which results in an efficiency of more than 99\% for the logical OR of the three RPCs in the bottom tracking layer.

\begin{figure}[ht]
\centering
\includegraphics[width=0.45\textwidth]{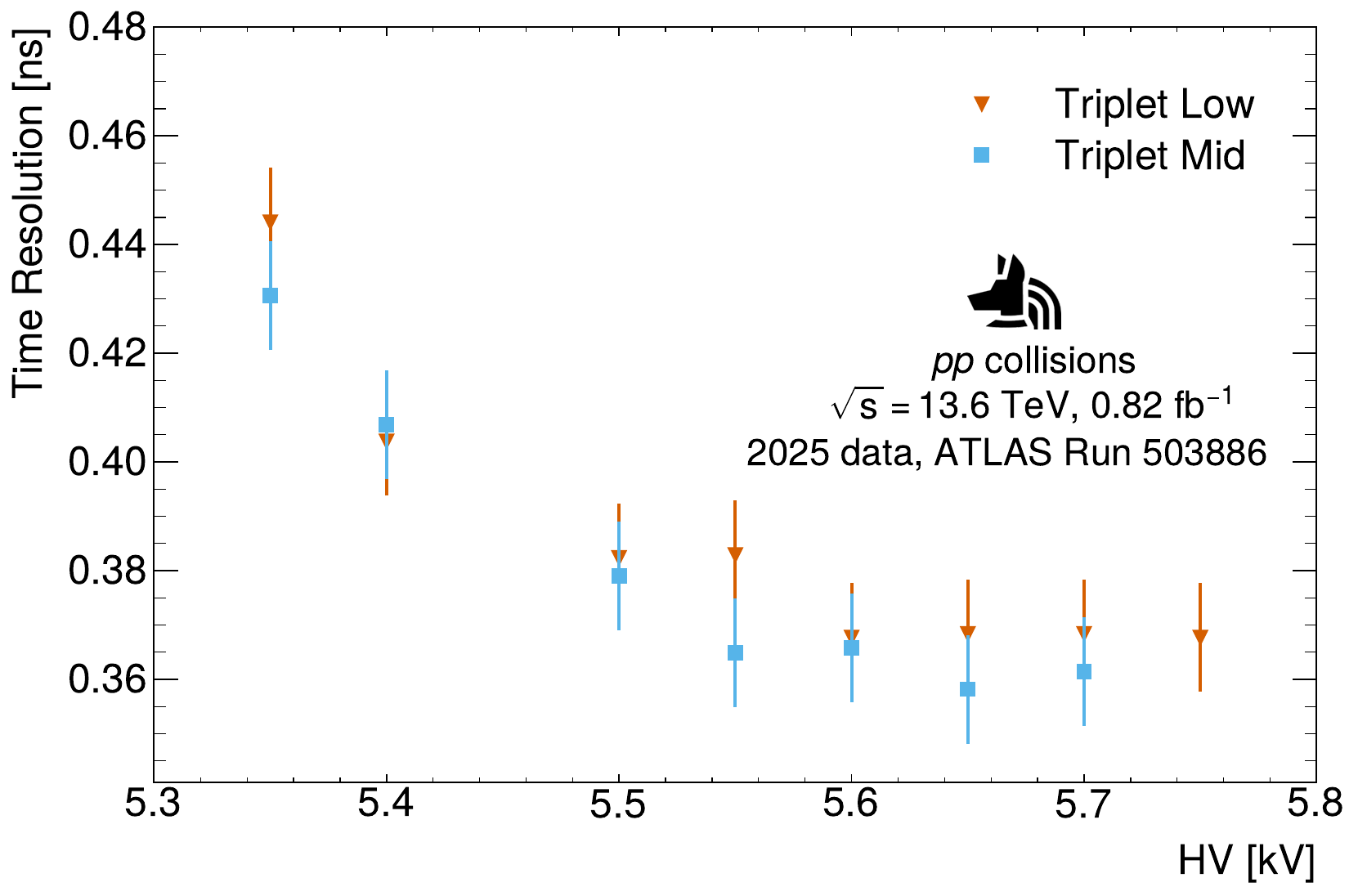}
\caption{Measured time resolution for two \proanubis RPC planes as a function of the high voltage applied.} 
\label{fig:timeResHVScan}
\end{figure}

The time resolution of the RPCs was determined by comparing the arrival times of hits in adjacent RPC planes with the expected time of flight for particles travelling at the speed of light.
The resolution of each individual plane was obtained from all pairwise combinations of the three planes in the bottom layer, by solving for the contribution of each plane.
The time resolutions are shown as a function of the applied voltage in Figure~\ref{fig:timeResHVScan}.
The measured resolution of each plane is below the target of 500~ps at all applied voltages and reaches a plateau of about 350~ps.

\begin{figure}[ht]
\centering
\includegraphics[width=0.45\textwidth]{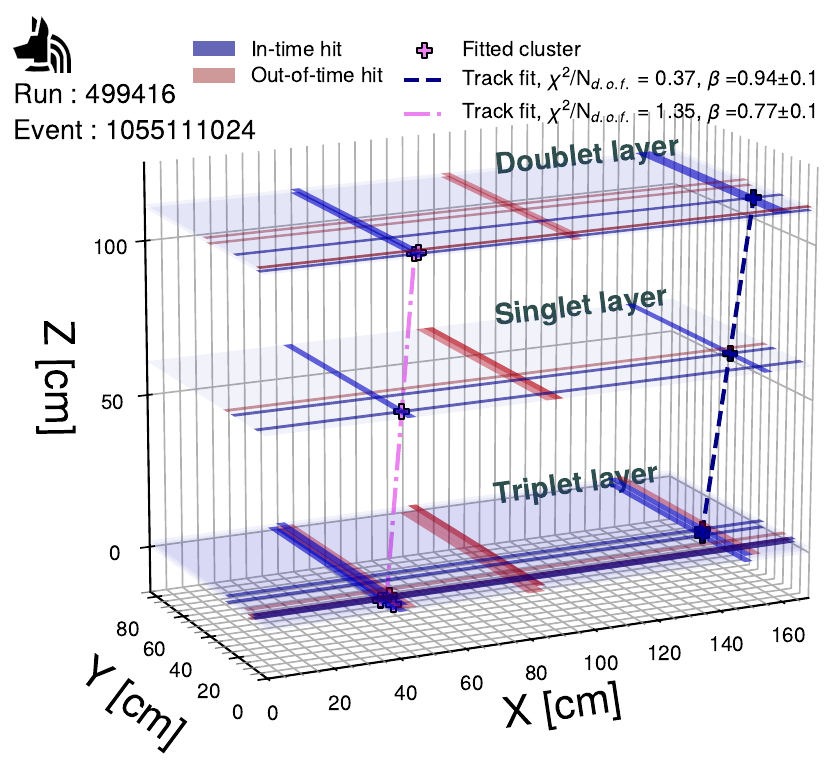}
\caption{A \proanubis event display for a two-track event in $pp$ collisions.} 
\label{fig:twoMuEventDisplay}
\end{figure}

Beyond the performance of individual RPCs, the collected data were used to study the reconstruction of events with two tracks, as required for the identification of LLP decays into charged particles.
An example of a two-track event is shown in Figure~\ref{fig:twoMuEventDisplay}.
Two questions relevant to the full \anubis detector were investigated: whether a central tracking layer, also known as the singlet, is necessary to pair the signals in two-track events, and the absolute rate of vertices during \proanubis operation.
Both studies used data collected during LHC $pp$ collisions in August 2025.

When two particles traverse the same module in the same event, two signals are measured in each of the top and bottom tracking layers.
To pair these signals into two tracks, the central layer is used as a spatial reference, and the combination that best matches the measured signals is chosen.

Without the central layer, the pairing is determined as follows.
All possible combinations of tracks are compared, and the one that maximises the number of RPCs with a signal, summed over all of its tracks, is chosen.
In case of a tie, the $\chi^2$ values and the projected flight speeds of the candidates, together with a term favouring vertices outside rather than inside the detector, are used to decide between the options.

To quantify the success rate of this approach, a reference sample is defined.
It consists of events in which all three layers measure both coordinates of exactly two tracks, which are both successfully fitted without sharing any hits and whose straight-line fits have a probability of at least $10^{-4}$.
Furthermore, for at least one of the tracks, the signal in the central layer must lie at least 5~cm away from both tracks when the pairing is swapped.
This ensures that the central layer has a decisive impact on the choice, so that the two methods do not agree merely because the available information does not discriminate between the pairings.
The reference events are then reconstructed again without the central layer.
In 98.7\% of these events, the algorithm assigns the pairing correctly.
Relative to the reference, only 0.2\% additional vertices are found, while 1.7\% of the reference vertices are lost.

Using the central tracking layer, the absolute rate of vertices in the detector during LHC $pp$ collisions was then measured, independently of the \proanubis trigger.
This rate includes real vertices from neutral particles scattering or decaying in the cavern, but also fake vertices from noise or from combinations of unrelated real tracks.
Since both would cause a trigger signal, no distinction is made between these sources in the following.

The quality requirements on the vertices include the $\chi^2$ of the track fits, timing, angular (relative to the IP) and spatial criteria, and a minimum number of signals on each track.
As in the study of the need for the central layer, the rate depends on how strictly the central layer is required to confirm the pairing.
The following values apply to \proanubis and have not been extrapolated to the full \anubis layout.

Without any requirement on the central layer, the rate is $86.1\pm5.9$~mHz, which constitutes an upper estimate.
Requiring at least one signal in the central layer, which must lie at least 5~cm away from both tracks when the pairing is swapped, gives an intermediate estimate of $40.8\pm4.1$~mHz.
The rate drops to $12.6\pm2.4$~mHz if the central layer is required to have measured signals on both tracks.
Since the efficiency of the central layer is below 100\%, this requirement also removes real vertices, so that this rate is to be understood as a lower estimate.
These measured values have been extrapolated to the area of the full \anubis detector to determine its trigger rate in Section~\ref{subsec:trig_rates}.

%% file: sections/structure.tex
The \anubis detector consists of two tracking layers with a separation of about one metre, to be installed above as much of the \atlas detector as possible, as indicated schematically in Figure~\ref{fig:loc_fiducial}.
In practice, this requires two different support mechanisms: one for the detector elements attached to the vault of UX15, and one for those that plug the PX14 and PX16 shafts, which are labelled in Figure~\ref{fig:loc_cavern}.

In both areas, the primary challenge is to find a structural solution that balances the detector design goals (module overlaps, acceptance, layer separation) with cost, complexity, ease and speed of installation, and future maintainability of both the detector and the structure itself.

Some challenges are unique to the vault.
Notably, the installation on the vault takes place above the level of any crane and therefore cannot be crane-assisted.
This limits the components to those that can be installed with manual or mobile tooling, albeit working from scaffolding or other support structures built up from the top of the bridge crane, which has been moved into position and is stationary.
This is less of an issue for the shaft plugs, which are modular and made from strip-shaped units that are suspended into the shaft individually and rest on steel beam stubs protruding from the shaft walls, similar to the established solution for the shaft plugs during LHC operation.
In contrast, the unique challenge for the plugs is to fix them within their shafts, with access to services, in such a way that they can be removed quickly during shutdowns.\footnote{The mechanism must also not conflict with the concrete blocks sealing the shaft tops during LHC runs.}

The baseline design proposed below is a minimum viable solution, which could be optimised in many ways subject to funding.
Notably, the shaft plugs of the baseline design use the same panel types as the cavern vault.
This minimises the number of panel types to be manufactured, but comes at the cost of a lower acceptance, as rectangular panels are not the best way of filling circles.

\subsubsection*{Baseline design}

\begin{figure}[tb]
\centering
\includegraphics[width=0.95\linewidth]{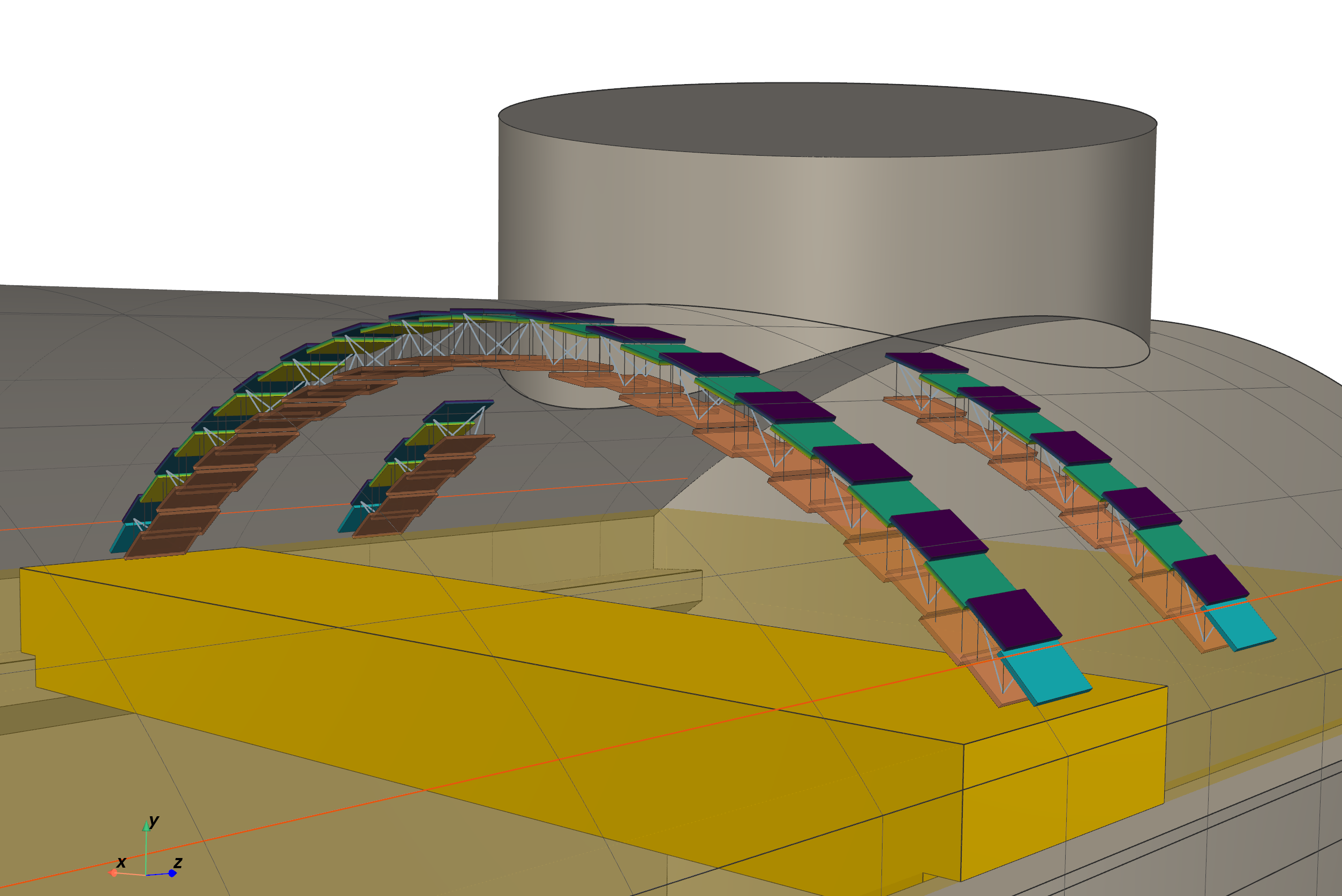}
\caption{
\label{fig:archesZeroAndFour}
Two vault arches: one is a full arch, while the other is interrupted by its intersection with a shaft.
Apart from the interruption, both arches would share common attachment mechanisms to the ceiling, and both would be supplied with services from the springing line of the vault.
The crane is shown where it might be positioned during the installation or maintenance of the uninterrupted arch.}
\end{figure}

\begin{figure}[tb]
\centering
\includegraphics[width=0.95\linewidth]{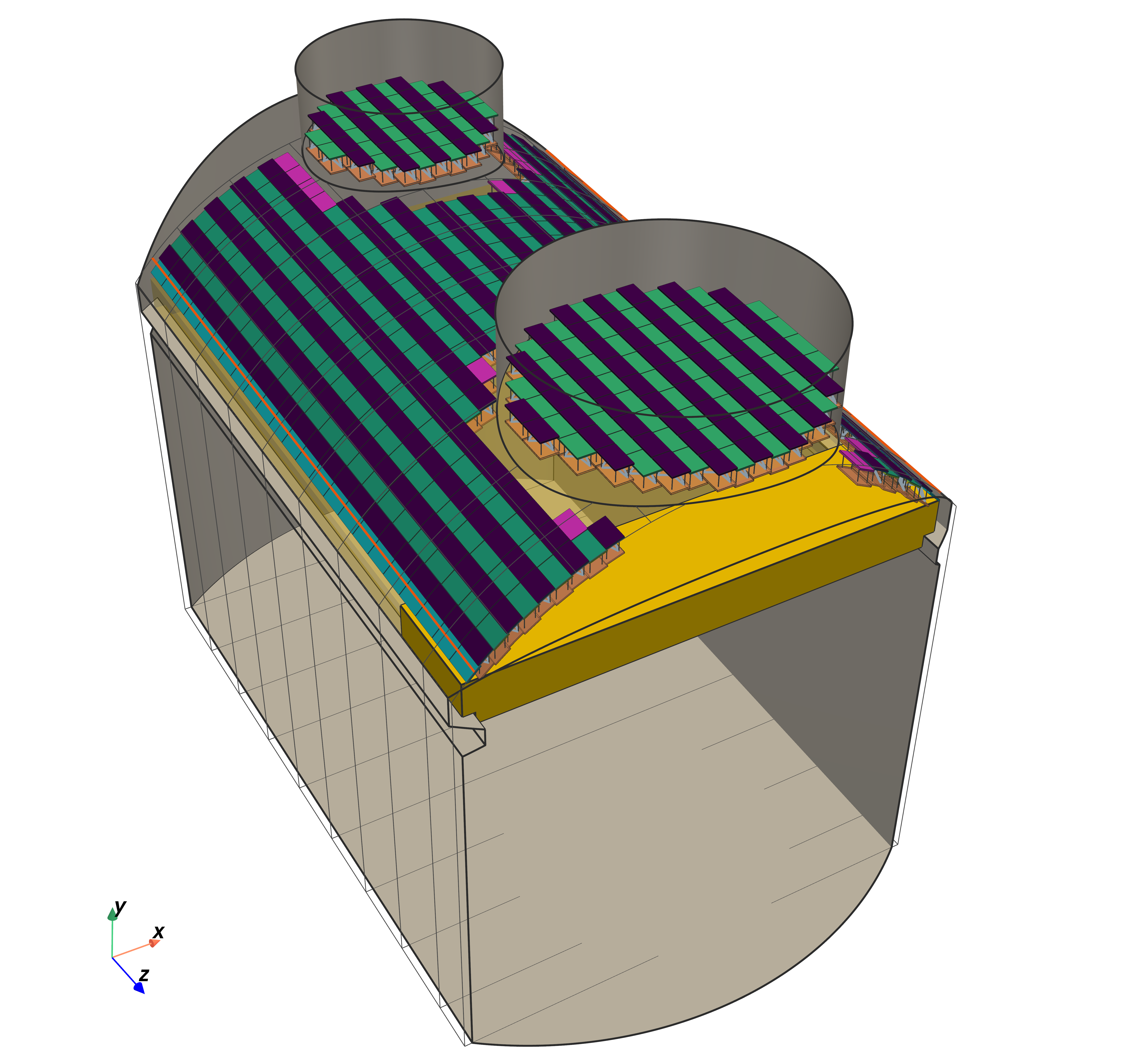}
\caption{
\label{fig:baselinestructure}
The baseline proposal.}
\end{figure}

\begin{figure}[tb]
\centering
\includegraphics[width=0.95\linewidth]{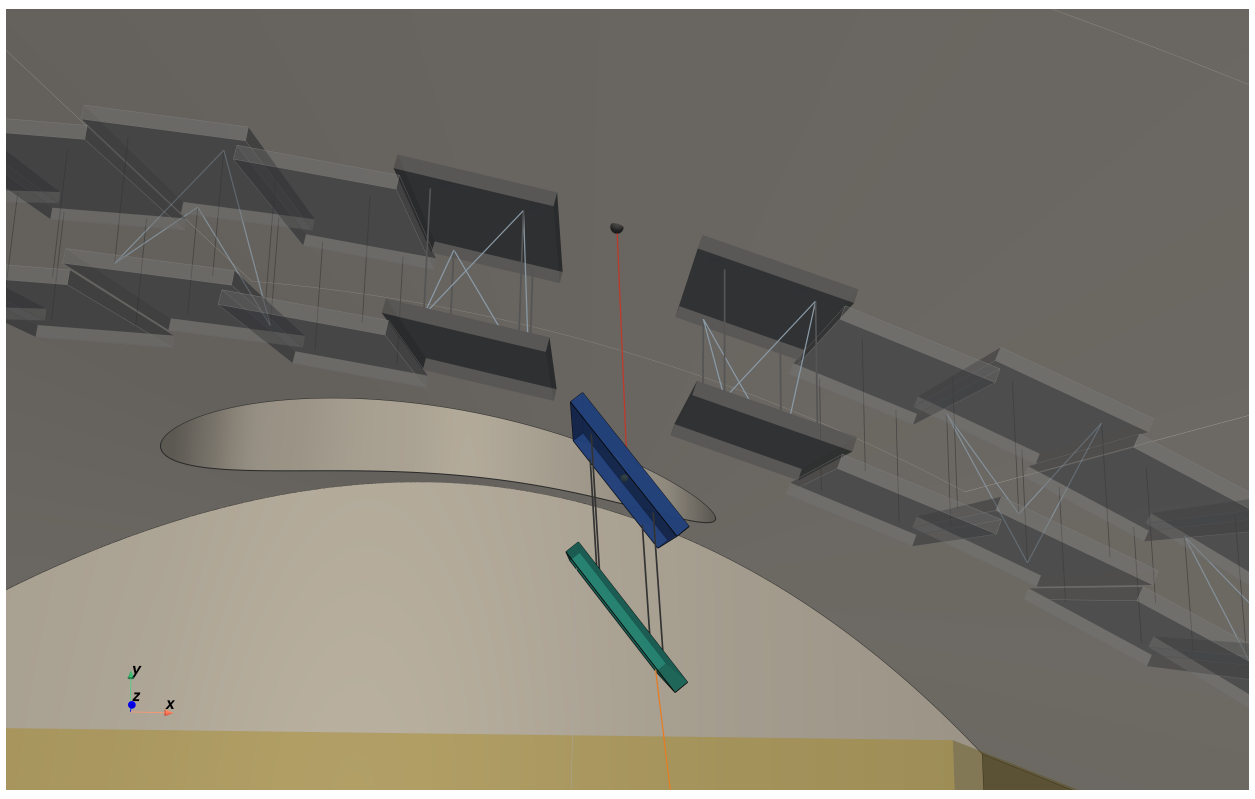}
\caption{
\label{fig:raisingfoldable}
Raising a foldable rig between two rigid rigs.
The services and the lightweight support frame on the ceiling are not shown.}
\end{figure}

It is proposed that the detector elements attached to the vault be arranged as a series of independent arches, two of which are shown in Figure~\ref{fig:archesZeroAndFour}.
The repeating units in each arch are termed \emph{rigs}.
Each rig corresponds to a tracking module, and each of its two panels to an RPC multiplet (Section~\ref{sec:performance}).

Each rig consists of a pair of panels containing RPCs, one hung below the other.\footnote{At the extreme ends of each arch, owing to space constraints, there would be special half-rigs (shown in cyan in Figure~\ref{fig:archesZeroAndFour}), which lack the lower panel of a standard rig.}
Each rig would be raised and installed on the ceiling as a single unit weighing about 100~kg.
It would be raised, or lowered for maintenance, by a cable and pulley system, one per rig, permanently mounted on a lightweight service frame that is attached directly to the vault.
This arrangement ensures that only lightweight components, such as thin Dyneema\texttrademark\ cables, small pulleys, cable guides, short lengths of aluminium beam or cable tray, and mounting plates, need to be lifted by hand.
Once fully raised, each rig could be held in place by long-term tension on its lifting cable, since Dyneema\texttrademark\ can be specified for long-term loads without creep.
Nonetheless, the upper panel would likely also use additional latching fixings on the vault for stability, redundancy and safety.
The lifting cables would likely be ducted to the springing line of the vault, where they could be locked in position by suitable cleats, and where windlasses or other motive power, as well as the electrical and gas services, would be supplied.

A single rig is shown being raised into place on an arch in Figure~\ref{fig:raisingfoldable}.
This figure also illustrates the design principle that the rigs on each arch come in two types: \emph{rigid rigs} and \emph{foldable rigs}.
The two types are interleaved with a small vertical stagger, which provides overlap across the arch along the long sides of each panel.
As far as the active elements are concerned, both types of rig are identical; the main difference is that the rigid rigs have cross bracing that transmits the position reference from the upper panel and its vault reference to the panel suspended below.
This is necessary because the rigs between the rigid rigs have to be foldable if they are to overlap with their neighbours and still be extractable for maintenance.\footnote{If overlap were required but extractability were not, the installation could in principle proceed from the centre of the arch outwards, with all rigs being rigid. This might allow considerable simplifications of the positioning mechanism, albeit at the expense of maintainability.}
After unfolding, the foldable rigs would take their position reference from the rigid rigs on either side, and could be aligned in situ using tracks.

Since each rig has two panels, upper and lower, and comes in two types, rigid for locking into position against the ceiling or foldable for locking into position against the neighbouring panels, the panels need to be manufactured in four types.
These are referred to as type-1 to type-4, with type-1 panels closest to the vault and type-4 panels furthest from it.
Rigid rigs therefore contain type-1 and type-3 panels, and foldable rigs type-2 and type-4 panels.
Wherever possible, the panel types would share common external features, with specialisations only for their different roles in the suspension and in the transfer of the position reference.

The only items fixed directly to the vault would be the elements of a lightweight aluminium frame, installed by hand from a working platform mounted on top of the bridge crane.
All such components must be liftable and installable by a small team with appropriate tooling.
The frame is not shown in the renderings, but would consist of longitudinal elements running along each arch, together with transverse links between the arches.
The longitudinal elements would carry the services from the springing line, and would also support, or if more convenient be supported by, the transverse elements.\footnote{Both transverse and longitudinal links are likely to be necessary, since a single lifting cable would have to attach to the centre of the upper panel of a rig, while stable seating likely requires fixtures near the panel edges. If the rigs were instead raised with two cables, one at each end as in theatre rigging systems, it might be possible to dispense with the transverse links, since the lifting pulleys would then lie in the same vertical plane as the panel seating. Whether the benefits of the resulting simpler ceiling installation and redundant load paths would outweigh the increased complexity of the winching system remains to be studied.}

Although the services would reach the rigs via the ceiling runs, the rigs could have their service connection points either on their tops, with the services forwarded to the panels below, or on their undersides, with the services forwarded to the panels above.

If the mating connections for the rig services were not made when the raised rig latches into its home position, the foldable rigs would need to receive their services via spools or other compliant mechanisms able to extend and contract by at least 1.5~\metre, so that these rigs can be raised.\footnote{Rigid rigs may not \textit{require} such spools, since they would be raised before their neighbours, so that their tops could in principle be accessed from a working platform.}
If such complexity is undesirable, the service connection point of each rig would need to be on its lower panel and be designed to accept fixed-length service lines hanging in the gap between the arches.\footnote{The baseline design models the gap between the arches as a 5~cm gap for clearance only. Increasing this gap by 3~cm to allow the services to use it as well would only amount to a 1.5\% loss of acceptance.}
The attachment points would then be accessible during the raising and lowering of the rigs.
The choice between a uniform approach, with services connected to the rigs independently of their type, and a mixed approach, \eg service reels for the foldable rigs and fixed connections for the rigid rigs, is an optimisation that needs to balance cost, installation time, reliability and accessibility.

In the baseline design, the shaft plugs are modelled as a planar version of the same alternation between raised and lowered rigs as in the vault arches.
Since the strip-shaped units of the plugs are suspended into the shafts individually and can be withdrawn for maintenance, none of the rigs needs to fold.
What is required, however, are dismountable service connections for each plug and a means of supporting them while the shaft is closed.

The larger shaft, PX14, already has steel beam stubs at its base that can support the weight of a plug.
Similar stubs would have to be installed in PX16 with wall anchors.
The services could reach the plugs via the vault.
Since the arch services do not need to cross the apex of the vault, two dedicated service lines, one for each plug, would run along the apex of the cavern from the nearest \atlas HO service structure to the shaft in question, or run from the SX15 surface building down to the bottom of the shafts.
The service connections in PX14 could be mated and unmated from the existing service platform within the shaft.
For the PX16 plug, this would have to be done using a remote-controlled mechanism, with the option of manual intervention and maintenance (i) from the cavern floor with a cherry picker, if its reach permits, (ii) by abseiling down the shaft, or (iii) from a dedicated access structure mounted on top of one of the cavern cranes and manoeuvred below the shaft.
Option~(i) is by far the simplest and is preferred.

\subsubsection*{Other structural considerations}

Part of the cross section of PX14 is occupied by air ducts passing through it.
These will presumably have to remain, but are not modelled in the baseline structure.
Including these ducts would likely remove 6--12 rigs from the PX14 plug.
The ducts could be covered with a fixed tracking installation at the expense of maintainability.

The present cavern ceiling contains extensions of these air ducts and a fire suppression system, both of which occupy space that the \anubis detector would need.
It is not yet known what limitations apply to the repositioning of the ductwork.
The most favourable scenario would be that the ducts can be rerouted to the HO end service structure on the A side, with the network on the C side removed.
Otherwise, since the ducts cannot pass below the \anubis detector without making its installation and servicing impractical, the next-best scenario would be to leave a gap of two or three rigs, either at the apex of the cavern between the shafts, for a loss of acceptance of about 5\%, or near one of the springing lines, for a loss of about 10\%, so that the ducts can run there instead.
Also in this case, the ducts could be covered with a fixed tracking installation at the expense of maintainability.

\subsubsection*{Baseline part count}
The baseline design of Figure~\ref{fig:baselinestructure} would require:
\begin{itemize}
\item 442 rigs on the vault, raised on cables;
\item 46 half-rigs on the vault, with one type-2 panel each, affixed manually without pulley systems;
\item 165 rigs in the shaft plugs, 113 in PX14 and 52 in PX16.
\end{itemize}
This would require the manufacture of at least 1260 panels, before spares, comprising:
\begin{itemize}
\item 316 type-1 panels (upper, rigid rig): 233 on the vault, 57 in PX14 and 26 in PX16;
\item 337 type-2 panels (upper, foldable rig): 255 on the vault, of which 46 in half-rigs, 56 in PX14 and 26 in PX16;
\item 316 type-3 panels (lower, below type-1): 233 on the vault, 57 in PX14 and 26 in PX16;
\item 291 type-4 panels (lower, below type-2): 209 on the vault, 56 in PX14 and 26 in PX16.
\end{itemize}
Each panel is assumed to have a bounding box of $1.8\times1.1\times0.1~\metre^3$ and a mass of about 40~kg.



%% file: sections/detector.tex
The \anubis baseline described in earlier studies used tracking layers formed from triplets of RPCs with $1\,\mathrm{mm}$ gas gaps~\cite{ANUBIS:sensitivity2025}.
This choice followed the technology developed for the BIS78 project during the \atlas Phase~I upgrade, in which three independently instrumented RPC singlets form a compact chamber~\cite{Pizzimento:2020BIS78}.
The \proanubis demonstrator inherited the BIS7 detector and front-end technology and deployed six gas gaps in the \atlas cavern, arranged as a triplet, a singlet and a doublet (Section~\ref{sec:proanubis})~\cite{ANUBIS:2025mme,ANUBIS:2025inn}.
This established technology provided a practical basis for demonstrating the \anubis concept and measuring its operating environment.

Subsequent detector studies motivate a different balance for the full \anubis detector.
The approximately $300\,\mathrm{ps}$ single-gap timing available with $1\,\mathrm{mm}$ gaps offers more precision than the $0.5\,\mathrm{ns}$ target requires, while the limited gaseous target penalises efficiency, particularly with light, CO$_2$-rich mixtures~\cite{Aielli:RPC2026}.
A triplet provides $3\,\mathrm{mm}$ of total gaseous target thickness and can recover high combined efficiency through a two-out-of-three majority.
If one gap fails, preserving efficiency may require accepting one hit out of the two remaining gaps, with a correspondingly weaker coincidence requirement.

For the same instrumented area, strip pitch and readout scheme, moving from two to three gaps requires $50\%$ more readout channels.
The additional gap also increases detector mass, mechanical complexity, cost and trigger-logic complexity.
Its consequences extend through the full system: front-end and downstream readout electronics, signal cables, high- and low-voltage distribution, gas lines and their supports.
Routing and installing these cables and services on the ceiling of the \atlas cavern is particularly demanding.

The design strategy is therefore to obtain high detection efficiency from each instrumented gas gap, making effective use of the investment in its electronics, cabling and services.
This motivates the baseline examined below: a symmetric doublet of independently read out $2\,\mathrm{mm}$ single-gap RPCs, with sensitive front ends and a CO$_2$-rich mixture.
The following discussion develops the physical basis for this evolution from the earlier triplet proposal and identifies the measurements needed to establish its operating margin.

\subsection{Detector response and choice of gas-gap architecture}
\label{subsec:det_optimisation}

The detector design follows from the performance requirements in Section~\ref{sec:performance} and supplies the response assumed in the trigger study of Section~\ref{sec:trigger}.
Charged particles from a displaced decay must provide enough independent position and time measurements to identify two trajectories and their common vertex.
Efficient detection of the incoming particle that precedes a cosmic-ray-induced vertex is also essential for rejecting such backgrounds.
The design must provide these measurements over a large area, using a gas mixture with a high CO$_2$ fraction and a construction compatible with installation and maintenance on the ceiling of the \atlas cavern.

These requirements must be satisfied together.
Reducing the gas-gap thickness can improve timing, but reduces the available ionisation target.
Replacing a dense fluorocarbon with CO$_2$ changes both primary ionisation and avalanche development.
A low-noise, sensitive front end is therefore an integral part of the detector choice: it makes smaller avalanches usable and can enlarge the operating region in which efficiency and timing are obtained at an acceptable delivered charge.
The preferred architecture is the simplest one that meets the required response with the selected mixture.

\paragraph{Gaseous target, useful gap and effective maximum efficiency.}
The first design choice is the total sensitive gas thickness served by an efficient readout unit.
It fixes the available primary-ionisation target.
For a particle traversing a gas path $\ell$, a primary-cluster density $\lambda_{\rm cl}$ gives the probability that the particle produces no primary cluster,
\begin{equation}
 P_0=\exp(-\lambda_{\rm cl}\ell).
 \label{eq:det_primary}
\end{equation}
The corresponding ceiling $1-P_0$ counts every primary cluster as a potential detection opportunity.
Increasing the total gas path reduces the probability that a track leaves no ionisation.

Detection also requires sufficient avalanche development.
The \emph{useful gas gap} is the subset of a gas gap in which a primary cluster can develop, over the remaining drift distance to the anode, an avalanche that produces a front-end signal above the discrimination threshold $V_{\rm th}$.
At fixed gas composition and electric field, the useful gas-gap thickness $d_{\rm useful}$ depends on both the physical gap thickness $d$ and the sensitivity of the coupled front end.
A cluster close to the anode has less multiplication distance available and requires more sensitive electronics to be detected.
In a schematic exponential-growth description, $n(x)=n_0\exp(\alpha_{\rm eff}x)$, where $\alpha_{\rm eff}$ is the effective Townsend coefficient, \ie the first Townsend coefficient reduced by the attachment coefficient, the distance needed to reach the effective avalanche population $n_{\rm th}$ corresponding to this detection threshold $V_{\rm th}$ is
\begin{equation}
 x_{\rm th}\simeq\frac{\ln(n_{\rm th}/n_0)}{\alpha_{\rm eff}},
 \qquad d_{\rm useful}\simeq\max(0,d-x_{\rm th}).
 \label{eq:det_useful}
\end{equation}
A more sensitive front end lowers $n_{\rm th}$ and increases $d_{\rm useful}$.
Actual response also includes avalanche fluctuations, transport, space-charge effects and the interaction of avalanches initiated by different clusters.

The second concept is the \emph{effective maximum efficiency}.
It is the highest efficiency practically attainable while maintaining stable avalanche multiplication, including its saturated regime, at the specified threshold and irradiation conditions.
Operationally,
\begin{equation}
 \epsilon_{\max}^{\rm eff}
 =\max_{V\in\mathcal A}\,\epsilon(V;d,\mathrm{gas},V_{\rm th}),
 \label{eq:det_effectivemax}
\end{equation}
where $V$ is the voltage applied to the gas gap and $\mathcal A$ denotes the acceptable avalanche operating range.
Its definition includes control of streamer production and delivered charge.
Reaching a desired efficiency requires a sufficiently large fraction of the avalanche-signal distribution to lie above the discrimination threshold before leaving this range.
The detection efficiency may still rise at higher voltage while the operating regime becomes unsuitable.
The effective maximum is therefore an operational property of the gas, gap and front end together.
The primary target and its useful fraction must be distinguished when choosing the detector thickness.

\paragraph{Gas mixture: avalanche stability and the cost of CO$_2$ substitution.}
The gas mixture determines how far stable multiplication can be extended towards the gain required by the threshold.
R134a-rich mixtures combine a dense ionisation target with the transport, attachment and quenching properties needed for avalanche operation.
The gas strategy adopted here follows the progressive replacement of R134a with CO$_2$ developed for the \atlas RPCs, with performance and longevity assessed together~\cite{Simsek:2025Longevity,Simsek:2024wnt}.

For the installed \atlas RPCs, the strategy is to retain R134a at a reduced concentration and to increase the CO$_2$ fraction within the operating range of the existing chambers and electronics.
This preserves the established ageing behaviour without introducing a hydrofluoroolefin (HFO) as the primary gas.
Performance measurements, long-term irradiation tests and operation in \atlas validate each step before further substitution~\cite{Simsek:2025Longevity,Simsek:2024wnt}.
This choice addresses a central limitation of HFO-rich primary-gas replacements: some tested mixtures have developed increasing currents and instabilities under prolonged irradiation~\cite{Abbrescia:2025LongTerm}.

New detectors can pursue the same CO$_2$-substitution strategy much further by adopting more sensitive, low-noise front ends.
Transferring part of the amplification from the gas to the electronics makes smaller avalanches detectable and lowers the required gas gain~\cite{Aielli:2016Rate}.
The additional operating margin opens the possibility of reducing the R134a content to nearly zero, or eliminating it, while using CO$_2$ as the main component and retaining stable avalanche operation~\cite{Aielli:RPC2026}.
This is the approach considered for \anubis.
It combines the gas and electronics choices, rather than replacing R134a with another heavy primary gas.

For \anubis, both fluorinated components of the standard RPC mixture are addressed, in line with the plans for the \atlas muon system.
CO$_2$ replaces R134a, owing to its GWP of~1, its reasonable ionisation potential, and its self-quenching properties~\cite{Simsek:2025Longevity,Shah:2025luk,RPCEcoGasGIF:2025gqp}.
For comparison, the GWP of R134a is 1530; all GWPs quoted here are 100-year values from the Sixth Assessment Report of the Intergovernmental Panel on Climate Change~\cite{IPCC:2021AR6ch7}.
SF$_6$ is used at the sub-percent level as an electronegative quenching gas to suppress streamers, but its GWP of 25\,200~\cite{IPCC:2021AR6ch7} makes it a significant contributor to the environmental impact of the mixture.
SF$_6$ is to be replaced by chloro-trifluoropropene (C$_3$H$_2$ClF$_3$, HFO-1233zd(E)), a strongly electronegative HFO with a GWP of about~4~\cite{IPCC:2021AR6ch7}, which is under study as the SF$_6$ substitute for the \atlas RPCs at the HL-LHC~\cite{Proto:2025SF6}.
Matching the streamer suppression of SF$_6$ requires about twice its concentration, \eg 0.6\% instead of 0.3\%~\cite{Proto:2025SF6}.
The ultimate goal for \anubis is a mixture free of both SF$_6$ and R134a, or at least one with an R134a fraction of at most 15\%.

\begin{figure*}[t]
 \centering
 \includegraphics[width=0.86\textwidth]{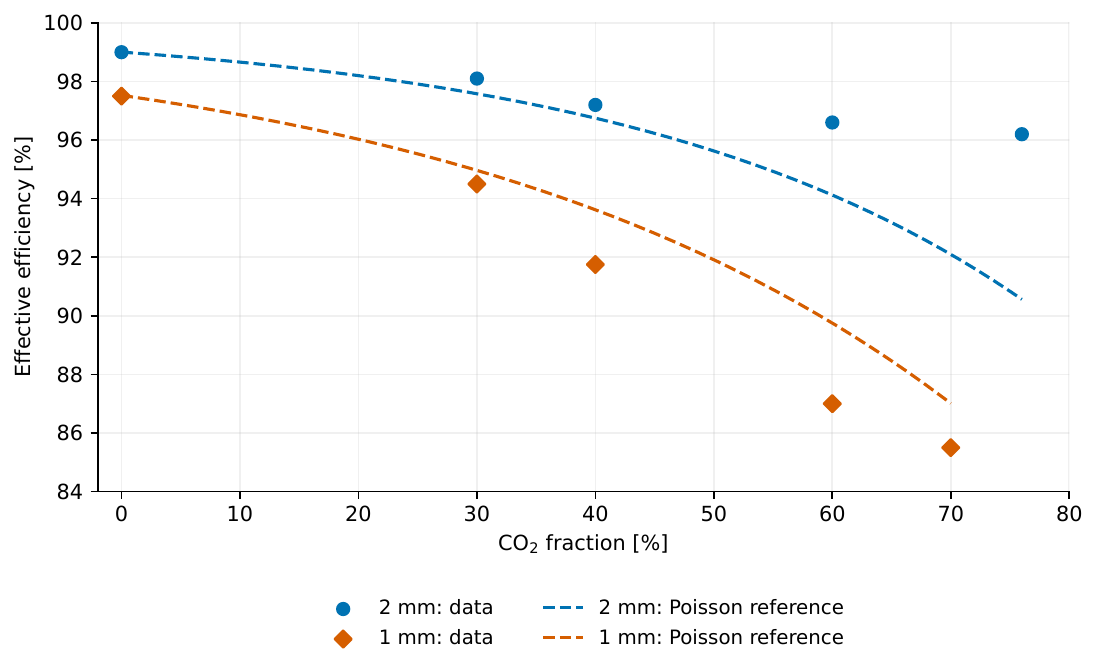}
 \caption{Effective efficiency versus CO$_2$ fraction for $2\,\mathrm{mm}$ R134a-based and $1\,\mathrm{mm}$ HFO-based configurations~\cite{Aielli:RPC2026}.
Symbols are measurements; dashed curves are the normalised Poisson references.
The zero-CO$_2$ points are standard-gas benchmarks.
Gas families and additives differ between the sequences.
The larger gas target preserves high efficiency over a wider CO$_2$ range.}
 \label{fig:det_co2}
\end{figure*}

Choosing a light, CO$_2$-rich mixture instead of replacing R134a with another heavy primary gas carries a primary-ionisation cost: the number of targets per unit gas path is reduced.
This makes high detection efficiency harder to maintain as the gap becomes thinner.
Figure~\ref{fig:det_co2} compares measured $1$ and $2\,\mathrm{mm}$ efficiencies with normalised Poisson references~\cite{Aielli:RPC2026}.
The curves illustrate the expected loss from the reduced primary target, using the standard-gas calibration.
The $2\,\mathrm{mm}$ detector retains high efficiency as the CO$_2$ fraction increases, whereas the $1\,\mathrm{mm}$ measurements show a substantially larger loss.
Departures from the Poisson references reflect several physical processes, including cluster mortality, discussed in Ref.~\cite{Aielli:RPC2026}.
This motivates retaining a sufficiently large gas target and using sensitive electronics to make it effective.

\begin{figure*}[t]
 \centering
 \includegraphics[width=0.86\textwidth]{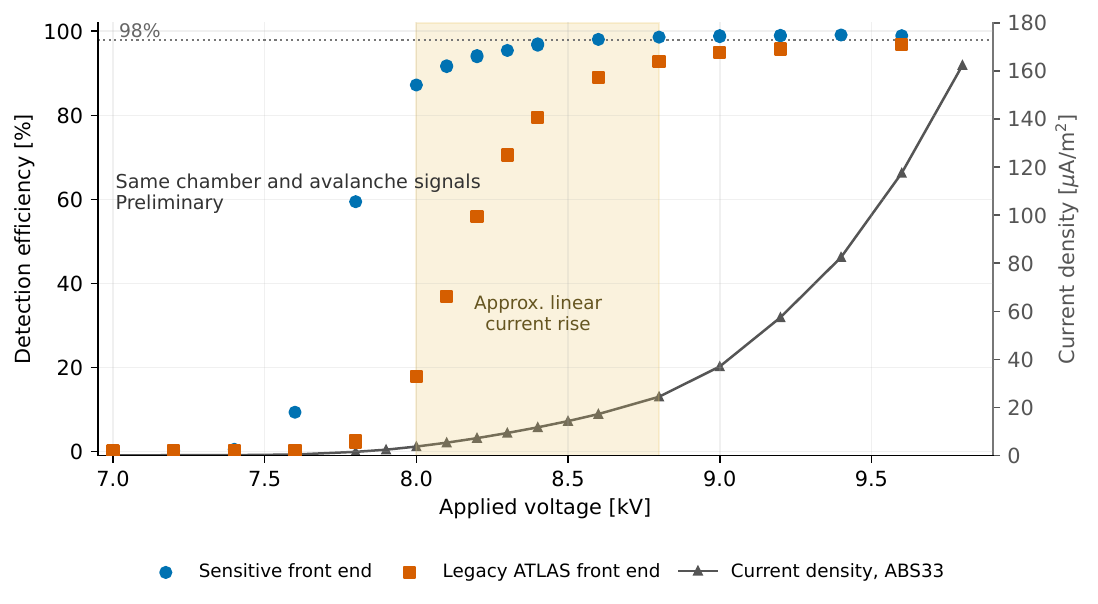}
 \caption{Preliminary simultaneous-readout comparison for the $2\,\mathrm{mm}$ test chamber with an R134a-free mixture~\cite{Aielli:RPC2026}.
Circles and squares give the efficiencies of the sensitive and legacy \atlas front ends on the left axis.
The grey curve gives the common chamber current density under gamma irradiation at the ABS33 absorber setting on the right axis.
The dotted line marks $98\%$ efficiency.
The shaded $8.0$--$8.8\,\mathrm{kV}$ interval highlights approximately linear current growth in the saturated-avalanche region.
The current shape illustrates the onset of more rapid charge growth as streamer production is approached; its absolute scale depends on the source intensity.}
 \label{fig:det_frontends}
\end{figure*}

\paragraph{Front-end sensitivity: experimental access to the required efficiency.}
Figure~\ref{fig:det_frontends} shows the experimental consequence of moving amplification to the electronics.
A $2\,\mathrm{mm}$ chamber operated with an R134a-free mixture is read simultaneously through legacy \atlas electronics and a high-sensitivity analogue preamplifier on orthogonal pickup views.
Both readouts therefore see the same gas target and the same avalanche magnitudes: changing the electronics changes the voltage at which the signals become detectable, while the chamber has a single current-versus-voltage response.

The measurement is performed under gamma irradiation, which makes the intrinsic dark-current contribution negligible.
At fixed gamma flux, the voltage dependence of the irradiation current reflects the photon-detection probability and the mean charge per converted photon.
The approximately linear current rise identifies the saturated-avalanche operating region; the subsequent faster rise indicates the approach to streamer production.

At $8.6\,\mathrm{kV}$ the high-sensitivity front end records about $98\%$ efficiency, compared with $89\%$ for the legacy readout electronics, while the chamber current remains in its approximately linear growth region.
The efficiency with the legacy electronics remains below $98\%$ throughout the measured scan, including at higher voltages where the current rises more rapidly.
The high-sensitivity front end therefore makes the R134a-free mixture usable at the required efficiency within the saturated-avalanche range.

The result demonstrates that a $2\,\mathrm{mm}$ target and R134a-free operation can provide high efficiency at a sufficiently low gas gain to retain the saturated-avalanche regime.
While this establishes the feasibility of the general approach, the concrete R\&D task is to optimise the mixture and electronics together so that the useful operating point has a robust margin against excessive charge and the avalanche-to-streamer transition.
A large usable signal-to-noise ratio supports this margin by allowing small prompt signals to be detected reliably.

\begin{figure*}[t]
 \centering
 \includegraphics[width=0.86\textwidth]{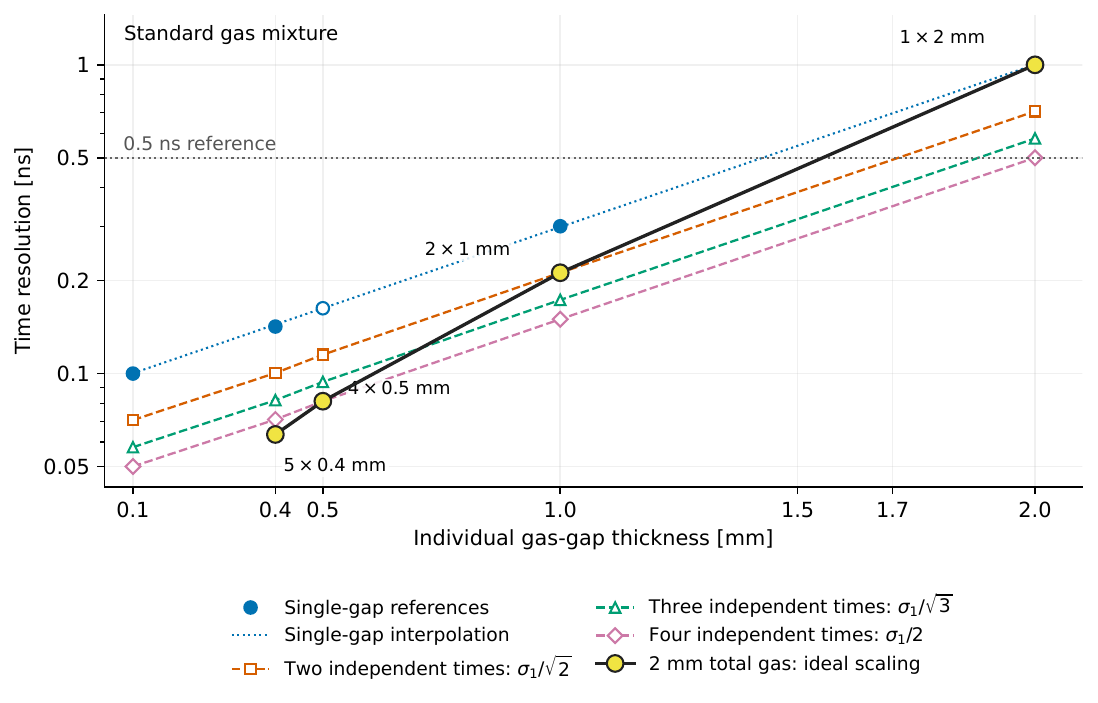}
 \caption{Timing comparison at fixed gas composition, using standard-mixture references~\cite{Aielli:RPC2026}.
Filled blue points are single-gap inputs; the open point at $0.5\,\mathrm{mm}$ follows the empirical interpolation $\sigma_1(d)=0.0888\exp(1.2117d)\,\mathrm{ns}$, with $d$ in mm, shown by the dotted blue curve.
Open coloured markers give $1/\sqrt N$ projections for the mean of two, three or four independent times.
Yellow-filled markers highlight $2\,\mathrm{mm}$ total gas thickness, including five $0.4\,\mathrm{mm}$ gaps.
The grey line marks $0.5\,\mathrm{ns}$.
For analogue summation the scaling is an ideal reference; the realised benefit depends on the avalanche contributions and the readout response.
Timing with a high-CO$_2$ mixture requires measurement.}
 \label{fig:det_timing}
\end{figure*}

\paragraph{Timing at a fixed total gas target.}
The efficiency results establish $2\,\mathrm{mm}$ of sensitive gas as the reference target. 
At fixed gas composition, splitting this target into thinner gaps offers faster individual responses while preserving the primary-ionisation target.
Figure~\ref{fig:det_timing} compares single-gap timing references in the standard mixture with projections for combinations of thinner gaps~\cite{Aielli:RPC2026}.
The constant-target sequence contains one $2\,\mathrm{mm}$ gap, two $1\,\mathrm{mm}$ gaps, four $0.5\,\mathrm{mm}$ gaps or five $0.4\,\mathrm{mm}$ gaps.
The $0.5\,\mathrm{mm}$ single-gap value is interpolated from the reference curve.

For $N$ equal independent measurements with uncertainty $\sigma_{\rm ind}$ and a common contribution $\sigma_{\rm com}$, an average has
\begin{equation}
 \sigma_{t,\rm mean}^2=\frac{\sigma_{\rm ind}^2}{N}
 +\sigma_{\rm com}^2.
 \label{eq:det_meantime}
\end{equation}
The $1/\sqrt N$ scaling applies to independently digitised times with equal, uncorrelated uncertainties.
For bi-gap structures, in which two gas volumes are coupled to a common pickup and their signals are summed in analogue form, the same factor is only an ideal reference: uneven avalanche contributions can reduce the timing benefit.
The shared-pickup response must therefore be measured.
All projections in the figure refer to the same gas mixture; changing the mixture provides a separate means of improving the response.

The assembled single-hit response includes the gas, electronics, digitisation, propagation and calibration contributions,
\begin{equation}
 \sigma_{t,\rm hit}^2\simeq\sigma_{\rm gas}^2+
 \sigma_{\rm FE}^2+\sigma_{\rm digit}^2+
 \sigma_{\rm prop}^2+\sigma_{\rm cal}^2,
 \label{eq:det_timebudget}
\end{equation}
with correlated terms treated through their covariance.
These contributions must be included when assessing the timing available from a complete chamber.

\paragraph{Intermediate gap thickness as a second R\&D route.}
If timing optimisation at $2\,\mathrm{mm}$ proves insufficient, a modest reduction to approximately $1.7\,\mathrm{mm}$ can be considered.
The standard-mixture interpolation in Figure~\ref{fig:det_timing} gives approximately $0.7\,\mathrm{ns}$ for a $1.7\,\mathrm{mm}$ single gap and hence approximately $0.5\,\mathrm{ns}$ for the mean of two independent measurements at that precision.
The timing projection follows the empirical single-gap references; its application to the selected CO$_2$-rich mixture and readout requires validation.

A preliminary estimate of the impact on the efficiency of reducing $d$ from $2\,\mathrm{mm}$ to $1.7\,\mathrm{mm}$ is provided below.
A thickness scaling at fixed effective detected-cluster density $\kappa$ gives
\begin{equation}
 \epsilon(d)=1-\exp(-\kappa d),\qquad
 \epsilon_{1.7\,\mm}=1-(1-\epsilon_{2\,\mm})^{1.7/2},
 \label{eq:det_extrapolation}
\end{equation}
where $\epsilon_{2\,\mm}$ and $\epsilon_{1.7\,\mm}$ are the efficiencies for $d=2\,\mathrm{mm}$ and $d=1.7\,\mathrm{mm}$, respectively.
Table~\ref{tab:det_effestimate} provides illustrative scenarios.
A $98\%$ reference at $2\,\mathrm{mm}$ scales to about $96.4\%$ at $1.7\,\mathrm{mm}$, while $99\%$ scales to about $98\%$.
This approximation keeps the detected response per unit thickness fixed.
The $15\%$ thickness reduction makes changes in the useful gas fraction relevant: a fixed ineffective thickness $a$ would replace the exponent by $(1.7-a)/(2-a)$.

Increasing the field can recover part of the useful gap by reducing the avalanche-development distance needed to cross the discrimination threshold.
The estimates make no allowance for this recovery and are therefore conservative.
The field and gain needed to recover efficiency must remain inside the stable avalanche range.
Matched measurements of efficiency, timing and charge after field optimisation will establish the attainable response and its operating margin.

\begin{table}[t]
 \centering
 \caption{Preliminary conservative efficiency scenarios from Eq.~\ref{eq:det_extrapolation}, with $\kappa$ held fixed.
The scaling omits possible recovery of the useful gap through field optimisation.
The achievable efficiency in stable avalanche operation requires measurement.}
 \label{tab:det_effestimate}
 \begin{tabular}{cc}
 \toprule
 $\epsilon_{2\,\mm}$ & Estimated $\epsilon_{1.7\,\mm}$\\
 \midrule
 $96.0\%$ & $93.5\%$\\
 $98.0\%$ & $96.4\%$\\
 $99.0\%$ & $98.0\%$\\
 $99.5\%$ & $98.9\%$\\
 \bottomrule
 \end{tabular}
\end{table}

\paragraph{Detector structures selected by the response.}
The \anubis baseline is a doublet of independently read out $2\,\mathrm{mm}$ single-gap RPCs, operated with a light, CO$_2$-rich mixture and sensitive electronics in the saturated-avalanche regime.
This preserves the gas target needed for high efficiency and allows separate timing measurements to be combined while retaining a simple chamber structure.
A bi-gap divides each $2\,\mathrm{mm}$ target into two thinner gas volumes coupled to a common pickup.
Separately instrumenting these volumes adds electronic channels; a common pickup increases construction complexity and requires its own timing qualification.

The first R\&D priority is to establish whether the faster response observed with CO$_2$ enrichment, combined with sensitive electronics, provides the required timing at $2\,\mathrm{mm}$~\cite{Aielli:RPC2026}.
Efficiency, timing and charge must be measured together to consolidate the operating margin.
A reduction to approximately $1.7\,\mathrm{mm}$ is reserved for a demonstrated timing need, with its efficiency assessed after field optimisation.

A dedicated bi-gap prototype would allow timing scans versus gas gain and discrimination threshold.
These measurements would test how the avalanche regime affects the benefit of analogue summation and determine whether the improvement is substantial across a sufficiently broad operating range to justify the additional construction complexity.
Table~\ref{tab:det_options} summarises the baseline and the alternatives that these measurements would qualify.

\begin{table*}[t]
 \centering
 \caption{Baseline and conditional alternatives for one tracking layer.
The two measurements of a doublet are read out independently.
Each bi-gap measurement uses analogue summation of two gas volumes.}
 \label{tab:det_options}
 \small\setlength{\tabcolsep}{5pt}
 \begin{tabular}{lll}
 \toprule
 Option per tracking layer & Motivation & Required result\\
 \midrule
 \shortstack[l]{Baseline: two $2\,\mathrm{mm}$\\single-gap units} & \shortstack[l]{High-efficiency reference and\\simple chamber construction} & \shortstack[l]{Timing with CO$_2$-rich gas and sensitive\\electronics; robust operating margin.}\\[5pt]
 \shortstack[l]{Fallback: two $1.7\,\mathrm{mm}$\\single-gap units} & \shortstack[l]{Faster response if the\\baseline timing is insufficient} & \shortstack[l]{Acceptable efficiency after field optimisation,\\within stable avalanche operation.}\\[5pt]
 \shortstack[l]{Alternative: two bi-gap\\units of $1+1\,\mathrm{mm}$} & \shortstack[l]{Thinner gas volumes with\\$2\,\mathrm{mm}$ target per unit} & \shortstack[l]{Substantial measured timing improvement\\over the single-gap baseline.}\\
 \bottomrule
 \end{tabular}
\end{table*}

The trigger study of Section~\ref{sec:trigger} uses two tracking layers separated by $1\,\mathrm{m}$, each consisting of one doublet, with an $8\,\mathrm{cm}$ reference spacing between its two independently read out gas gaps.
It assumes $98\%$ single-hit efficiency and $0.5\,\mathrm{ns}$ uncertainty for each measurement.
Combining the two times improves the time precision per tracking layer according to Eq.~\ref{eq:det_meantime}; the trigger inputs retain the individual-hit uncertainties and their correlations.
Measured single-hit efficiencies, time distributions, missing-hit patterns and the selected assembly geometry will be incorporated into the trigger and reconstruction studies.

\begin{figure*}[t]
 \centering
 \includegraphics[width=0.98\textwidth]{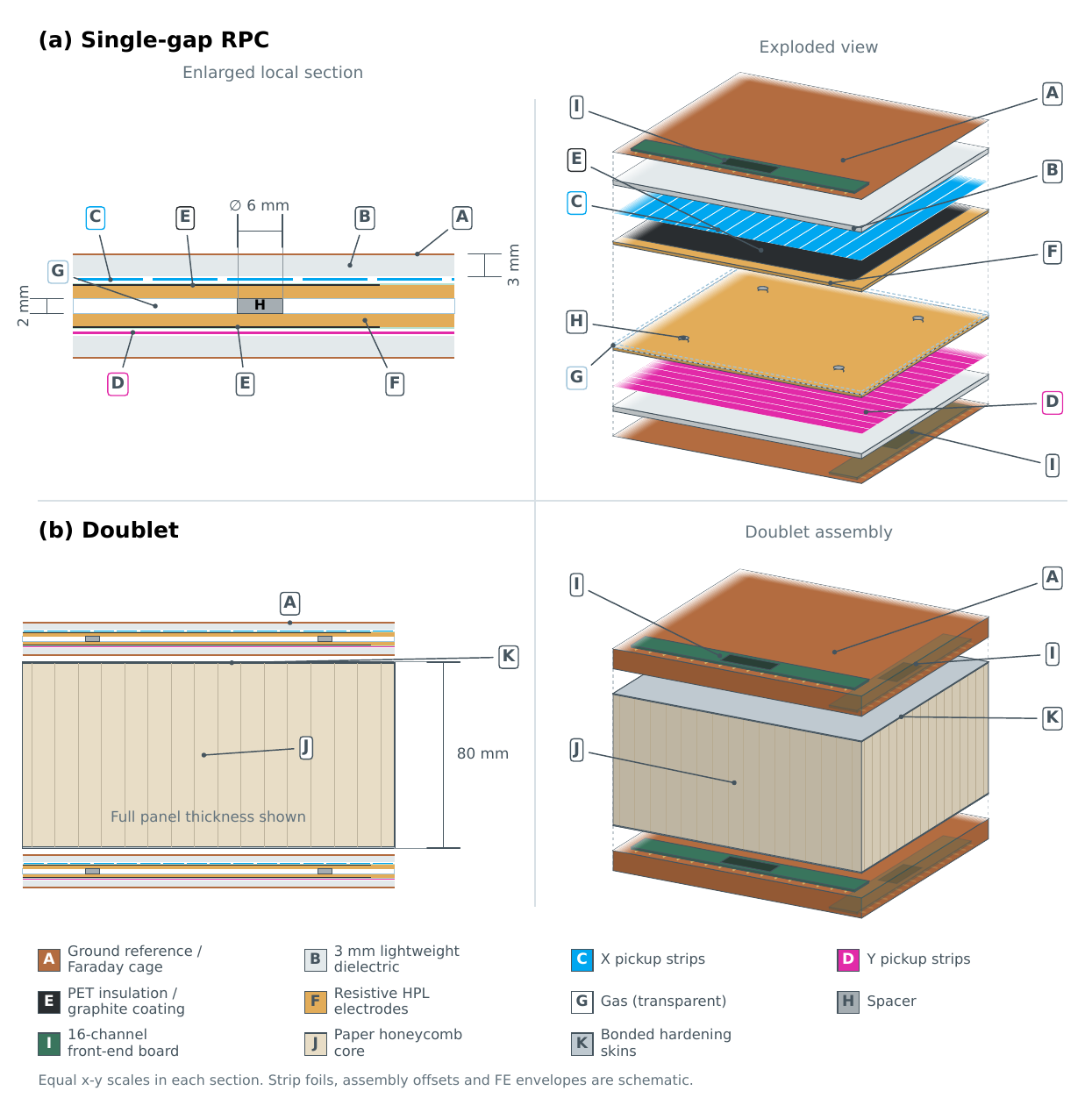}
 \caption{Single-gap RPC (a) and doublet (b).
The sections on the left use equal horizontal and vertical scales, with the single-gap detail enlarged.
Strip foils are emphasised and separated graphically for readability.
On the right, the single-gap layers are exploded around a transparent gas volume; the doublet view shows closed RPCs and a separated support panel.
The letters identify the parts in the legend.
Front ends lie on the external ground-reference shields; the Y-side boards are shown through the shields.
The $80\,\mathrm{mm}$ dimension spans the full separation panel, including its skins.
Exploded offsets, thin films and front-end envelopes are schematic.}
 \label{fig:det_stratigraphy}
\end{figure*}

Figure~\ref{fig:det_stratigraphy} shows the single-gap chamber and doublet construction.
Each orthogonal pickup panel consists of a strip layer and an external copper backplane sandwiching a $3\,\mathrm{mm}$ lightweight dielectric, typically foam or paper honeycomb.
The backplanes provide the ground reference for the strip transmission lines and the main armour of the Faraday cage, with the front-end boards mounted on their external surfaces.
The outer faces of the high-pressure laminate (HPL) electrodes carry graphite coatings, insulated from the pickups by polyethylene terephthalate (PET) films; the graphite leaves a $10\,\mathrm{mm}$ uncoated border along the two edges forming the front-end corner.
The $5\,\mathrm{mm}$-diameter spacers follow a $10\,\mathrm{cm}$ square grid.
The illustrated segment contains 16 strips per view at an indicative $1\,\mathrm{cm}$ pitch, with one 16-channel front end per view.

The doublet uses an $80\,\mathrm{mm}$ separation panel made from paper honeycomb with bonded stiffening skins.
Adding the chamber layers between this panel and the gas volumes gives a gas-midplane lever arm of about $10\,\mathrm{cm}$, somewhat larger than the $8\,\mathrm{cm}$ reference spacing assumed in the trigger study.
The final layer thicknesses and service clearances will determine the assembled measurement-plane coordinates.

\begin{figure*}[!tp]
 \centering
 \includegraphics[width=0.90\textwidth]{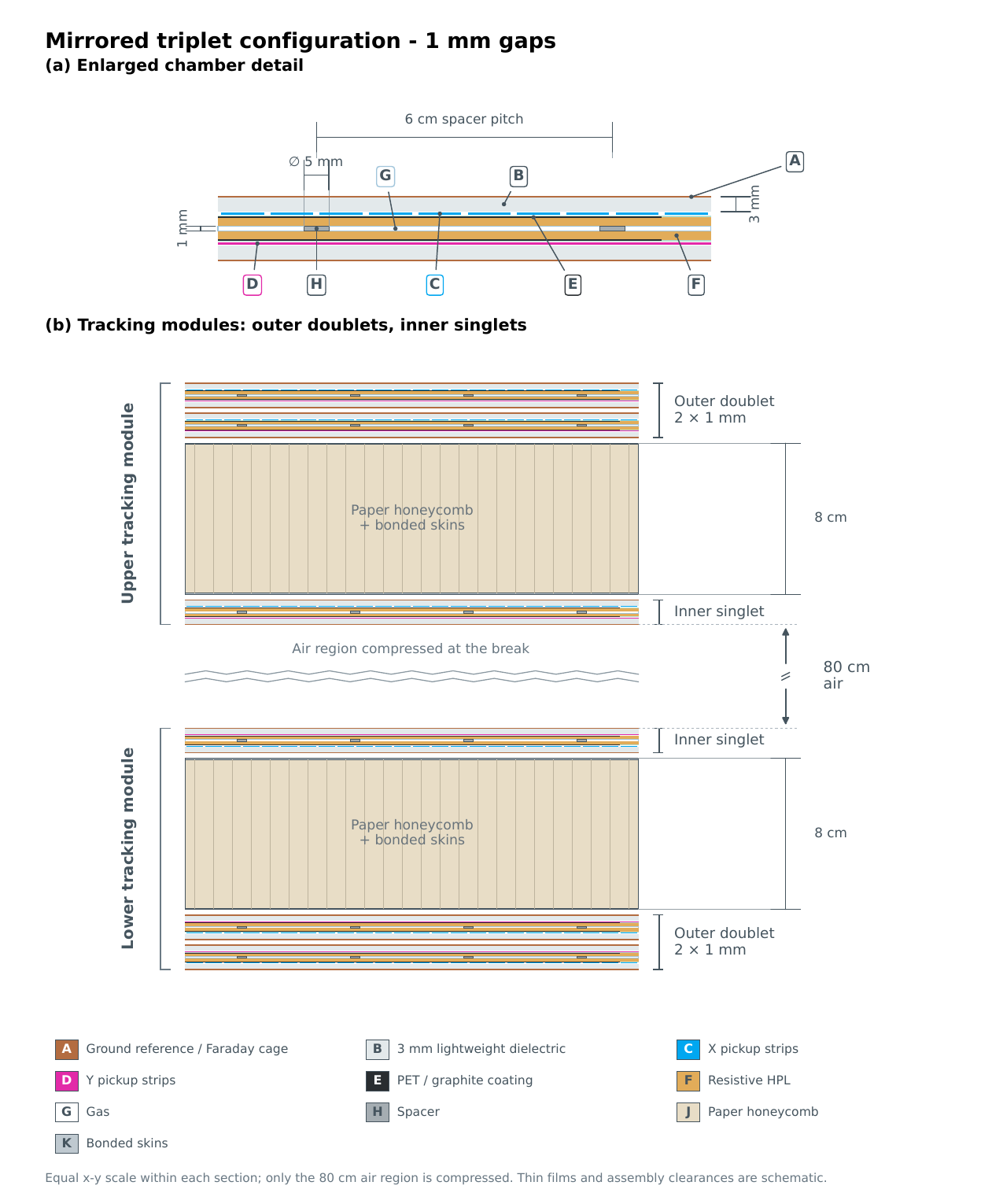}
 \caption{Earlier triplet configuration, shown for comparison with the proposed doublet baseline.
Each of the two mirrored triplets uses three independently instrumented $1\,\mathrm{mm}$ gas gaps.
The doublets lie on the outer sides and the singlets face the central air region.
Panel (a) enlarges the chamber section, with $5\,\mathrm{mm}$-diameter spacers at a $6\,\mathrm{cm}$ pitch.
Panel (b) shows the two $8\,\mathrm{cm}$ separation panels and the $80\,\mathrm{cm}$ air clearance between the inner singlets.
The air region is compressed at the explicit break.
Passive layers, thin films and assembly clearances are illustrative.}
 \label{fig:det_triplet}
\end{figure*}

\subsection{Readout and interface to the trigger}
\label{subsec:det_readout}

The front-end electronics developed for the \atlas RPC Phase~II upgrade provide a baseline for the \anubis electronics~\cite{Cardarelli:2019Performance,Pizzimento:2019FrontEnd}.
The recent implementation for the new barrel-inner (BI) RPCs combines a discrete silicon bipolar-junction-transistor preamplifier with a custom SiGe BiCMOS application-specific integrated circuit (ASIC) containing discrimination, time digitisation and serial output.
It reports an effective sensitivity of $1$--$2\,\mathrm{fC}$ and a time-to-digital converter (TDC) resolution below $100\,\mathrm{ps}$.
Leading and trailing times provide the hit timestamp and time over threshold (ToT), which can be calibrated as a charge observable~\cite{Pizzimento:2024BI}.
These capabilities support both the low-charge operation established in Section~\mbox{\ref{subsec:det_optimisation}} and the track-association methods of Section~\mbox{\ref{subsec:det_separation}}.

A new version of the digitisation and serial-readout ASIC would add daisy-chain transmission to these capabilities.
The published BI implementation provides a separate serial output for each channel~\cite{Pizzimento:2024BI}; the \anubis development would aggregate the hit records while preserving timestamps, ToT and channel identifiers.
A common clock and calibrated channel delays would maintain the relative timing precision throughout the readout.

The baseline uses orthogonal X and Y strip panels, each read at one end, giving comparable precision in both coordinates.
With 96 strips per metre per coordinate, a square reference chamber with an area of $1~\metre^2$ has 192 electronic channels; a doublet has 384, and a tracking module of the same area, consisting of one doublet in each of the two tracking layers (Section~\ref{sec:trigger}), has 768.
The costing in Section~\ref{sec:costing} assumes larger RPC singlets of $1.1\times1.8\,\mathrm{m}^2$ with a comparable strip density.
A single strip panel read at both ends also has 192 channels: the strip position gives one coordinate, the time difference gives the other, and the mean time permits a propagation correction.
This option is relevant to a bi-gap structure with a common pickup, but gives asymmetric coordinate resolution.
The orthogonal-strip arrangement supports the preferred single-gap baseline.

Full-length strip tests will qualify the characteristic impedance, front-end input impedance and terminations, including propagation, dispersion and reflections in the timing budget of Eq.~\ref{eq:det_timebudget}.
The fast trigger benchmark groups 8, 16 or 24 strips in each view.
OR16 spans $16.67\,\mathrm{cm}$ and defines 36 virtual-pad addresses per square-metre chamber.
Individual-strip times and ToT supply the precision information used to resolve tracks within these coarse regions.

Integrating discrimination, digitisation and serial transmission close to the chamber reduces external cabling.
Local zero suppression and data concentration organise the output into a fast stream carrying the hit addresses, times and pad multiplicity required by the trigger.
Processing, buffering and serial links must sustain the expected burst traffic within the trigger latency budget.
The readout chain should be divided into sections that allow faulty modules to be isolated while the remaining detectors continue operating.
The physical groups and transmission filter in Section~\ref{sec:trigger} determine the records needed to recover tracks crossing module boundaries; accepted events also require the precision-hit record and precursor-veto history.

Wireless streaming between neighbouring concentrators is a further readout R\&D option, motivated by millimetre-wave developments for high-energy physics detectors~\cite{Aziz:2025WADAPT}.
Each concentrator could continuously broadcast its locally timestamped records to its immediate neighbours for the overlapping trigger search.
Qualification would establish simultaneous reception of neighbouring streams, bounded latency under load and compatibility with the high-sensitivity front end.
Transmission-error tolerance would be assessed through hit losses and residual corrupt records, with error detection enabling damaged records to be discarded.

\subsection{Spatial precision and separation of nearby tracks}
\label{subsec:det_separation}

Two-track separation and single-hit spatial resolution are separate performance requirements, both influenced by strip pitch.
Two-track separation is governed by the discrete granularity of the readout: two particles must produce distinguishable strip-hit patterns.
Spatial resolution describes how precisely each hit is localised, using the available digital and analogue information.

Assigning each hit to the centre of one strip of pitch $p$, with uniform illumination and no additional position information, gives the binary reference $\sigma_x=p/\sqrt{12}$.
A $10\,\mathrm{mm}$ pitch gives $2.89\,\mathrm{mm}$, while 96 strips per metre correspond to $p=10.42\,\mathrm{mm}$ and $\sigma_x\simeq3.01\,\mathrm{mm}$.

Charge sharing supplies additional position information.
For cluster size $\mathrm{CS}>1$, where CS counts adjacent fired strips, calibrated strip charges can be used to reconstruct a charge-weighted centroid~\cite{Aielli:2013PrecisionTracking}.
The charge estimates can be obtained from the calibrated ToT described in Section~\ref{subsec:det_readout}.
Cluster size itself also carries position information: a $\mathrm{CS}=1$ event preferentially identifies the central region of a strip, while hits near a strip boundary have an appreciable probability of producing $\mathrm{CS}=2$.
Combining these classes with calibrated charge interpolation can therefore improve the spatial resolution substantially beyond the approximately $3\,\mathrm{mm}$ binary reference for centimetre-pitch strips.
The achieved precision depends on the charge-sharing response, threshold and noise, and will be measured with the selected chamber and front end.

For two comparable position measurements separated by $L$, the projected small-angle uncertainty is approximately $\sigma_\theta=\sqrt{2}\sigma_x/L$.
At $L=1\,\mathrm{m}$, the binary reference $\sigma_x=3\,\mathrm{mm}$ gives $4.2\,\mathrm{mrad}$ per projection; improved hit localisation sharpens this angular measurement.
The two-track requirement depends on the projected separation of the hits.
For a symmetric pair at normal incidence and a vertex a distance $h$ below a plane, the separation is $s=2h\tan(\omega/2)\simeq h\omega$.
Small openings or decays close to the lower tracking layer can therefore give overlapping signals even when single-track precision is adequate.
The angular samples and their geometric acceptance are treated in Section~\ref{subsec:trig_collection}.

The probability of resolving two physical signals depends on their separation, relative arrival time and incidence angle.
Induced cluster width, crosstalk and the recovery time of the front end determine whether nearby signals remain distinct.
Once clusters are resolved in each strip view, their X--Y association can exploit the time and charge information supplied by the electronics.

Two tracks firing separate X and Y clusters generate both genuine and ghost strip crossings.
The 2D meantimer method tests each candidate crossing against its propagation-time relation~\cite{Aielli:2017FiveD}.
If $\tau_X(x,y)$ and $\tau_Y(x,y)$ are the calibrated propagation delays to the two readout ends, a genuine association gives a residual
\begin{equation}
 r_{XY}=[t_X-\tau_X(x,y)]-[t_Y-\tau_Y(x,y)]
 \label{eq:det_xyassociation}
\end{equation}
consistent with zero.
Both signals reproduce the same avalanche, so the intrinsic RPC timing fluctuation cancels in this comparison.
The residual width is governed by relative readout precision, including TDC resolution, differential electronic jitter and calibration.
At a propagation speed near $20\,\mathrm{cm/ns}$, $100\,\mathrm{ps}$ corresponds to $2\,\mathrm{cm}$ of propagation distance, setting a characteristic scale for the association test.
The measured uncertainty of the relative time determines the realised discrimination power; both spatial coordinates retain the strip-based precision.

An incorrect association combines different avalanches.
Their independent fluctuations in threshold-crossing time broaden the timing residual, while their different charges supply a complementary discriminator.
Calibrated ToT measurements provide charge estimates whose compatibility can be tested between the two views.
The substantial avalanche-charge spread therefore helps identify the correct pairing, including for nearby clusters with similar propagation delays~\cite{Aielli:2017FiveD}.
Reading both ends provides additional consistency relations.
A single-panel, double-ended implementation instead derives one coordinate from timing, introducing the asymmetric spatial resolution discussed in Section~\ref{subsec:det_readout}; this contributes to the preference for the single-gap X--Y baseline.

The combined space-time and charge criteria will be qualified with realistic two-particle signals, measuring merged-hit and ghost-association probabilities versus separation and relative time.
The trigger study assumes that local multiplicity information identifies two particles within one coarse virtual pad.
Folding the measured response into its opening-angle samples and boundary-sharing logic will establish the corresponding trigger and reconstruction efficiencies.

\subsection{Services, integration and reproducible operation}
\label{subsec:det_services}

The services design must support reproducible detector operation within the installation and maintenance scheme of Section~\ref{sec:engineering}.
Standardised modules should combine manageable mass, accessible connections and independent isolation of power, high voltage, gas and data.
Module borders and service attachments must be included in the active-area and material maps used by reconstruction.

Power distribution must provide stable voltage rails while preserving front-end noise, threshold and timing performance.
Local DC--DC conversion is a possible means of reducing current in long supply cables; its efficiency, thermal load and electromagnetic compatibility require qualification in the assembled detector.
Monitoring and recovery should support operation during periods of restricted access.

Gas services must maintain controlled composition, flow and pressure through branches that can be monitored and isolated.
Gas-tight chambers and reliable fittings limit consumption and environmental impact.
Selection of the complete mixture requires performance, material-compatibility and longevity measurements, together with its full GWP and gas-handling requirements, including gas recuperation if R134a is retained (Section~\ref{sec:costing}).

Full-size module tests must demonstrate that the service implementation preserves efficiency, time distributions and tails, charge, noise and two-track response at a common operating point under the expected background and environmental conditions.
These measurements will provide the calibrated response and inactive-region maps for the trigger and reconstruction studies.

%% file: sections/trigger.tex
\subsection{Detector geometry and trigger objectives}
\label{subsec:trig_inputs}

The detector geometry and fiducial volume are described in Sections~\ref{sec:location} and~\ref{sec:performance}. The trigger study considers two identical horizontal tracking layers, each covering $1600~\unit{m^2}$. Each tracking layer of Section~\ref{sec:performance} is implemented as an RPC doublet. The tracking layer midplanes are separated by $L=1~\unit{m}$, and the gas gaps within each doublet by $d=8~\unit{cm}$. We label the gaps G1--G4 in order of increasing height: G1 and G2 belong to the lower tracking layer, and G3 and G4 to the upper tracking layer. The reference signal volume lies below both tracking layers; the metre between the tracking layers is excluded.

\begin{figure*}[t]
 \centering
 \includegraphics[width=0.98\textwidth]{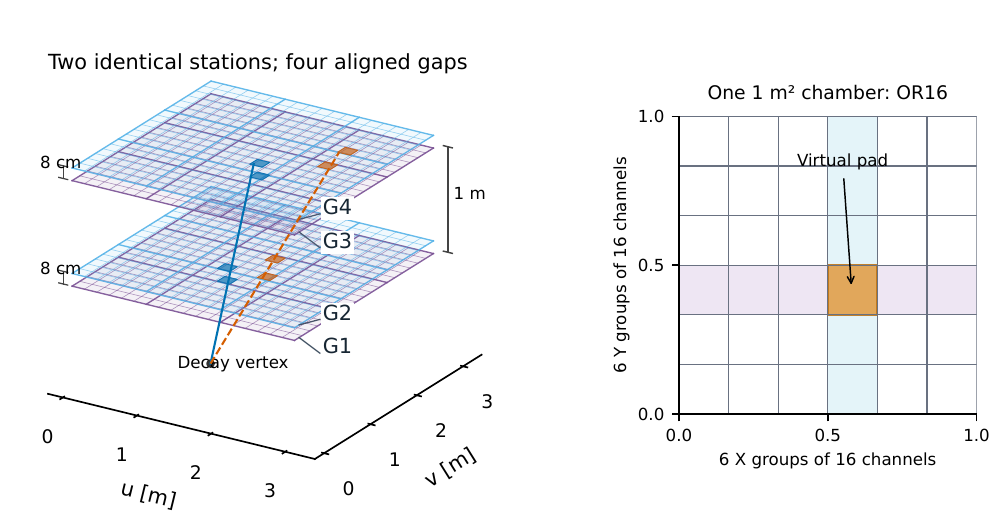}
 \caption{Detector and readout geometry. Left: two identical tracking layers over $3\times3$ tracking modules, with gaps G1/G2 below and G3/G4 above. Each doublet has a violet lower gap and light-blue upper gap. Heavy and fine lines delimit chambers and virtual pads; daughter tracks illuminate crossed pads. Gap separations are enlarged; labels give physical distances. Right: OR16 forms six channel groups per coordinate and 36 virtual pads per chamber.}
 \label{fig:trig_apparatus}
\end{figure*}

The calculation uses a local right-handed frame with $\hat{\mathbf u}=-\hat{\mathbf x}$, $\hat{\mathbf v}=\hat{\mathbf z}$ and upward $\hat{\mathbf w}=\hat{\mathbf y}$, relative to the \atlas coordinates in Section~\ref{sec:loc_geometry}. The $u,v$ coordinates span a tracking layer. Taking G1 at $w=0$, the gap heights are
\begin{equation}
 (w_{G1},w_{G2},w_{G3},w_{G4})=(0,0.08,1.00,1.08)~\unit{m}.
 \label{eq:trig_planes}
\end{equation}
The study uses a reference tracking module (Section~\ref{sec:location}) with a footprint of $1~\unit{m^2}$, which contains four aligned square-metre chambers, one in each gap, \ie one doublet in each tracking layer. Figure~\ref{fig:trig_apparatus} shows a $3\times3$ group of tracking modules and the fast spatial readout.

The front end provides individual strip signals and a fast OR for each group of channels in both coordinates. With 96 strips per metre, OR8, OR16 and OR24 have group widths $b=8.33$, $16.67$ and $25.0~\unit{cm}$, respectively. The crossing of an X group and a Y group defines a square \emph{virtual pad}. OR16 supplies $6\times6=36$ virtual-pad addresses in a square-metre chamber. The study assumes additional local multiplicity information that distinguishes two particles crossing the same virtual pad.

The trigger selects events with at least two charged daughters while keeping its final output near $100~\unit{Hz}$. A signal-acceptance target near $99\%$ at the nominal detector working point guides the choice of conditions. Geometric collection, gap inefficiency and primitive transmission determine how much of the signal reaches that decision. Their combined effect is evaluated over the same generated events.

The fast trigger uses coarse positions, timestamps and multiplicity. The \anubis response target is $5~\unit{\mu s}$; the latency requirements for integration with \atlas are discussed in Section~\ref{sec:performance}. For an accepted trigger, the complete precision-hit record is retained over the event window and its preceding history. Offline reconstruction uses all available measurements for a common vertex in three spatial dimensions and time, identifies descending cosmic precursors, and correlates the candidate with the event recorded by \atlas. Its target is fewer than one accepted instrumental or misidentified cosmic event over the live exposure. Collision-related backgrounds and \atlas information are discussed in Section~\ref{sec:backgrounds}.

Table~\ref{tab:trig_inputs} collects the detector inputs and operating targets used below.

\begin{table}[t]
 \centering
 \caption{Detector and operating inputs used in the trigger comparison. The gap efficiency and independent background counting density are scanned.}
 \label{tab:trig_inputs}
 \begin{tabular}{lc}
 \toprule
 Quantity & Value\\
 \midrule
 Instrumented area per tracking layer & $1600~\unit{m^2}$\\
 Chamber area & $1~\unit{m^2}$\\
 Tracking-layer separation & $1~\unit{m}$\\
 Gas-gap separation in a doublet & $8~\unit{cm}$\\
 Single-gap efficiency & $85$--$99\%$; reference $98\%$\\
 Hit-time uncertainty & $0.5~\unit{ns}$\\
 Precision position uncertainty & $3~\unit{mm}$; $1~\unit{mm}$ option\\
 Fast OR widths & 8 / 16 / 24 channels\\
 Physical group & $3\times3$ tracking modules\\
 Logical searches compared & $3\times3$ and $4\times4$\\
 Logical-search step & One chamber\\
 Reference gap counting density & $100~\unit{Hz/cm^2}$ at $\epsilon=1$\\
 Descending-muon flux & $0.544~\unit{m^{-2}s^{-1}}$\\
 Reference density scan & $2$--$120~\unit{Hz/cm^2}$\\
 Central-output reference line & $100~\unit{Hz}$\\
 \anubis latency target & $5~\unit{\mu s}$\\
 Pretrigger precision history & $0.5~\unit{\mu s}$\\
 \bottomrule
 \end{tabular}
\end{table}

\subsection{Chamber groups and the spatial search}
\label{subsec:trig_collection}

The reference physical chamber group, or cluster, contains $3\times3$ tracking modules. It collects the four-gap data over $9~\unit{m^2}$ for optical transmission. Each cluster sends every qualified doublet to the central trigger processor. A qualified doublet consists of one hit in each gap of a tracking layer satisfying the spatial and temporal coincidence defined in Section~\ref{subsec:trig_timing}. Single hits from the same cluster are also sent within a collection interval around any qualified doublet. These fast records, called trigger primitives, carry coarse pad positions, timestamps and multiplicity information. The interval includes earlier and later hits that could contribute to an accepted event.

\begin{figure*}[t]
 \centering
 \includegraphics[width=0.97\textwidth]{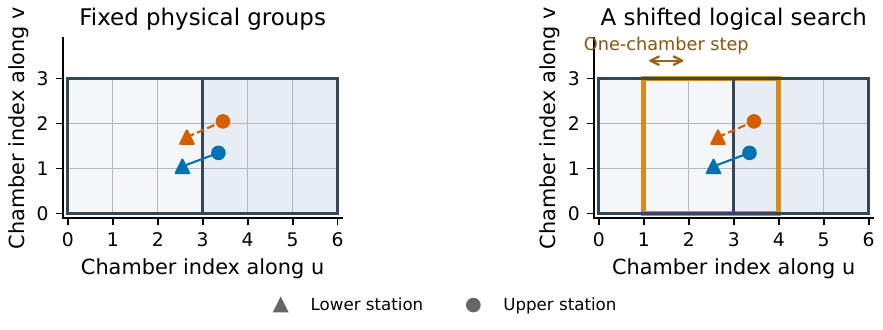}
 \caption{Physical collection and logical correlation, viewed from above. Heavy boundaries delimit fixed $3\times3$ chamber groups. Triangles and circles denote lower- and upper-layer doublets; colour and line style identify the daughter. The event crosses two physical groups (left), while a logical search shifted by one chamber contains all four primitives (right). Central duplicate merging counts an accepted event once.}
 \label{fig:trig_search}
\end{figure*}

The central processor combines these primitives across physical-group boundaries. Its \emph{logical search region} contains $n\times n$ consecutive chamber positions and is moved by one chamber in each horizontal coordinate. Thus, physical grouping determines where data are collected, while the logical region determines which transmitted primitives can enter one event candidate. Figure~\ref{fig:trig_search} illustrates this distinction. Locally recognised candidates may accompany the data as tags; the final decision uses the combined information at the centre.

\begin{figure}[t]
 \centering
 \includegraphics[width=\linewidth]{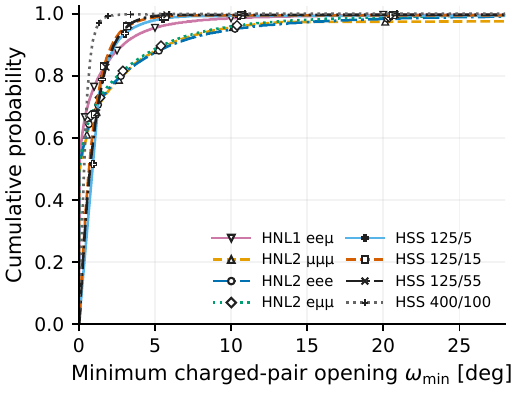}
 \caption{Angular inputs to the geometric acceptance study: cumulative distributions of the minimum opening among charged decay products in the HNL and scalar-portal samples. HSS labels give $(m_H,m_S)$ in GeV; HNL lifetimes and final states are specified in Table~\ref{tab:openAngles}. Openings are sampled from these CDFs; separate broad and IP--vertex bisector hypotheses complete the angular model.}
 \label{fig:anglesCDF}
\end{figure}

To choose $n$, we fold the opening-angle distributions of the HNL and scalar-portal samples of Sections~\ref{sec:phys_hnl} and~\ref{sec:phys_higgs} with the detector geometry. For charged-product unit directions $\mathbf n_i$, define $\omega_{\min}=\min_{i<j}\arccos(\mathbf n_i\cdot\mathbf n_j)$. Figure~\ref{fig:anglesCDF} gives its cumulative distribution function (CDF), $F(\omega)=P(\omega_{\min}\leq\omega)$; Table~\ref{tab:openAngles} lists the 90th percentiles. These percentiles illustrate the collimation, while the complete CDFs determine the sampled openings, including the wider-angle tail.

\begin{figure*}[t]
 \centering
 \includegraphics[width=0.98\textwidth]{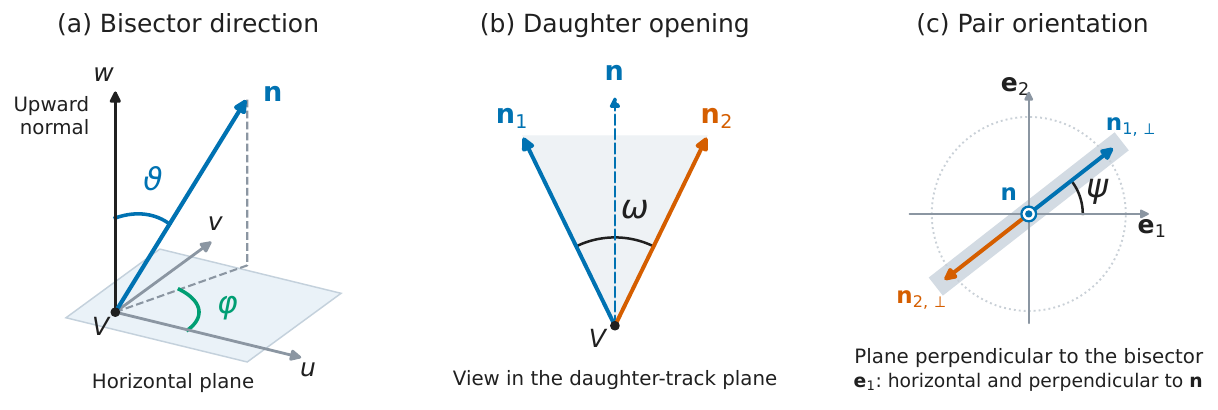}
 \caption{Angular parameters of a charged daughter pair. Unit directions $\mathbf n_1,\mathbf n_2$ define the geometric bisector $\mathbf n=(\mathbf n_1+\mathbf n_2)/|\mathbf n_1+\mathbf n_2|$. (a) Its polar angle $\vartheta$ is measured from the upward normal to the tracking layers; its azimuth $\varphi$ from $+u$ towards $+v$ in the horizontal plane. (b) The opening $\omega$ is measured between the daughters in their common plane. (c) The orientation $\psi$ is measured around the bisector from $\mathbf e_1$ to the transverse projection of daughter 1, with $\mathbf e_1=(-\sin\varphi,\cos\varphi,0)$ and $\mathbf e_2=\mathbf n\times\mathbf e_1$. The shaded band marks the intersection of the daughter plane with the transverse plane.}
 \label{fig:trig_angles}
\end{figure*}

The numerical comparisons use HNL2 $eee$ and HSS $125/15$ from Table~\ref{tab:openAngles}. Their sampled minimum opening defines a representative two-track benchmark. Full-event acceptance requires applying the selection to the correlated momenta of all daughters. Figure~\ref{fig:trig_angles} defines its opening $\omega$, local bisector angles $\vartheta,\varphi$ and orientation $\psi$.

\begin{table}[t]
 \centering
 \caption{Ninetieth percentiles $\omega_{90}=F^{-1}(0.9)$ of the opening distributions in Figure~\ref{fig:anglesCDF}. HSS denotes the decay of a scalar $H$ into two long-lived scalars $S$, labelled by $(m_H,m_S)$ in GeV. Complete CDFs enter the geometric fold.}
 \label{tab:openAngles}
 \begin{tabular}{lc}
 \toprule
 Sample & $\omega_{90}$ [deg]\\
 \midrule
 HNL1, $c\tau=1000~\unit{mm}$, $ee\mu$ & 2.88\\
 HNL2, $c\tau=300~\unit{mm}$, $\mu\mu\mu$ & 5.83\\
 HNL2, $c\tau=1000~\unit{mm}$, $eee$ & 6.10\\
 HNL2, $c\tau=1000~\unit{mm}$, $e\mu\mu$ & 5.54\\
 HSS $(125,5)$ & 2.39\\
 HSS $(125,15)$ & 2.16\\
 HSS $(125,55)$ & 2.30\\
 HSS $(400,100)$ & 0.88\\
 \bottomrule
 \end{tabular}
\end{table}

For the geometric benchmark, the arrangement in Figure~\ref{fig:trig_apparatus} is extended to four flat, aligned $40\times40~\unit{m^2}$ gas-gap surfaces, with G1 at height $H=21.8~\unit{m}$ above the beam. This reference geometry approximates the curved, staggered layout of Section~\ref{sec:performance}. A vertex at depth $h$ below G1 has coordinates $(u_V,v_V,-h)$ and projection $\mathbf r_V=(u_V,v_V)$ onto G1. The benchmark samples vertices uniformly over $0.5\leq h\leq15~\unit{m}$, with $\sqrt{u_V^2+(H-h)^2}\geq9.5~\unit{m}$ following the beam-distance exclusion of Section~\ref{sec:loc_fiducial}. The sampled depth interval defines the reference ensemble used below.

Two bisector hypotheses test the sensitivity of signal collection to the pair direction. The broad-direction hypothesis draws the local geometric angles $\vartheta,\varphi$ uniformly over directions intercepting G1, normalised separately for each vertex. The IP--vertex hypothesis aligns the bisector with the line from the interaction point to the decay vertex. In both ensembles $\psi$ is uniform around the bisector and the opening follows the chosen CDF. Their relative weight in a physical signal requires the full charged-pair direction distribution.

For daughter $j$, the horizontal slope is $\mathbf q_j=\mathbf n_{j,\perp}/n_{j,w}$. Here $n_{j,w}$ and $\mathbf n_{j,\perp}$ are the upward and horizontal components of the daughter unit direction. Its intersection points with the four gas-gap planes are
\begin{equation}
 \mathbf r_{g,j}=\mathbf r_V+(h+w_g)\mathbf q_j.
 \label{eq:trig_impacts}
\end{equation}
The scalar distances between the two daughter intersection points at G1 and G3 are $s_L=h|\mathbf q_2-\mathbf q_1|$ and $s_U=(h+L)|\mathbf q_2-\mathbf q_1|$. Both distances are measured within their respective horizontal sensitive surfaces.

The reference events have both daughters crossing all four instrumented surfaces. This common denominator includes events distributed across physical groups. Let $i,j$ index the chambers containing the eight ideal intersection points along $u,v$. The consecutive chamber counts needed to contain them are
\begin{equation}
 N_u=i_{\max}-i_{\min}+1,\qquad N_v=j_{\max}-j_{\min}+1.
 \label{eq:trig_chambers}
\end{equation}
A logical $n\times n$ search moved by one chamber contains every configuration with $N_u\leq n$ and $N_v\leq n$. Counting includes both extreme chambers: impacts in chambers 1 and 4 require four consecutive chambers. We distinguish events contained in one physical group from events distributed across several groups.

\begin{figure*}[t]
 \centering
 \includegraphics[width=0.98\textwidth]{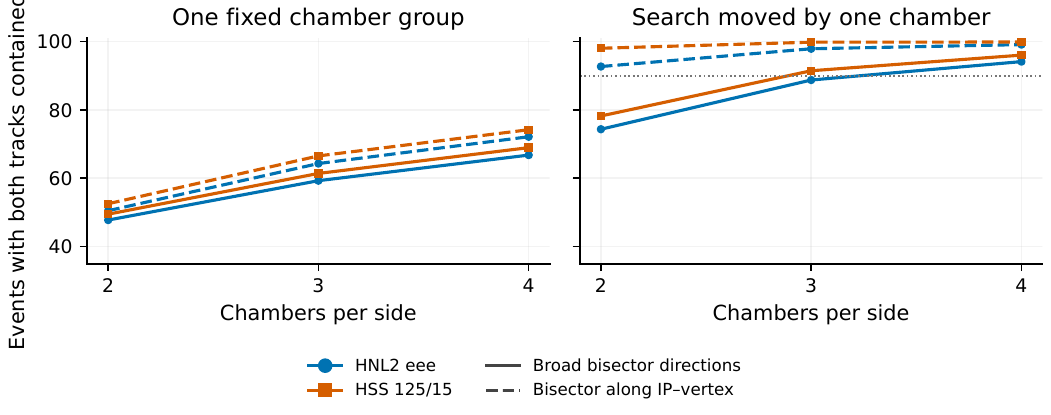}
 \caption{Ideal geometric acceptance for containing all eight impact points in a fixed chamber group (left) or a logical region moved by one chamber (right). All points are assumed available. Pairs cross all four surfaces; vertices are uniform from 0.5 to 15 m below G1. Origins are averaged over chamber-grid alignments. Complete trigger response follows in Section~\ref{subsec:trig_rates}.}
 \label{fig:trig_boundary}
\end{figure*}

Table~\ref{tab:trig_boundary} and Figure~\ref{fig:trig_boundary} give the ideal geometric acceptance for collecting all eight impact points, assuming full information availability. They guide the search size before detector response and trigger selection. Group boundaries align across all four planes; fractions average over integer offsets on the chamber grid.

\begin{table*}[t]
 \centering
 \caption{Geometric containment of all eight ideal intersection points. Depth is measured below G1. The first four rows use vertices uniform from 0.5 to 15 m; the remaining rows use fixed depths. The fixed-group column refers to one physical $3\times3$ group. Sliding searches move by one chamber. Angles are defined in Figure~\ref{fig:trig_angles}.}
 \label{tab:trig_boundary}
 \begin{tabular}{lllccc}
 \toprule
 Sample & Bisector & Depth [m] & Fixed $3\times3$ & Sliding $3\times3$ & Sliding $4\times4$\\
 \midrule
 HNL2 $eee$ & Broad & $0.5$--$15$ (uniform) & $59.3\%$ & $88.8\%$ & $94.1\%$\\
 HSS $125/15$ & Broad & $0.5$--$15$ (uniform) & $61.4\%$ & $91.5\%$ & $96.0\%$\\
 HNL2 $eee$ & IP--vertex & $0.5$--$15$ (uniform) & $64.3\%$ & $97.9\%$ & $99.1\%$\\
 HSS $125/15$ & IP--vertex & $0.5$--$15$ (uniform) & $66.5\%$ & $99.8\%$ & $99.9\%$\\
 HNL2 $eee$ & Broad & 5 (fixed) & $58.4\%$ & $88.2\%$ & $93.9\%$\\
 HSS $125/15$ & Broad & 5 (fixed) & $60.5\%$ & $89.9\%$ & $95.1\%$\\
 HNL2 $eee$ & Broad & 15 (fixed) & $62.3\%$ & $90.8\%$ & $94.9\%$\\
 HSS $125/15$ & Broad & 15 (fixed) & $64.6\%$ & $96.2\%$ & $98.9\%$\\
 \bottomrule
 \end{tabular}
\end{table*}

The gain from moving the logical search motivates central correlation across physical-group boundaries; increasing its size recovers more extended events.

The fixed-depth rows of Table~\ref{tab:trig_boundary} quantify collection at 5 and 15 m. Small daughter opening angles favour compact searches; a common pair inclination can still carry the trajectories across physical-group boundaries.

The baseline transmission rule can discard a single hit whose physical group contains no qualified doublet. It can also miss a doublet whose two hits lie on opposite sides of a physical-group boundary. The $8~\unit{cm}$ separation makes the latter effect sensitive to inclination. Sharing boundary-chamber data between the relevant concentrators can recover these coincidences; diagonal contacts require the corresponding corner information. A mesh of neighbouring concentrators can also perform the logical search locally. Central collection avoids this mesh at the cost of the additional event loss quantified in Section~\ref{subsec:trig_rates}.

\subsection{Timing requirements and coincidence windows}
\label{subsec:trig_timing}

All timestamps use a calibrated common clock. The timing requirements first establish a local doublet, then test upward flight between tracking layers, and finally test whether the selected contributions can belong to one decay. Their application depends on the available hit pattern; the acceptance conditions are specified in Section~\ref{subsec:trig_selection}.

For a relativistic daughter from a decay at time $t_0$, the expected arrival time is
\begin{equation}
 t_{g,j}=t_0+\frac{h+w_g}{c\,n_{j,w}}.
 \label{eq:trig_truth}
\end{equation}
Time differences remove the unknown decay time.

\begin{figure*}[t]
 \centering
 \includegraphics[width=0.99\textwidth]{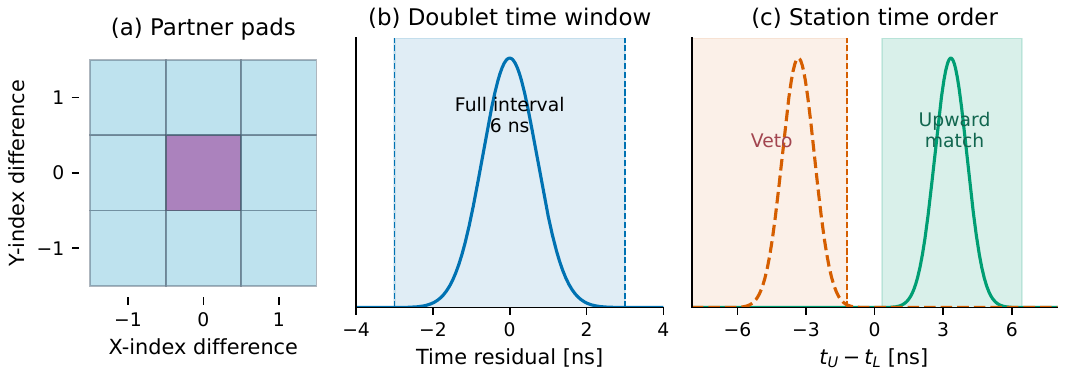}
 \caption{Local and layer-to-layer timing. (a) The neighbour pattern includes the corresponding partner pad and its eight neighbours. (b) The local time residual after subtracting $0.267~\unit{ns}$ lies within $\pm3~\unit{ns}$. (c) Solid and dashed curves represent upward and downward tracks. Vertical G1--G3 flight gives mean time differences $+3.34~\unit{ns}$ upwards and $-3.34~\unit{ns}$ downwards. The upward interval includes pad-position uncertainty and a $3~\unit{ns}$ allowance. Every tested comparison must pass the descending veto; at least one must match upward flight.}
 \label{fig:trig_localtiming}
\end{figure*}

\paragraph{Doublet coincidence.}
Figure~\ref{fig:trig_localtiming}(a,b) illustrates the virtual-pad matching pattern and the local time window. The corresponding-pad option requires equal X and Y group indices in the two gaps. The neighbour option also accepts an index difference of one in either coordinate, including diagonal combinations. It therefore covers the corresponding virtual pad and its eight neighbours. Comparisons can cross a chamber boundary within the physical group.

For either pattern, the measured times obey
\begin{equation}
 |t_{\rm upper\ gap}-t_{\rm lower\ gap}-d/c|\leq3~\unit{ns},
 \qquad d/c=0.267~\unit{ns}.
 \label{eq:trig_localtime}
\end{equation}
The full window is $6~\unit{ns}$, centred on the nominal upward flight time across the doublet. The uncertainty of this time difference is $\sqrt2\sigma_t=0.707~\unit{ns}$. This local condition forms the primitives transmitted by the clusters.

\paragraph{Selecting upward-going candidates.}
We compare the hit times in the lower and upper tracking layers to select upward-going candidates from decays below the detector and reject downward-going cosmic-ray tracks. Figure~\ref{fig:trig_localtiming}(c) shows the two time orders. A complete daughter supplies two doublets, one per tracking layer. A candidate with one doublet and a single hit in the other tracking layer also supplies a layer-to-layer timing measurement. Contributions with only two hits are validated together with the other daughter at event level.

The virtual-pad addresses bound the distance between a lower-layer hit and an upper-layer hit. For an index difference $\Delta i$ in one coordinate, the horizontal separation lies between $b\max(|\Delta i|-1,0)$ and $b(|\Delta i|+1)$. Combining the two coordinates with the known vertical separation gives a shortest flight time $\tau_{\min}$ and a longest flight time $\tau_{\max}$. These intervals can be stored or evaluated arithmetically.

Writing $\Delta t_{UL}=t_U-t_L$, an upward match satisfies
\begin{equation}
 \tau_{\min}-3~\unit{ns}\leq\Delta t_{UL}\leq\tau_{\max}+3~\unit{ns}.
 \label{eq:trig_uptime}
\end{equation}
At least one tested comparison must satisfy this condition. Every tested comparison must also satisfy $\Delta t_{UL}>-\tau_{\min}+3\sqrt2\sigma_t$, which vetoes a time order compatible with a descending track. At normal incidence, a relativistic particle takes $3.34~\unit{ns}$ to travel between G1 and G3. The tests use coarse positions and time differences before a precision trajectory fit is available.

\begin{figure*}[t]
 \centering
 \includegraphics[width=0.99\textwidth]{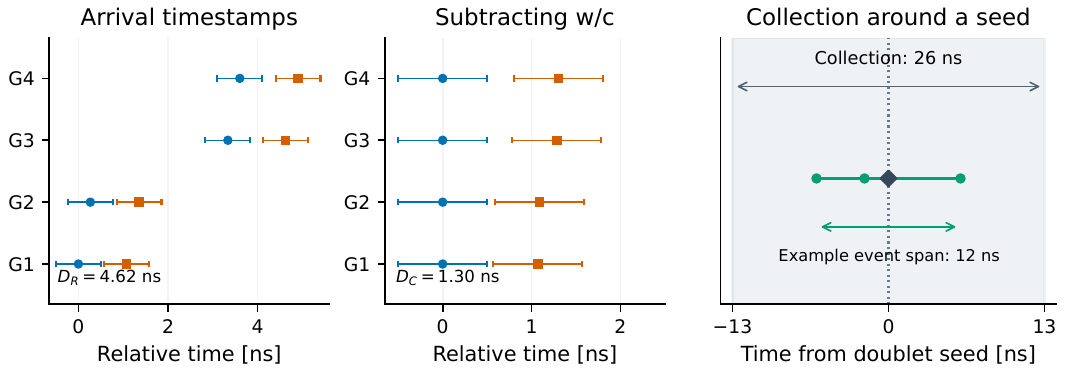}
 \caption{Arrival times for a common vertex $5~\unit{m}$ below G1 and daughter inclinations $0^\circ$ and $20^\circ$, before and after subtracting $w_g/c$ (first two panels). Circles and squares identify the daughters; bars show $0.5~\unit{ns}$ single-hit resolution. $D_R$ uses doublet lower-gap times; $D_L,D_U,D_C$ use corrected hit times. The final panel is a separate example: a seed at zero (diamond) collects hits within $\pm13~\unit{ns}$, while the selected event spans at most $13~\unit{ns}$.}
 \label{fig:trig_eventtiming}
\end{figure*}

\paragraph{Two daughters from one decay.}
The selected contributions must also fit within an event-time interval that allows different daughter flight paths. Figure~\ref{fig:trig_eventtiming} shows an asymmetric example. For hit-based conditions, we subtract the known normal-flight offsets, $t'_i=t_i-w_i/c$, and form the ranges $D_L$ in the lower tracking layer, $D_U$ in the upper tracking layer and $D_C$ over all selected hits. The event requires $D_L\leq S_L$, $D_U\leq S_U$ and $D_C\leq W_C$. For a condition based on doublet coincidences, each doublet supplies its lower-gap timestamp; the range of those references is $D_R$, required to satisfy $D_R\leq W_R$.

\begin{table}[t]
 \centering
 \caption{Common fixed timing settings for both logical search sizes. Event spans are maximum-minus-minimum values; the collection interval is centred on a doublet seed.}
 \label{tab:trig_gates}
 \begin{tabular}{lc}
 \toprule
 Timing condition & Setting\\
 \midrule
 Local doublet residual & $\pm3~\unit{ns}$\\
 Doublet-reference event span $W_R$ & $16~\unit{ns}$\\
 Lower-layer corrected span $S_L$ & $8~\unit{ns}$\\
 Upper-layer corrected span $S_U$ & $9~\unit{ns}$\\
 Complete corrected span $W_C$ & $13~\unit{ns}$\\
 Single collection about a doublet & $\pm13~\unit{ns}$ (full $26~\unit{ns}$)\\
 \bottomrule
 \end{tabular}
\end{table}

Spatial reach and timing acceptance are linked by path length. Within a tracking layer, a common vertex gives $|\Delta t|\leq s/c$ for daughter separation $s$. An $n\times n$ chamber region admits $s\leq\sqrt2 n~\unit{m}$: conservative bounds of $14.2$ and $18.9~\unit{ns}$ for $n=3$ and $4$, before timing uncertainty. Table~\ref{tab:trig_gates} fixes the same working gates for both sizes; the signal calculation includes their asymmetric-path losses. Full use of a larger geometric envelope requires reassessing these gates. A cluster collects single hits with $|t'_i-t'_{\rm seed}|\leq13~\unit{ns}$, where $t'_{\rm seed}$ is the corrected lower-gap timestamp of any qualified doublet in that cluster. The full collection interval is therefore $26~\unit{ns}$. It covers all selected event times with span $W_C\leq13~\unit{ns}$ whenever a selected doublet supplies the local seed.

The doublet coincidence is formed locally. The upward-flight test can be evaluated within a cluster when its necessary hits are present and completed centrally for distributed trajectories. The two-daughter condition is evaluated after combining the cluster data. The final central output is subject to the $100~\unit{Hz}$ reference line and the $5~\unit{\mu s}$ \anubis latency target.

\subsection{Event acceptance conditions}
\label{subsec:trig_selection}

The preceding windows define temporal compatibility. We now specify which physical hit combinations are tested. A four-hit daughter provides two doublets and a three-hit daughter one doublet plus a confirmation in the other tracking layer. The six-hit event patterns include $3+3$ and $4+2$. In a $4+2$ pattern, the less complete contribution is assessed together with the four-hit daughter; its two hits may form a doublet or lie in different tracking layers if those single hits have been transmitted.

Three alternative conditions quantify the trade between efficiency and accidental rate. Each condition is evaluated separately:
\begin{enumerate}
 \item \textit{Three doublets}: three disjoint doublet coincidences using the neighbour matching pattern, with two in one tracking layer and one in the other. Their lower-gap reference timestamps supply the upward-flight comparisons and the event range $D_R$.
 \item \textit{Six hits, corresponding-pad pairs}: six distinct hits, with at most two in each gap, containing at least two disjoint corresponding-pad doublet coincidences. All six timestamps supply the upward-flight comparisons and the ranges $D_L,D_U,D_C$.
 \item \textit{Six hits, neighbour-pad pairs}: the same six-hit condition, with the neighbour matching pattern for its two required doublets.
\end{enumerate}
Requiring six distinct hits with at most two per gap selects a two-particle hit pattern. Hit membership and the additional pad-multiplicity information prevent a single trajectory from being counted twice. The upward-flight comparisons test all selected lower--upper pairs from the specified timestamp set: at least one must match upward flight and every pair must pass the descending-track veto. This is a fully specified coincidence algorithm on coarse information.

Each tested six-hit combination must fit in a logical $n\times n$ region. Seven- and eight-hit events are accepted when any eligible six-hit subset satisfies the condition. Running time buffers permit the comparisons to be made as primitives arrive. Copies of a primitive in neighbouring search regions retain the same identifiers, allowing duplicate event candidates to be merged. Physical-cluster track tags are optional.

\subsection{Signal acceptance, data flow and accidental event rates}
\label{subsec:trig_rates}

We now apply the complete trigger response to the geometric reference in Figure~\ref{fig:trig_boundary}. Acceptance $\eta_{\rm event}$ retains the denominator of pairs crossing all four surfaces. We apply gap inefficiency, local doublet formation, transmission, logical search, timing and multiplicity conditions. Each six-, seven- or eight-hit registration pattern receives its exact probability $\epsilon^k(1-\epsilon)^{8-k}$. Pad addresses follow the intersection points. Each hit receives an independent Gaussian time error with $\sigma_t=0.5~\unit{ns}$, reused across its tests.

\begin{table}[t]
 \centering
 \caption{OR16 six-hit neighbour-pad acceptance at $98\%$ gap efficiency and common timing settings, for vertices uniform from 0.5 to 15 m. Broad bisectors sample directions over G1 acceptance; IP--vertex bisectors follow the line from the interaction point to the decay vertex. Each entry gives central acceptance $\eta_{\rm event}$ [\%] / additional transmission loss $\Delta\eta$ [percentage points]. Full detected-hit availability gives $\eta_{\rm event}+\Delta\eta$.}
 \label{tab:trig_transport}
 \begin{tabular}{lcc}
 \toprule
 Sample / bisector & $n=3$ & $n=4$\\
 \midrule
 \shortstack[l]{HNL2 $eee$\\Broad} & $91.17$ / $1.22$ & $93.46$ / $1.82$\\
 \shortstack[l]{HSS $125/15$\\Broad} & $94.09$ / $0.74$ & $96.06$ / $1.05$\\
 \shortstack[l]{HNL2 $eee$\\IP--vertex} & $97.89$ / $0.90$ & $98.49$ / $0.89$\\
 \shortstack[l]{HSS $125/15$\\IP--vertex} & $99.25$ / $0.57$ & $99.27$ / $0.58$\\
 \bottomrule
 \end{tabular}
\end{table}

Table~\ref{tab:trig_transport} compares the broad-direction and IP--vertex bisector hypotheses defined in Section~\ref{subsec:trig_collection}. For each sample it gives central acceptance and the additional loss from transmission filtering, relative to full detected-hit availability. The suppressed-hit set is independent of $n$; a larger search changes the event loss by admitting new events or alternative surviving combinations.

For the six-hit neighbour-pad condition in Table~\ref{tab:trig_transport}, the extra loss is about $0.6$--$1.8$ percentage points across the four benchmarks. The three-doublet condition is more sensitive to this loss: OR16 gives about $2.7$ and $3.1$ percentage points for the broad HNL sample with $n=3$ and $4$, respectively. Boundary sharing can recover acceptance; increasing $n$ uses the surviving information at a cost in accidental rate, processing and latency.

\begin{figure*}[t]
 \centering
 \includegraphics[width=0.98\textwidth]{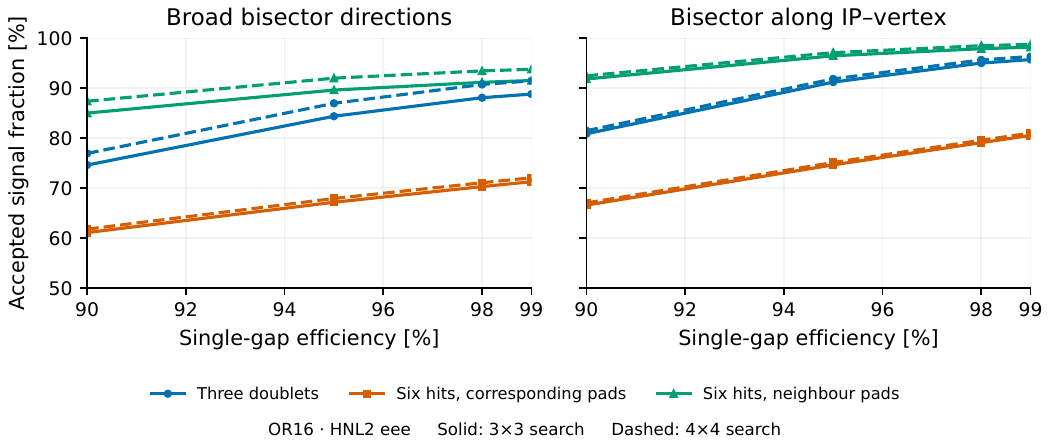}
 \caption{Complete trigger acceptance for OR16 and HNL2 $eee$, including gap response, transmission and coincidences. The geometric reference in Figure~\ref{fig:trig_boundary} requires all eight impact points; the trigger accepts any valid six-hit subset (Section~\ref{subsec:trig_selection}). Vertices are uniform from 0.5 to 15 m. Left: uniform geometric bisector angles over G1-intercepting directions; right: IP--vertex alignment. Denominators contain pairs crossing all four surfaces. Fixed physical $3\times3$ groups transmit neighbour-pad doublets and associated singles within $\pm13~\unit{ns}$ of a local doublet in corrected time, using their own chamber data. Central correlation combines transmitted data across group boundaries: solid $3\times3$, dashed $4\times4$ logical regions, shifted by one chamber. Three-doublet and six-hit neighbour-pad curves allow the corresponding pad and its eight neighbours. Six-hit corresponding-pad curves require equal pad indices for at least two pairs on the same transmission stream. Timing gates are in Table~\ref{tab:trig_gates}; six-hit neighbour-pad values at $\epsilon=98\%$ match the HNL2 entries of Table~\ref{tab:trig_transport}.}
 \label{fig:trig_gapeff}
\end{figure*}

Figure~\ref{fig:trig_gapeff} shows the resulting acceptance under both bisector hypotheses. Any valid six-hit subset can trigger, including events whose eight-point pattern extends beyond the search region. Transmission filtering and spatial and temporal coincidences remain active at unit gap efficiency. All curves include central correlation across physical-group boundaries.

At low occupancy, independent-hit accidentals scale as $(\epsilon r_0)^6$; the simulation also includes additional hits and merged proposals.

\paragraph{Traffic from a physical chamber group.}
We estimate the transmitted random-hit contribution to assess the required bandwidth. The per-gap reference counting density is $r_0=100~\unit{Hz/cm^2}$ at unit efficiency, and the detected density is $r_{\rm bg}=\epsilon r_0$. Photon- and neutron-induced counts are treated as independent Poisson processes. With a nine-pad matching pattern, a doublet of area $A_D$ has rate
\begin{equation}
 R_D\simeq A_D(9b^2)r_{\rm bg}^2W_D,\qquad W_D=6~\unit{ns}.
 \label{eq:trig_primitive_rate}
\end{equation}
A group serving both tracking layers has raw rate $R_{\rm raw}=4(9~\unit{m^2})r_{\rm bg}$. Its unrelated single-hit payload per doublet seed is approximately $R_{\rm raw}(26~\unit{ns})$, before merging repeated intervals.

Table~\ref{tab:trig_primitives} gives the resulting doublet and collected-single rates per physical group, with an illustrative payload.

\begin{table}[t]
 \centering
 \caption{Nominal random-hit traffic per full $3\times3$ physical group serving both tracking layers, at $r_0=100~\unit{Hz/cm^2}$ and $\epsilon=98\%$. The doublets use the neighbour pattern and full $6~\unit{ns}$ window. Extra singles use the full $26~\unit{ns}$ collection interval. Payload estimates illustrate 128 bits per doublet and 64 bits per extra single, before line overhead.}
 \label{tab:trig_primitives}
 \begin{tabular}{cccc}
 \toprule
 OR & Doublets [kHz] & Extra singles [kHz] & Payload [Mbit/s]\\
 \midrule
 8 & $6.48$ & $5.95$ & $1.21$\\
 16 & $25.93$ & $23.79$ & $4.84$\\
 24 & $58.34$ & $53.52$ & $10.89$\\
 \bottomrule
 \end{tabular}
\end{table}

\paragraph{Final accidental-trigger rate.}
The output rate counts accepted events. All proposals sharing hit identifiers are joined into one event, including proposals from neighbouring search regions and alternative six-hit subsets of a larger hit record. The trigger decision for that event is the logical OR of its proposals. Thus,
\begin{equation}
 R_P=\frac{N_{\rm accepted\ events}}{T_{\rm live}}.
 \label{eq:trig_eventrate}
\end{equation}
The independent photon- and neutron-hit model is a stationary Poisson process with detected density $r_{\rm bg}=\epsilon r_0$. We simulate complete hit records, apply the baseline transmission filter and the event conditions of Section~\ref{subsec:trig_selection}, and count each accepted event once. The spatial test uses the chamber-index extents of Eq.~\ref{eq:trig_chambers}, which directly tests containment in any one-chamber-step logical region.

Importance sampling generates rare coincidences efficiently, restores the surrounding Poisson background and weights each accepted event by the inverse number of sampled coincidences that could generate it. The reference points use two million trials per configuration. The event logic agrees in 38\,850 comparisons with the signal-selection implementation; the weighted rate agrees within one standard error with a direct Poisson-stream simulation in an accelerated reference sample. Additional-hit contributions and repeated six-hit proposals are included.

Correlated descending-muon contributions are assigned a separate allowance $U_\mu$, using $F_\mu=0.544~\unit{m^{-2}s^{-1}}$ and the Gaussian time-order veto. This allowance bounds the configurations in which a cosmic muon supplies selected hits together with random counts or other muons. It gives an upper limit on their additional accepted-event rate. The reference comparison uses $R_P+U_\mu$, with the Monte Carlo statistical margin stated separately. Precision cosmic rejection is discussed in Section~\ref{subsec:trig_offline}.

\begin{figure*}[t]
 \centering
 \includegraphics[width=0.90\textwidth]{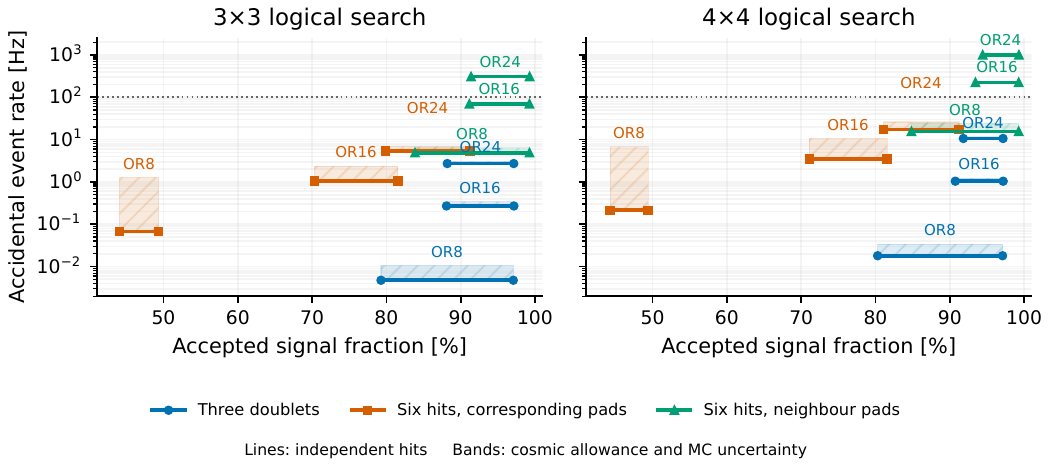}
 \par\smallskip
 \includegraphics[width=0.90\textwidth]{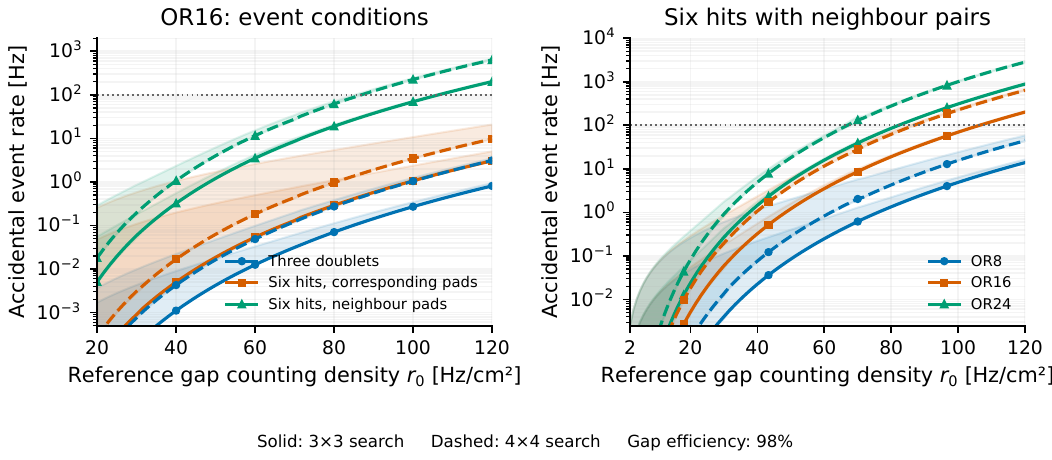}
 \caption{Final-trigger figures of merit with baseline transmission, one-chamber search steps and the timing gates of Table~\ref{tab:trig_gates}. Lines and markers give the simulated independent-hit event rate $R_P$; bands extend from $R_P-3\sigma_{\rm MC}$ to $R_P+3\sigma_{\rm MC}+U_\mu$, where $U_\mu$ is the descending-muon allowance at $F_\mu=0.544~\unit{m^{-2}s^{-1}}$. Each accepted event is counted once. Top: $\epsilon=98\%$ and $r_0=100~\unit{Hz/cm^2}$; horizontal segments give minimum--maximum signal acceptance across HNL2 $eee$ and HSS $125/15$ with both bisector hypotheses, at fixed trigger settings. Panels compare $3\times3$ and $4\times4$ searches. Bottom left: density scan for the three OR16 conditions. Bottom right: six-hit neighbour-pad condition with OR8/16/24, extending to $2~\unit{Hz/cm^2}$. Solid and dashed curves use $3\times3$ and $4\times4$ searches; gap efficiency remains $98\%$. Dotted lines mark $100~\unit{Hz}$.}
 \label{fig:trig_performance}
 \label{fig:trig_density}
\end{figure*}

Table~\ref{tab:trig_rateeff} and Figure~\ref{fig:trig_performance} report event-level rates and signal acceptance. For each fixed trigger configuration, the $\eta_{\rm event}$ range gives the minimum and maximum across HNL2 $eee$ and HSS $125/15$, each evaluated with broad or IP--vertex bisectors. This range describes variation among the four signal benchmarks at fixed gap efficiency.

\begin{table*}[t]
 \centering
 \caption{Final-trigger figures of merit at $98\%$ gap efficiency and $r_0=100~\unit{Hz/cm^2}$. Each $\eta_{\rm event}$ range is the minimum--maximum acceptance of HNL2 $eee$ and HSS $125/15$, with broad and IP--vertex bisectors, after baseline transmission. $R_P$ is the simulated independent-hit event rate, counting all proposals from an event once; its Monte Carlo statistical error is below $0.7\%$. $U_\mu$ is a separate upper allowance for correlated descending-muon contributions at $F_\mu=0.544~\unit{m^{-2}s^{-1}}$. The total accidental rate is bounded by $R_P+U_\mu$, with the Monte Carlo error added separately. All rates are in Hz and use Table~\ref{tab:trig_gates}.}
 \label{tab:trig_rateeff}
 \small\setlength{\tabcolsep}{3pt}
 \begin{tabular}{lccccccc}
 \toprule
  & & \multicolumn{3}{c}{$3\times3$ logical search} & \multicolumn{3}{c}{$4\times4$ logical search}\\
\cmidrule(lr){3-5}\cmidrule(lr){6-8}
Condition & OR & $\eta_{\rm event}$ & $R_P$ & $U_\mu$ & $\eta_{\rm event}$ & $R_P$ & $U_\mu$\\
 \midrule
 Three doublets & 8 & $79.3$--$97.1\%$ & \num{0.00475} & \num{0.00584} & $80.3$--$97.1\%$ & \num{0.018} & \num{0.0157}\\
  & 16 & $88.1$--$97.2\%$ & \num{0.27} & \num{0.0716} & $90.7$--$97.2\%$ & \num{1.04} & \num{0.186}\\
  & 24 & $88.2$--$97.1\%$ & \num{2.72} & \num{0.343} & $91.8$--$97.2\%$ & \num{10.6} & \num{0.881}\\
 Six hits, corresponding pads & 8 & $44.1$--$49.4\%$ & \num{0.0673} & \num{1.21} & $44.3$--$49.4\%$ & \num{0.215} & \num{6.54}\\
  & 16 & $70.3$--$81.5\%$ & \num{1.06} & \num{1.28} & $71.1$--$81.6\%$ & \num{3.45} & \num{7.2}\\
  & 24 & $79.9$--$91.2\%$ & \num{5.42} & \num{1.4} & $81.0$--$91.2\%$ & \num{17.6} & \num{8.31}\\
 Six hits, neighbour pads & 8 & $83.8$--$99.3\%$ & \num{4.88} & \num{1.4} & $84.8$--$99.3\%$ & \num{15.6} & \num{8.31}\\
  & 16 & $91.2$--$99.2\%$ & \num{69.4} & \num{2.07} & $93.5$--$99.3\%$ & \num{225} & \num{14.4}\\
  & 24 & $91.4$--$99.3\%$ & \num{308} & \num{3.2} & $94.4$--$99.3\%$ & \num{1.00e3} & \num{24.6}\\
 \bottomrule
 \end{tabular}
\end{table*}

\paragraph{Comparison with the vertex rate measured with \proanubis.}
The rate of reconstructed vertices measured with the \proanubis demonstrator during LHC $pp$ collisions (Section~\ref{sec:proanubis}) provides a complementary, data-driven estimate of the trigger rate, which also includes real vertices from neutral particles scattering or decaying in the cavern.
Scaling the measured rates linearly with the instrumented area, from $0.99\times1.78~\unit{m^2}$ for \proanubis to $1600~\unit{m^2}$ for the full \anubis detector, \ie by a factor of about 900, the upper, intermediate and lower estimates of $86.1$, $40.8$ and $12.6~\unit{mHz}$ correspond to about $78$, $37$ and $11~\unit{Hz}$, respectively.
All three values lie below the $100~\unit{Hz}$ reference line.
The linear scaling assumes a uniform vertex rate per unit area over the ceiling and does not account for the differences in geometry, layer configuration and selection between \proanubis and the \anubis trigger, which will be assessed in a dedicated study.

\subsection{Recovery near an inactive chamber}
\label{subsec:trig_fault}

An isolated inactive gas-gap chamber removes one potential hit when crossed by one daughter and two when crossed by both. The surrounding chambers remain active. If both daughters cross the same inactive gap, only two doublet coincidences remain possible, both in the other tracking layer. The three-doublet condition then has zero efficiency, while a six-hit condition can accept the remaining $3+3$ pattern.

Table~\ref{tab:trig_fault} compares the three conditions when one or both daughters cross an isolated inactive chamber.

\begin{table*}[t]
 \centering
 \caption{Contained diagnostic for an isolated inactive G1 chamber, OR16 and $98\%$ gap efficiency. The vertex is $5~\unit{m}$ below G1; daughter inclinations are $0^\circ$ and $6.1^\circ$. Both trajectories lie inside one physical group and the search is $3\times3$, using the common timing settings. Entries give trigger efficiency / trigger efficiency with a recorded layer-to-layer lever arm on each daughter. This recorded pattern requires at least three hits per daughter, or $4+2$ with one of the two hits in each tracking layer.}
 \label{tab:trig_fault}
 \begin{tabular}{lcc}
 \toprule
 Condition & One daughter affected & Both affected\\
 \midrule
 Three doublets & $88.6\%$/$86.8\%$ & $0.0\%$/$0.0\%$\\
 Six hits, corresponding pads & $98.5\%$/$96.8\%$ & $83.7\%$/$83.7\%$\\
 Six hits, neighbour pads & $99.2\%$/$97.4\%$ & $88.6\%$/$88.6\%$\\
 \bottomrule
 \end{tabular}
\end{table*}

The availability of six recorded hits in this latter case is $\epsilon^6$, or $88.58\%$ at $\epsilon=98\%$. Actual trigger acceptance additionally requires the pad, timing and transmission conditions. In particular, the single hits in the affected tracking layer must reach a cluster with a qualified seed doublet or be recovered through data sharing.

Six-hit neighbour-pad recovery can be enabled in logical regions containing a known inactive chamber, together with the three-doublet selection over the full detector. The combined decision accepts an event once when either condition is satisfied. The increment is calculated with zero activity in the inactive G1 chamber and counts only events that the three-doublet condition would reject. For one interior inactive G1 chamber, OR16 and $3\times3$ searches, the added independent-hit event rate is $0.447\pm0.004~\unit{Hz}$ at the reference working point. A descending-muon allowance of at most $0.013~\unit{Hz}$ accompanies this increment. The useful signal recovery and its accepted offline background determine whether to enable this option.

\subsection{Precision reconstruction and background rejection}
\label{subsec:trig_offline}

Offline reconstruction fits all recorded fine-coordinate and time measurements to daughter trajectories sharing a three-dimensional vertex and a decay time. For a path length $\ell_i$ from the vertex, the predicted hit time is $t_i=t_0+\ell_i/(\beta_i c)$. The fit uses hit uncertainties, active-region maps and all available measurements, including seventh and eighth hits. It compares the possible hit associations and determines the vertex uncertainty used to set a material-separation margin.

\begin{figure}[t]
 \centering
 \includegraphics[width=\linewidth]{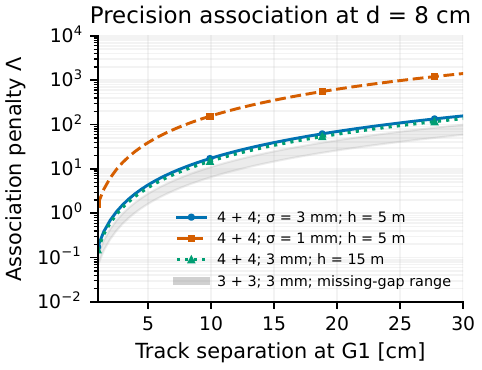}
 \caption{Spatial information for resolving hit associations at $d=8~\unit{cm}$. The residual penalty for a swapped association grows with daughter separation. Curves compare four hits per daughter at $h=5$ and $15~\unit{m}$ and position resolutions of $3$ and $1~\unit{mm}$. The hatched band spans the three-hit missing-gap choices at $5~\unit{m}$. These penalties enter the full vertex fit along with timing and the common-vertex constraint.}
 \label{fig:trig_offline}
\end{figure}

The $8~\unit{cm}$ internal separation provides a local slope measurement with uncertainty $\sqrt2\sigma_x/d$: $0.053$ at $3~\unit{mm}$ position resolution and $0.018$ at $1~\unit{mm}$. This information helps distinguish alternative associations between tracking layers. Figure~\ref{fig:trig_offline} shows the residual penalty $\Lambda=\sum r_i^2/\sigma_x^2$ incurred by a swapped association for symmetric daughter trajectories. Here $r_i$ is the residual of an assigned point from its fitted straight trajectory, and $\sigma_x$ is the single-hit coordinate uncertainty. For a $5~\unit{cm}$ separation at G1, four hits per daughter give $\Lambda=4.37$ at vertex depth $5~\unit{m}$ and $3.80$ at $15~\unit{m}$ with $3~\unit{mm}$ resolution. Reducing the uncertainty to $1~\unit{mm}$ multiplies these penalties by nine. The common-vertex and timing constraints provide additional information.

For an assigned trajectory, the precision direction test fits
\begin{equation}
 t_i=t_*+q\ell_i/c,\qquad
 \sigma_q=\frac{c\sigma_t}{\sqrt{\sum_i(\ell_i-\bar\ell)^2}},
 \label{eq:trig_toffit}
\end{equation}
where $\ell$ increases upwards. Relativistic upward and downward tracks have $q=+1$ and $-1$. At normal incidence, $\sigma_q=0.149$ for four hits and $0.176$--$0.191$ for three. An upward $3\sigma_q$ acceptance band gives a two-track signal efficiency of $99.46\%$ in the independent Gaussian timing model. Its single-track downward survival is $1.45\times10^{-25}$ for four hits and $3.3\times10^{-17}$--$3.6\times10^{-14}$ for the three-hit subsets. Measured timing tails, additional associations and the full event selection determine the final rejection.

The outward daughter directions and fitted decay time also support a causal association with an \atlas interaction. The parent hypothesis allows an unknown velocity and momentum carried by additional charged or neutral products. The vertex fit and \atlas isolation and kinematic information are combined as described in Section~\ref{subsec:col_bkg}.

\paragraph{Cosmic-induced upward vertices.}
\label{subsec:trig_cosmicveto}
The non-collision backgrounds are introduced in Section~\ref{subsec:noncol_bkg}. A real upward secondary vertex has the signal time order. The fitted vertex must lie in air, separated from solid material by a margin set by its reconstruction uncertainty. An earlier descending parent provides a causal veto. The recorded history includes hits outside the firing search region. Two space--time-compatible precursor hits can identify an incoming trajectory to the reconstructed vertex.

For a parent crossing $m$ active gaps, the probability of recording fewer than two hits is
\begin{equation}
 p_{<2}=(1-\epsilon)^m+m\epsilon(1-\epsilon)^{m-1}.
 \label{eq:trig_precursor}
\end{equation}
At $98\%$ efficiency, the corresponding ideal hit-availability suppression is $3.17\times10^4$ for four crossed gaps and $845$ for three. The full veto also includes recognition of the causal trajectory and the probability that the parent crosses instrumented area. A vertical vertex $15~\unit{m}$ below G1 gives about $104~\unit{ns}$ between the G4 parent crossing and the returning daughter signal. The reference $0.5~\unit{\mu s}$ history accommodates the flight times in the benchmark cavern geometry.

The accepted cosmic contribution is obtained by integrating air-interaction production, upward-daughter acceptance and precursor-veto survival over the same volume, including parents that enter outside the apparatus or cross inactive regions. The instrumental contribution similarly requires the full fit over candidate associations and measured response tails. For live exposure $T_{\rm live}$, the design condition is $R_{\rm off}T_{\rm live}<1$. Ten continuous years require $R_{\rm off}<3.17\times10^{-9}~\unit{Hz}$; a $10^8~\unit{s}$ live exposure requires $10^{-8}~\unit{Hz}$. These event-level requirements guide the validation with simulation and data control regions.

\subsection{Design choices supported by the study}
\label{subsec:trig_outlook}

OR16 with a $3\times3$ search supports the six-hit neighbour-pad trigger at the reference point $r_0=100~\unit{Hz/cm^2}$ and $\epsilon=98\%$. Its independent-hit rate is $69.4~\unit{Hz}$; including the full descending-muon allowance and three Monte Carlo standard errors keeps the final rate below $73~\unit{Hz}$ (Table~\ref{tab:trig_rateeff}). Central signal acceptance spans $91.2$--$99.2\%$ across the declared angular benchmarks. The three-doublet alternative has an independent-hit rate of $0.270~\unit{Hz}$ with lower acceptance and less recovery near an inactive chamber.

The $4\times4$ search increases the six-hit independent rate to $224.5~\unit{Hz}$ while recovering additional signal. With the same gates and gap efficiency, the $100~\unit{Hz}$ budget is reached at $r_0\simeq85.3~\unit{Hz/cm^2}$ for $4\times4$ and $105.7~\unit{Hz/cm^2}$ for $3\times3$, including the same allowance and statistical margin (Figure~\ref{fig:trig_density}, lower panels).

%% file: sections/interface.tex
As outlined in Section~\ref{sec:intro}, the \anubis detector relies on the full event information recorded by the \atlas detector to suppress backgrounds and to reach its full sensitivity (Sections~\ref{sec:phys_strengths} and~\ref{sec:backgrounds}).
This results in a close relationship between the \atlas and \anubis detectors.
The \anubis detector would make use of all \atlas data recorded in the bunch crossings of interest and provide a trigger to the \atlas experiment, while also receiving triggers from it (Section~\ref{sec:trigger}).
The \anubis detector is therefore conceived as an integral extension of the \atlas experiment, rather than as a stand-alone experiment.

\subsection{Benefits of the integration}
\label{sec:int_symbiosis}

The integration of the \anubis detector into the \atlas experiment is beneficial for both.

For the \atlas experiment, the \anubis detector effectively extends the muon spectrometer by two tracking layers about 23~\metre above the beamline, covering a large fraction of the solid angle above the \atlas detector.
This allows the \atlas experiment to:
\begin{itemize}
\item reconstruct muons falling into the acceptance of the \anubis detector with an additional, longer lever arm;
\item improve the reconstruction of displaced vertices in the outer layers of the muon spectrometer, whose decay products are also recorded by the \anubis detector;
\item measure the time of flight (ToF) of charged particles with a precision of $\order{0.5~\text{ns}}$ over a distance of more than 23~\metre, corresponding to a precision on the velocity $\beta$ of $\order{0.5\%}$, which can be exploited, \eg, in searches for heavy, slowly moving charged particles.
\end{itemize}

For the \anubis detector, the \atlas experiment is essential to fully exploit its physics potential:
\begin{itemize}
\item 
the \atlas detector acts as a passive shield against SM particles produced in the collisions;
\item 
the event information of the \atlas detector provides an active veto.
Prompt objects, jets and \met reconstructed by the \atlas detector are used to reduce the backgrounds from hadronic interactions with material that are induced by SM LLPs (Section~\ref{sec:backgrounds});
\item 
tracks and vertices help to improve the vertex reconstruction efficiency of the \anubis detector.
\end{itemize}

Triggering the \atlas experiment on DV candidates in the \anubis detector, and vice versa, provides access to the full event for every LLP candidate, which is essential to characterise a potential discovery.

\subsection{Relationship with the ATLAS Collaboration}
\label{sec:int_relationship}

The \anubis initiative has enjoyed a strong and fruitful relationship with the \atlas Collaboration since its inception.
The \atlas Technical Coordination (TC) team has been highly helpful in facilitating the installation of the \proanubis demonstrator in the \atlas cavern, including the provision of services, and the \atlas Collaboration has provided storage resources on the CERN EOS system for the \proanubis data.
The \anubis Collaboration is very grateful for this support.

The construction, commissioning, calibration and performance of \proanubis have been documented in Refs.~\cite{ANUBIS:2025mme,ANUBIS:2025inn,ANUBIS:2026jxg}, and further publications are in preparation.
As discussed in Section~\ref{sec:proanubis}, the \proanubis demonstrator recorded $pp$ collision data corresponding to 104~\invfb in 2024 and 73~\invfb in 2025 from its position close to the cavern ceiling, and demonstrated that the RPC technology meets the efficiency and time-resolution requirements of the \anubis detector, with tracking-layer efficiencies above 99\% and a time resolution of about 350~ps per RPC plane.

The \atlas Collaboration is generally supportive of the \anubis initiative, since it expands the physics potential of the \atlas experiment.
The \anubis Collaboration is fully committed to the full physics exploitation of the \atlas experiment, as evidenced by the contribution of several ANUBIS institutes to the Phase~II upgrade of the \atlas RPC system for LS3~\cite{CERN-LHCC-2017-017}.
Further contributions from \anubis institutes to the \atlas experiment are expected in the future.

The next step to formalise the relationship between the \atlas and \anubis Collaborations is to make the \anubis detector a Phase~III upgrade project of the \atlas experiment, to be installed during LS4, with the option of starting the installation of its services and mechanical support system in the YETS of Run~4 if this can be accommodated.
Initial discussions have been constructive, and the formal process of making \anubis a Phase~III upgrade project is expected to start in early 2027, with an Initial Design Review foreseen for spring 2027.

Discussions with the \atlas TC team and CERN engineers have not revealed any fundamental showstoppers for the installation of the \anubis detector~\cite{bib:ludo}.
They highlighted, however, that a detailed engineering proposal for the mechanical and service integration of the \anubis detector into the \atlas cavern is needed, covering the mechanical structure, the retractable shaft plugs, and the provision of power, gas, cooling and readout services (Sections~\ref{sec:engineering} and~\ref{sec:design}).
This proposal will form the backbone of a Technical Design Report (TDR), and requires the support of the \atlas TC team through regular consultations.
At present, the \atlas TC team is fully focused on the successful delivery of the Phase~II upgrade.
There is, however, a general interest in supporting the \anubis initiative along the lines above once engineering capacity becomes available in the \atlas TC team.
The \anubis Collaboration therefore aims to complete a TDR for the detector and electronics by summer 2028, and a separate TDR for the mechanics and services in 2031, after the consultation with the \atlas TC team, as detailed in Section~\ref{sec:timeline}.

\subsection{A demonstrator system at the end of LS3}
\label{sec:int_ls3}

To prepare for the installation of the tracking modules during LS4, the \anubis Collaboration considers installing $\order{10}$ tracking modules across the ceiling towards the end of LS3, if and only if this can be accommodated in the LS3 schedule of the Phase~II upgrade.
These modules would form the \emph{Slice Test}, a demonstrator representing a slice of the full \anubis detector.
They would be built from the prototypes developed in the R\&D towards the final prototype rather than from pre-production modules, and their installation may involve a higher fraction of manual labour and makeshift fixes than that of the full detector.
These would be installed in a location that is representative of the final \anubis detector.
If the installation cannot be accommodated in LS3, the Slice Test would instead be carried out with tracking modules installed manually in accessible positions on the HO-A gantry in the \atlas cavern, which is not ideal but still sufficient for a proof of concept of the integration of the \anubis detector into the \atlas detector control system (DCS) and data-acquisition (DAQ) system.
Such a system would allow the development of the DAQ, including the external L0 trigger path to the \atlas experiment, and of the DCS under realistic conditions during Run~4 (Section~\ref{sec:trigger}).
When the tracking modules of the full \anubis detector are installed in LS4, these systems would then only need to be extended to the full detector.

\subsection{Research and development of common interest}
\label{sec:int_rnd}

The R\&D for the \anubis detector is carried out within the \atlas RPC community, to which several of the institutes involved in the \anubis project belong, so that its results benefit both the \anubis detector and the existing \atlas RPC system.
A prominent example is the search for low-GWP gas mixtures for RPC detectors~\cite{Simsek:2024wnt,Shah:2025luk,Ahmadi:2025pjv,RPCEcoGasGIF:2025gqp}, which is of direct relevance for the operation of the \atlas RPC system at the HL-LHC as well as for the \anubis detector.
Further examples include the development of quality assurance techniques, such as the accelerated ageing of RPC gas gaps using thermal cycling, which has informed the re-design of the gas gaps for the Phase~II upgrade of the \atlas RPCs~\cite{Shah:2025pcy}.

\subsection{Integration into the ATLAS infrastructure}
\label{sec:int_infrastructure}

The \anubis detector would be installed in the \atlas experimental cavern UX15, and would therefore be integrated into the cavern space and services of the \atlas experiment.
The \anubis Collaboration is fully committed to fitting into the existing constraints without causing any destructive interference with the \atlas experiment.
The guiding principles are that the installation of the \anubis detector fits into LS4, which is foreseen from December 2033 to 2036, and, only if they can accommodate it, into the YETS of Run~4, and that the risk to the \atlas experiment and to the existing installations in UX15 is minimised.
Wherever possible, the \anubis detector is therefore designed to be self-sufficient, so that the interfaces to the \atlas infrastructure are few, well defined, and can be developed and validated without impact on \atlas operation.

{\em Cavern space.}
The ceiling of UX15 and the PX14 and PX16 shafts already host several installations, which constrain the layout of the \anubis detector:
\begin{itemize}
\item The ventilation pipework into and out of UX15 runs through PX14 and has a diameter of $\order{1~\metre}$.
In principle, these pipework runs could be altered, provided that they do not obstruct the crane.
\item Other pipework into and out of UX15 also runs through PX14.
It generally has a smaller diameter than the ventilation pipes, so that the resulting loss of acceptance could potentially be accommodated.
Any alterations would be discussed with the \atlas Collaboration.
\item The pipework and foam cannons of the emergency fire-extinguishing system are mounted close to the ceiling.
A relocation of individual elements may be possible, but is subject to strong constraints owing to its health and safety implications.
\end{itemize}

{\em Services.}
The services required by the \anubis detector are summarised below.
\begin{itemize}
\item {\em Mains power (230~V).}
The switchboard and the mains power supply into the cavern need to be extended.
This extension appears straightforward, with minimal impact on the \atlas experiment.
\item {\em Networking.}
The current network system needs to be extended, potentially requiring additional network switches in the \atlas service cavern USA15.
This extension also appears straightforward, with minimal impact on the \atlas experiment.
\item {\em Gas system.}
The improved design and physics capabilities of the \anubis RPCs, including their more sensitive readout, provide the flexibility to operate them with low-GWP gas mixtures (Section~\ref{sec:int_rnd}).
To benefit fully from this flexibility, a gas supply system separate from that of the \atlas RPCs is being considered.
The gas could be supplied directly from the surface building SX15 through PX14 or PX16 into UX15, without the detour through USA15 taken by the legacy and Phase~II \atlas RPC systems.
The environmental and cost impact would be minimised by using mixtures with a high CO$_2$ fraction of more than 80\%.
CO$_2$ is a self-quenching gas with a fast avalanche formation, which is beneficial for the time resolution, and has a GWP of~1~\cite{Shah:2025luk,RPCEcoGasGIF:2025gqp}.
Greenhouse gas emissions and operating costs could be further reduced by recovering fluorinated gases from the RPC exhaust for reuse, as recently demonstrated by the CMS RPC recuperation plant, which achieved an R134a recovery efficiency of approximately 80\%~\cite{Arena:2024recuperation}.
The gas would be distributed through a network installed on the cavern ceiling, running parallel to the beamline at a constant height.
A separate gas system minimises the impact on the \atlas experiment.
\item {\em Low voltage (LV).}
The LV would be generated locally from the mains power.
This avoids the installation of bulky LV crates in USA15 and long runs of thick and heavy cables, and results in a modular and therefore robust system.
The impact on the \atlas experiment is minimal, since the system is located away from USA15.
\item {\em High voltage (HV).}
The HV would also be generated locally from the mains power, potentially using lightweight HV power supplies of the CAEN A75xx family.
For example, a 12~V supply converted from the mains could power both the LV, through DC--DC converters, and compact HV power supplies providing up to 12~kV.
This avoids error-prone long HV cables from US15 into UX15, and has a minimal impact on the \atlas experiment since the system is located away from USA15.
\end{itemize}

\subsection{Interfaces to the ATLAS systems}
\label{sec:int_interfaces}

Beyond the infrastructure, the operation of the \anubis detector as an integral part of the \atlas experiment requires interfaces to the \atlas detector control, timing, data-acquisition and trigger systems.
These interfaces are summarised below.
\begin{itemize}
\item {\em Detector control system.}
The monitoring hardware of the \anubis DCS would be factorised from that of the \atlas experiment.
Ideally, the \anubis DCS would be integrated into the \atlas DCS framework and procedures, which requires a careful integration including a full validation using dry runs.
This represents a significant amount of work, but is in principle straightforward.
\item {\em LHC clock.}
The LHC clock would be received from the \atlas central trigger processor and DAQ system through an optical-fibre distribution, similar to the setup used for \proanubis~\cite{ANUBIS:2025inn}.
Following the \proanubis approach, this is relatively straightforward, with minimal impact on the \atlas experiment.
\item {\em Data acquisition.}
The \anubis DAQ system would be located in USA15 or US15.
Its computing needs, and hence its requirements for rack space and cooling, are modest: since the magnetic field inside the \anubis fiducial volume is small, the track reconstruction reduces to straight-line fits.
The integration with the \atlas DAQ is paramount, and requires significant development, a careful integration and a full validation.
During the development phase, the impact on the \atlas experiment is minimal.
\item {\em Trigger.}
The \anubis trigger would be interfaced to the \atlas L0 trigger through its external trigger path, within the latency envelope of the \atlas L0 trigger of $10~\unit{\mu s}$ (Section~\ref{sec:performance}).
A parallel trigger path in the reverse direction would allow the \atlas experiment to trigger the readout of the \anubis detector, which requires the buffer latency of the \anubis detector to exceed that of the \atlas experiment (Section~\ref{sec:trigger}).
Both trigger directions require a careful development and integration, and a full validation using dry runs with cosmic rays and $pp$ collisions.
During the development phase, the impact on the \atlas experiment is minimal.
\end{itemize}

%% file: sections/timeline.tex
\begin{figure*}[tb]
\centering
\includegraphics[width=\textwidth]{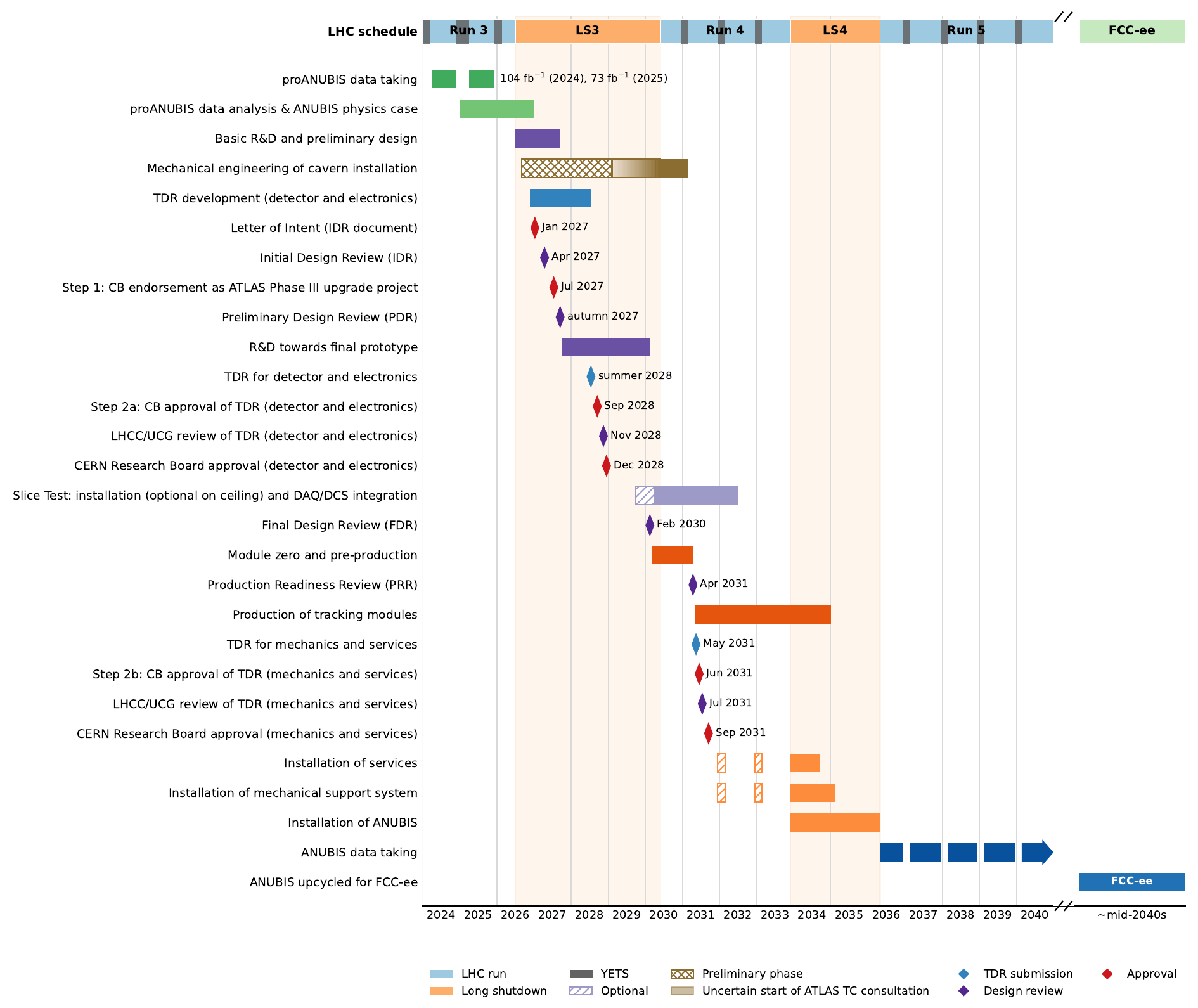}
\caption{
\label{fig:timeline}
Tentative timeline of the \anubis project from 2024 to 2040 in the context of the LHC schedule, including the LHC runs, long shutdowns and YETS.
Future YETS are assumed to last ten weeks, starting in the third week of December.
Design reviews (IDR: Initial Design Review, PDR: Preliminary Design Review, FDR: Final Design Review, PRR: Production Readiness Review), the submission of the TDRs and approvals are shown as milestones.
Optional items, namely the installation of the Slice Test on the cavern ceiling at the end of LS3 and the installation of the services and of the mechanical support system in the YETS of Run~4, are hatched.
The preliminary phase of the mechanical engineering is cross-hatched, and the shading indicates that the consultation with \atlas TC may start about two thirds into LS3 or only after its end.
\anubis data taking is interrupted during the YETS.
After the end of the HL-LHC programme, the \anubis detector could be upcycled for operation at the FCC-ee.}
\end{figure*}

The tentative timeline of the \anubis project is shown in Figure~\ref{fig:timeline} together with the schedule of the LHC.
Run~3 of the LHC ended in June 2026, and LS3 extends until the start of Run~4 in June 2030~\cite{Zerlauth:2026hllhc}.
LS4 is currently scheduled to begin in December 2033 and to run through 2034 and 2035, with beam commissioning for Run~5 expected to start around May 2036.
This reflects the recent one-year shift of the CERN long-term schedule, which extended LS3 and moved the start of Run~4 to 2030.
The YETS of 2024/25 and 2025/26 are shown with their actual dates; for future YETS, a duration of ten weeks starting in the third week of December is assumed.
The timeline is organised such that the installation work in the \atlas cavern is confined to LS4 and, optionally, to the YETS of Run~4, in line with the commitment to avoid any destructive interference with the \atlas experiment (Section~\ref{sec:interfacing}).

{\em Demonstrator and physics case (2024--2026).}
The \proanubis demonstrator recorded $pp$ collision data corresponding to 104~\invfb in 2024 and 73~\invfb in 2025, outside the YETS (Section~\ref{sec:proanubis}).
The analysis of these data, together with the development of the \anubis physics case (Section~\ref{sec:physics}), has been carried out in 2025 and 2026, and culminates in the submission of this Letter of Intent in January 2027.

{\em Project organisation and milestones.}
The planning follows the structure of the \atlas upgrade projects, in which design reviews and approvals are milestones, and the start of each phase is determined by a preceding milestone or by the completion of a preceding task.
At this stage, the timeline is organised as a base scheme of the main phases and milestones: an Initial Design Review (IDR), followed by basic R\&D leading to a Preliminary Design Review (PDR), the path to a final prototype concluded by a Final Design Review (FDR), the path to a module zero and pre-production concluded by a Production Readiness Review (PRR), and the production.
A detailed work breakdown structure for the individual systems, such as the RPCs, the readout electronics, the DAQ and the mechanics, will be developed for the TDR.
The schedule is ambitious, in particular for the production of the tracking modules, but it is designed to be flexible, so that gains and delays in individual tasks can be accommodated as the project proceeds.

{\em Initial Design Review and Step~1 (2027).}
The formalisation of the \anubis detector as an \atlas Phase~III upgrade project proceeds in two steps (Section~\ref{sec:int_relationship}), with the second step split into Step~2a for the detector and electronics and Step~2b for the mechanics and services.
Step~1 starts with the submission of an extended version of this Letter of Intent including ATLAS-internal information as the IDR document in January 2027, followed by the IDR, foreseen for April 2027, in line with the timeline of recent \atlas Phase~II upgrade projects.
The IDR is reviewed by a committee appointed by the \atlas Upgrade Coordinator, which verifies that the performance and technical requirements are clearly defined, that the initial design is technically sound and meets them, and that the integration and installation issues have been discussed with \atlas TC, and which examines the initial cost estimate and the schedule up to the TDR.
Shortly after the IDR, a kick-off meeting with all interested institutes defines the initial organisation of the upgrade project and an interim project leader.
This is followed by a recommendation by the \atlas Upgrade Steering Committee, the approval by the \atlas Executive Board and the endorsement by the \atlas Collaboration Board, foreseen for July 2027, which concludes Step~1 for the \anubis detector to become an \atlas Phase~III upgrade project.
Since the IDR includes the assessment of the integration and installation issues, this endorsement implies that the installation of the \anubis detector in the \atlas cavern is considered in principle feasible, pending the final verification of the mechanical integration in the TDR for the mechanics and services.
The endorsement is also a prerequisite for requesting resources from the funding agencies.

{\em Basic R\&D and Preliminary Design Review (2026--2027).}
The R\&D on the \anubis detector started in July 2026, building on the studies of the gas-gap architecture, gas mixture and front-end electronics described in Section~\ref{sec:design}.
It includes the qualification of 2~mm gas gaps with a low-GWP, \ie SF$_6$-free and R134a-free, gas mixture, or at least a mixture with a R134a fraction of at most 15\%.
The IDR provides recommendations on the R\&D options.
Based on these, initial studies and pre-prototypes built from existing components lead to a preliminary design with a reduced set of options, which is reviewed at the PDR, foreseen for autumn 2027.
The PDR releases the R\&D funds for the construction of real prototypes.
The readout scheme is to be agreed before the PDR, so that the funding for a multi-project wafer run can be secured and a slot at the selected foundry booked in advance, and the prototype ASICs, \eg for the front end and the receiver, can be ordered directly after the PDR.

{\em R\&D towards the final prototype and Final Design Review (2027--2030).}
After the PDR, the R\&D proceeds with real prototypes towards a final prototype of the tracking module.
This includes the development of the readout ASIC with an integrated TDC for 16 channels, which starts directly after the PDR and is an extension of a preliminary design developed for the \atlas Phase~II upgrade that already incorporated a TDC but requires further work, and of the corresponding front-end board.
The detector R\&D and the development of the readout electronics are closely connected, since it is the overall performance of the detector and its readout that is being optimised, \eg the choice of the gas-gap thickness and gas mixture together with the sensitivity of the front end.
The prototypes developed in this phase, which precede the pre-production version, are also used for the Slice Test described below.
The final prototype is reviewed at the FDR, foreseen for February 2030, which authorises the pre-production.

{\em Engineering of the cavern installation (2026--2031).}
The engineering of the installation in the \atlas cavern started in September 2026 with a preliminary phase, in which options for the mechanical support system and the services are developed based on assumptions that do not require detailed consultations with the engineering team of \atlas TC (Section~\ref{sec:engineering}).
The \atlas TC team is currently focused on the Phase~II upgrade, and will support the engineering of the \anubis installation once capacity becomes available, at the latest after the end of LS3, but, circumstances permitting, possibly about two thirds into LS3, \ie from early 2029 (Section~\ref{sec:int_relationship}).
In this phase, the developed options are solidified in consultation with \atlas TC, which is foreseen to be completed nine months after the end of LS3, \ie by February 2031.
The engineering of the cavern installation is therefore described in a separate TDR for the mechanics and services, foreseen for May 2031, so that it does not delay the TDR for the detector and electronics.
Since the mechanical support system and the services are installed in LS4, and optionally from the second YETS of Run~4, this schedule remains compatible with the installation.

{\em TDR for the detector and electronics, Step~2a and final approval (2026--2028).}
The development of the TDR for the detector and electronics starts in November 2026, in parallel with the preparation of the IDR and the basic R\&D, and its submission is foreseen for summer 2028, \ie about 15 months after the IDR and thus within the two years in which the TDR is normally completed for \atlas upgrade projects.
The TDR is based on the preliminary design reviewed at the PDR and on the results of the R\&D towards the final prototype available at that time, while the design is finalised at the FDR.
Since the engineering of the cavern installation is covered by a separate TDR for the mechanics and services, the TDR for the detector and electronics does not depend on the availability of engineering resources in \atlas TC during LS3.
The approval of the TDR releases the funds for the production of the detector, so that the funding agencies can commit these funds well before the start of the pre-production and production.
Both the development of the TDR and Step~2a are \atlas-internal processes.
Before its submission, the TDR undergoes an internal review organised by the \atlas Upgrade Coordinator, including a review of the cost, and is endorsed by the Upgrade Steering Committee and the Executive Board.
The approval of the TDR by the \atlas Collaboration Board, which concludes Step~2a, and the review of the TDR by the LHC Committee (LHCC) and the Upgrade Cost Group (UCG) then proceed in parallel.
Following the example of recent \atlas Phase~II upgrade projects, the Collaboration Board approval is foreseen for September 2028 and the LHCC and UCG review for November 2028.
In the event of a successful review, the LHCC and UCG recommend the approval of the \anubis detector as an \atlas Phase~III upgrade project to the CERN Research Board, whose approval, foreseen for December 2028, \ie about five months after the submission of the TDR, is the final step for the detector and electronics.
In parallel, a Memorandum of Understanding (MoU) is prepared, which defines the responsibilities and commitments of the participating institutes and their funding agencies, and is submitted to the funding agencies for signature after the approval by the Research Board.

{\em TDR for the mechanics and services, Step~2b and final approval (2031).}
The TDR for the mechanics and services builds on the engineering of the cavern installation in consultation with \atlas TC described above, and its submission is foreseen for May 2031.
It follows the same review and approval process as the TDR for the detector and electronics.
After the internal review and the endorsement by the Upgrade Steering Committee and the Executive Board, its approval by the \atlas Collaboration Board, foreseen for June 2031, concludes Step~2b.
The LHCC and UCG review is foreseen for July 2031 and the approval by the CERN Research Board for September 2031, well before the installation of the services and of the mechanical support system, which starts at the earliest in the second YETS of Run~4 in the winter of 2031/32.

{\em Slice Test (2029--2032).}
The Slice Test would comprise only $\order{10}$ tracking modules, which would be built from the prototypes developed in the R\&D towards the final prototype, \ie the version that precedes the pre-production (Section~\ref{sec:int_ls3}).
If, and only if, it can be accommodated in the LS3 schedule of the Phase~II upgrade, the Slice Test would be installed on the cavern ceiling between October 2029 and March 2030.
Otherwise, the Slice Test would be carried out with tracking modules installed manually in accessible positions, \eg on the HO-A gantry, in the \atlas cavern.
This is not ideal, but still sufficient for a proof of concept of the integration of the \anubis detector into the \atlas DCS and DAQ systems.
From March 2030 to June 2032, the Slice Test would be used during Run~4 to develop the full DAQ and DCS interface to the \atlas experiment, including the external L0 trigger path (Section~\ref{sec:trigger}).
This period of about two years corresponds to the time estimated for the full integration into the \atlas DAQ and DCS, after which the effort focuses on other tasks.
The experience gained with the prototypes in the \atlas cavern also informs the pre-production and the PRR.

{\em Module zero, pre-production and production (2030--2034).}
After the FDR, a module zero and the pre-production tracking modules are built and qualified from March 2030 to April 2031.
The PRR, foreseen for April 2031, authorises the production of the \anubis tracking modules, which starts in May 2031 and extends until the end of 2034, \ie into the first year of LS4, in parallel with the installation of the services and of the mechanical support system.

{\em Installation and data taking (from 2033).}
The \anubis detector is installed during LS4, starting in December 2033.
The installation of the services is completed in the first third of LS4, by about September 2034, and that of the mechanical support system in the first half of LS4, by about February 2035, while the tracking modules are installed until the end of LS4.
If, and only if, it can be accommodated in the schedule of the YETS of Run~4, the installation of the services and of the mechanical support system would already start in the second and third YETS of Run~4, \ie in the winters of 2031/32 and 2032/33, to reduce the workload during LS4.
The DAQ, DCS and trigger systems developed in the Slice Test then only need to be extended to the full detector.
The \anubis detector takes data together with the \atlas experiment from the start of Run~5 in 2036.
The sensitivity projections of Section~\ref{sec:physics} assume an integrated luminosity of 3~\invab in line with the conventions in the transverse detector community with the understanding that about 0.5~\invab expected for Run~4 could be missed.

{\em Beyond the HL-LHC.}
The \anubis detector is also a proof of concept for transverse LLP detectors proposed for future $e^+e^-$ colliders, such as LAYCAST~\cite{laycast} and HECATE~\cite{hecate}, which would instrument the caverns or surroundings of the FCC-ee experiments~\cite{Benedikt:2928193}.
The experience gained with the construction and operation of the \anubis detector, including its integration with a general-purpose detector, its trigger and DAQ interface and the operation of large-area RPCs with low-GWP gas mixtures, therefore represents an important stepping stone towards the FCC.
Moreover, should a future FCC-ee detector be hosted in the UX15 cavern, the \anubis detector could be upcycled and continue to take data at the FCC-ee, which is foreseen to start operation in the mid-2040s~\cite{Benedikt:2928193}.

%% file: sections/costing.tex
This section presents a preliminary estimate of the capital expenditure (CapEx) and the operational expenditure (OpEx) of the \anubis detector.
It follows the European costing model: existing positions are not accounted for, and only additional temporary personpower, such as technicians and contractors to be hired for the installation, is included.
All figures are given in 2026 Swiss francs (CHF) and do not include contingency, unless stated otherwise.
The estimate is given for two detector configurations.
The main scenario, which corresponds to the detector design of Section~\ref{sec:design}, uses tracking layers made of RPC doublets with 2~\mm gas gaps.
The reference scenario uses RPC triplets with 1~\mm gas gaps, similar to the RPCs of the \atlas Phase~II upgrade.

{\em Definitions.}
An \emph{RPC singlet} is a single RPC, consisting of a gas gap, two strip panels and the front-end electronics.
An \emph{RPC multiplet} is the set of RPC singlets forming one tracking layer of a tracking module, \ie a doublet in the main and a triplet in the reference scenario, together with its local infrastructure.
A \emph{tracking module} consists of two RPC multiplets, one in each of the two tracking layers.

\subsection{Detector}

The unit costs of an RPC singlet, an RPC multiplet and a tracking module are summarised in Table~\ref{tab:cost_units}.
In the main scenario, an RPC singlet has an area of $1.1\times1.8~\metre^2$, including the overlap with its neighbours, and 272 readout channels, \ie 96 strips in $\phi$ and 176 strips in $\eta$.
In the reference scenario, an RPC singlet has an area of $1.05\times1.05~\metre^2$ and 192 readout channels.
The cost of the gas gaps, the HPL and the strip panels is estimated as 550, 200 and 250~CHF per square metre, respectively.
The readout ASIC is assumed to cost 15~CHF and to serve 16 channels in the main and 8 channels in the reference scenario, and the front-end boards (FEBs) to cost 30 and 20~CHF, respectively.
The local infrastructure of an RPC multiplet comprises its mechanical support, cable and DCS supports, the HV/LV and gas distributors, and the receiver and serialiser that connect the RPC multiplet to the readout.
Overall, a tracking module, consisting of two RPC multiplets, costs about 16~kCHF in both scenarios.

\begin{table}[tb]
  \centering
  \caption{Unit costs of an RPC singlet, an RPC multiplet and a tracking module in the main (2~\mm doublets) and the reference (1~\mm triplets) scenario.}
  \label{tab:cost_units}
  \begin{tabular}{lrr}
    \toprule
    & Main & Reference \\
    \midrule
    \multicolumn{3}{l}{\em RPC singlet} \\
    Area incl.\ overlap [m$^2$] & 1.98 & 1.10 \\
    Number of readout channels & 272 & 192 \\
    ASIC costs [CHF] & 255 & 360 \\
    Front-end board costs [CHF] & 510 & 480 \\
    Strip panel costs (2 panels) [CHF] & 990 & 500 \\
    Gas gap and HPL costs [CHF] & 1\,485 & 750 \\
    RPC singlet total [CHF] & 3\,240 & 2\,090 \\
    \midrule
    \multicolumn{3}{l}{\em RPC multiplet} \\
    Number of RPC singlets & 2 & 3 \\
    RPC singlet costs [CHF] & 6\,480 & 6\,270 \\
    Mechanical support [CHF] & 50 & 50 \\
    Cable and DCS supports [CHF] & 100 & 88 \\
    HV/LV distributors (2) [CHF] & 200 & 200 \\
    Gas distributors (2) [CHF] & 100 & 100 \\
    Receiver and serialiser [CHF] & 1\,200 & 1\,200 \\
    RPC multiplet total [CHF] & 8\,130 & 7\,908 \\
    \midrule
    Tracking module [CHF] & 16\,260 & 15\,816 \\
    \bottomrule
  \end{tabular}
\end{table}

For the cost of the full detector, the 1600~m$^2$ of the cavern ceiling are assumed to be covered completely, which corresponds to 840 tracking modules with 3\,360 RPC singlets and about $0.9\times10^6$ readout channels in the main scenario, and to 1\,600 tracking modules with 9\,600 RPC singlets and about $1.8\times10^6$ readout channels in the reference scenario.
The resulting costs of 13.7 and 25.3~MCHF, respectively, are listed in Table~\ref{tab:cost_capex}.
The main scenario is less expensive because of the larger chambers and the smaller number of gas gaps and readout channels.

\subsection{Services}

The HV and LV power system, including its distribution to the tracking modules, is estimated at 2.5~MCHF, and the optical fibres for the readout to the counting room USA15 at 0.05~MCHF.
The gas system comprises a gas mixing system, estimated at 0.4~MCHF, and the gas distribution from the surface via the PX14 and PX16 shafts to the tracking modules, estimated at 0.5~MCHF.
The total gas volume amounts to about 13~m$^3$ in the main and 10~m$^3$ in the reference scenario.
If a gas mixture free of R134a can be implemented, as foreseen in the baseline design (Section~\ref{sec:design}), no gas recuperation system is required.
Should the mixture contain R134a, at a fraction of up to 15\%, a recuperation plant would be needed, which is estimated at 0.2~MCHF and is conservatively included in the estimate.
For lack of a dedicated estimate, the cost of the services is assumed to be the same in both scenarios, although the larger number of gas gaps in the reference scenario would increase the cost of the HV system.

\subsection{Mechanics}

The cost of the mechanical support system and its installation has been estimated bottom-up for the baseline design of Section~\ref{sec:engineering}, \ie 442 rigs and 46 half-rigs on the vault and 165 rigs in the PX14 and PX16 plugs.
The hardware, which amounts to 2.2~MCHF, comprises the service frame on the vault, the pulley and lifting-cable system, the rig mechanics outside the panels, the support of the services, the shaft plugs, the installation infrastructure and tooling, including a working platform on the bridge crane, as well as prototyping, type tests and certification, and spares.
The installation labour, which amounts to 3.9~MCHF, corresponds to about 3\,300 person-days of technicians, riggers and surveyors, assumed to be hired for the installation, together with the associated crane operation and supervision.
About 60\% of this labour is needed for the installation of the pulley system and the services, and the remainder for the assembly, raising and alignment of the rigs and for the shaft plugs.
The mechanical engineering and design in 2026--2031, estimated at about 0.8~MCHF, is assumed to be provided in kind by the participating institutes and is therefore not included.
The panel-level supports of the RPC multiplets are included in the detector costs above, so that there is no double counting.
The cost of the mechanics scales approximately with the number of rigs and is assumed to be the same in both scenarios.
With a contingency of 30\% on the hardware and 40\% on the labour, the cost of the mechanics would amount to 8.3~MCHF.

\begin{table*}[tb]
  \centering
  \caption{CapEx for the main (2~\mm doublets) and the reference (1~\mm triplets) scenario, without contingency.
  The detector costs assume a complete coverage of the 1600~m$^2$ of the cavern ceiling, and the cost of the mechanics corresponds to the baseline design of Section~\ref{sec:engineering}.
  The gas recuperation plant is included conservatively, although it would only be needed if the gas mixture contains R134a.}
  \label{tab:cost_capex}
  \begin{tabular}{lrr}
    \toprule
    & Main & Reference \\
    \midrule
    \multicolumn{3}{l}{\em Detector} \\
    Number of tracking modules & 840 & 1\,600 \\
    Number of RPC singlets & 3\,360 & 9\,600 \\
    Number of RPC multiplets & 1\,680 & 3\,200 \\
    RPC singlets [kCHF] & 10\,886 & 20\,064 \\
    \quad of which ASICs [kCHF] & 857 & 3\,456 \\
    \quad of which front-end boards [kCHF] & 1\,714 & 4\,608 \\
    \quad of which strip panels [kCHF] & 3\,326 & 4\,800 \\
    \quad of which gas gaps and HPL [kCHF] & 4\,990 & 7\,200 \\
    RPC multiplet infrastructure [kCHF] & 2\,772 & 5\,242 \\
    Detector subtotal [kCHF] & 13\,658 & 25\,306 \\
    \midrule
    \multicolumn{3}{l}{\em Services} \\
    HV and LV power, including distribution [kCHF] & 2\,500 & 2\,500 \\
    Gas mixing system [kCHF] & 400 & 400 \\
    Gas distribution [kCHF] & 500 & 500 \\
    Gas recuperation (only if R134a is used) [kCHF] & 200 & 200 \\
    Readout fibres to USA15 [kCHF] & 50 & 50 \\
    Services subtotal [kCHF] & 3\,650 & 3\,650 \\
    \midrule
    \multicolumn{3}{l}{\em Mechanics} \\
    Vault service frame [kCHF] & 267 & 267 \\
    Pulley and lifting-cable system [kCHF] & 511 & 511 \\
    Rig mechanics outside the panels [kCHF] & 279 & 279 \\
    Support of services [kCHF] & 378 & 378 \\
    Shaft plugs (PX14, PX16) [kCHF] & 335 & 335 \\
    Installation infrastructure and tooling [kCHF] & 205 & 205 \\
    Prototyping, type tests and certification [kCHF] & 140 & 140 \\
    Spares [kCHF] & 117 & 117 \\
    Installation labour: pulley system and services [kCHF] & 2\,389 & 2\,389 \\
    Installation labour: rigs, shaft plugs and other [kCHF] & 1\,478 & 1\,478 \\
    Mechanics subtotal [kCHF] & 6\,098 & 6\,098 \\
    \midrule
    Total CapEx [kCHF] & 23\,406 & 35\,054 \\
    \bottomrule
  \end{tabular}
\end{table*}

\subsection{Total capital expenditure}

The total CapEx amounts to about 23.4~MCHF in the main and 35.1~MCHF in the reference scenario, without contingency (Table~\ref{tab:cost_capex}).
It would be lower by 0.2~MCHF if a gas mixture free of R134a is implemented, so that no gas recuperation plant is needed.
In the main scenario, the detector accounts for about 58\%, the mechanics for about 26\% and the services for about 16\% of the total.
The use of the existing \atlas cavern, of the \atlas detector as shield and active veto, and of the existing infrastructure at Point~1 avoids any civil engineering, which keeps the cost substantially below that of a stand-alone detector of comparable acceptance.

\subsection{Operational expenditure}

The estimated OpEx is summarised in Table~\ref{tab:cost_opex}.
The power consumption of the front-end electronics is estimated at about 100~W per tracking module, \ie 84~kW in the main and 160~kW in the reference scenario, which results in electricity costs of 74 and 140~kCHF per year, respectively, at 0.10~CHF/kWh.
The cost of the fresh gas is estimated at 25~kCHF per year.
This is much lower than the gas cost of the \atlas RPC system of about 200~kCHF per year, which was reduced from about 300~kCHF per year since 2024 by the use of a gas mixture containing 30\% CO$_2$, for two reasons.
First, the \anubis detector will use a gas mixture with a dominant fraction of CO$_2$.
Second, and most importantly, the gas cost of the \atlas RPC system is dominated by leaks of about 1000~l/h, which are not expected for a newly built detector.
Instead, the required rate of fresh gas will be driven by the contamination with air, which can be made negligible by limiting the use of plastic pipes.
For a fresh-gas consumption of about 100~l/h, the gas cost is expected to be below 20~kCHF per year under conservative assumptions, so that the 25~kCHF per year assumed here includes a safety margin.
The maintenance of the detector, including spare parts, is estimated at 10~kCHF per year in the main and 20~kCHF per year in the reference scenario.
The periodic inspection of the lifting cables and latches, and the mating and unmating of the service connections of the shaft plugs at each YETS and long shutdown, add about 35~kCHF per year.
The operation of the detector, including shifts and on-call experts, is provided by the members of the \anubis Collaboration and is therefore not included.
Over ten years of operation, the OpEx amounts to about 1.4~MCHF in the main and 2.2~MCHF in the reference scenario.

\begin{table}[tb]
  \centering
  \caption{OpEx per year and over ten years of operation for the main (2~\mm doublets) and the reference (1~\mm triplets) scenario.
  Electricity is costed at 0.10~CHF/kWh.}
  \label{tab:cost_opex}
  \begin{tabular}{lrr}
    \toprule
    & Main & Reference \\
    \midrule
    Power consumption [kW] & 84 & 160 \\
    Electricity [kCHF/yr] & 74 & 140 \\
    Fresh gas [kCHF/yr] & 25 & 25 \\
    Maintenance and spares [kCHF/yr] & 10 & 20 \\
    Mechanics and plug operations [kCHF/yr] & 35 & 35 \\
    \midrule
    Total per year [kCHF/yr] & 144 & 220 \\
    Total over ten years [kCHF] & 1\,440 & 2\,200 \\
    \bottomrule
  \end{tabular}
\end{table}

%% file: sections/conclusion.tex
The \anubis detector is proposed to close a gap in the experimental coverage of long-lived particles at the HL-LHC: neutral LLPs produced at or above the electroweak scale with decay lengths of $\order{10~\metre}$ and beyond, which escape the general-purpose detectors and cannot be produced at sufficient rates in forward and beam-dump experiments.
It instruments the ceiling of the \atlas cavern UX15 and the bottom of the PX14 and PX16 access shafts with two layers of RPC tracking detectors about 1~\metre apart, each covering about 1600~m$^2$, and uses the air-filled volume between the \atlas detector and the cavern ceiling as decay volume (Section~\ref{sec:location}).
Together with the \atlas detector, it forms a continuous decay volume from the interaction point up to about 23~\metre.

{\em Physics potential.}
Following the PBC framework, the sensitivity has been evaluated for the scalar, neutrino and axion portals with \setanubis, assuming 3~\invab at the HL-LHC (Section~\ref{sec:physics}).
For exotic decays of the Higgs boson into long-lived scalars (BC5), the \anubis detector probes $\br(h\to ss)$ down to $\order{10^{-5}}$ in the nominal and $\order{10^{-6}}$ in the background-free projections, two to three orders of magnitude beyond the latest \atlas and \cms results, and extends the $c\tau$ reach by three to four orders of magnitude.
It surpasses the projections of \codex and \mathusla{}40 for this benchmark.
For heavy neutral leptons coupled to any of the three lepton flavours (BC6--BC8), it reaches mixings of $|V_{1\ell}|^2\approx2\times10^{-8}$ at masses of about 6~GeV, and is the only proposal considered with sensitivity to small couplings above about 5~GeV.
For axion-like particles, it improves on current \atlas limits for photophobic ALPs by two orders of magnitude, and reaches fermion couplings about forty times smaller than the main LHC experiments in the BC10h scenario.
These results demonstrate that the \anubis detector covers a region of parameter space that is complementary to forward, beam-dump and general-purpose experiments.

{\em Backgrounds.}
The dominant backgrounds arise from interactions of neutral SM LLPs, mainly $K_L$ and neutrons, with material (Section~\ref{sec:backgrounds}).
They are strongly suppressed by the air-filled decay volume, by the \atlas detector acting as a passive shield, and by the \atlas event information acting as an active veto.
The backgrounds are estimated in a data-driven way, by rescaling the background of an \atlas search for LLPs decaying in the muon spectrometer.
With these estimates, an exclusion at 95\% CL requires 36 signal events in the conservative and 17 in the nominal background scenario, while only four events are required for signatures with two DVs or with an associated prompt object, which are effectively background-free.
Cosmic rays are rejected by the time-of-flight measurement between the two tracking layers.

{\em Detector technology and design.}
The \proanubis demonstrator recorded 104~\invfb in 2024 and 73~\invfb in 2025 close to the ceiling of the \atlas cavern (Section~\ref{sec:proanubis}).
It showed that the RPC technology meets the requirements of the \anubis detector, with tracking-layer efficiencies above 99\% and a time resolution of about 350~ps per RPC plane, and provided first measurements of the rate of vertices in situ.
The detector design builds on the RPC technology of the \atlas Phase~II upgrade (Section~\ref{sec:design}).
It combines a CO$_2$-rich, low-GWP gas mixture, \ie free of SF$_6$ and of R134a or at least with an R134a fraction of at most 15\%, with a low-noise, sensitive front end, with which a 2~mm gas gap operated with an R134a-free mixture has reached an efficiency of about 98\%.
The trigger study shows that a six-hit coincidence between two RPC doublets can retain about 91--99\% of the signal at an accidental rate below 100~Hz, compatible with the latency envelope of the \atlas L0 trigger (Section~\ref{sec:trigger}).
A baseline mechanical design based on independent arches of standardised rigs on the cavern vault, and on modular shaft plugs made of strip-shaped units that are suspended into the shafts individually, provides a starting point for the engineering of the installation (Section~\ref{sec:engineering}).

{\em Integration with the \atlas experiment.}
The \anubis detector is conceived as an integral extension of the \atlas experiment rather than as a stand-alone experiment (Section~\ref{sec:interfacing}).
It relies on the full \atlas event information to reach its sensitivity, provides a trigger to the \atlas experiment and receives triggers from it, and in turn extends the \atlas muon system by two tracking layers about 23~\metre above the beamline.
The \anubis Collaboration is fully committed to fitting into the existing constraints of the \atlas cavern and the LHC schedule without causing any destructive interference with the \atlas experiment.

\subsection*{Outlook}

The next step is to formalise the \anubis detector as a Phase~III upgrade project of the \atlas experiment (Section~\ref{sec:timeline}).
The project would start with an Initial Design Review in spring 2027, building on this Letter of Intent, followed by a Preliminary Design Review in autumn 2027 and a Final Design Review in early 2030, with the detector R\&D, including the qualification of 2~mm gas gaps with a low-GWP gas mixture, and the development of the readout ASIC and front-end board in between.
A TDR for the detector and electronics would be submitted in summer 2028, and a separate TDR for the mechanics and services in 2031, after the engineering of the cavern installation in consultation with the \atlas TC team, which may start about two thirds into LS3 or only after its end.
The approval of the TDR for the detector and electronics by the \atlas Collaboration Board and, following the review by the LHCC, by the CERN Research Board is targeted for September and December 2028, respectively.
A Slice Test of $\order{10}$ tracking modules would be used to develop the DAQ, DCS and trigger interfaces to the \atlas experiment during Run~4.
It would be installed on the cavern ceiling at the end of LS3 if, and only if, the LS3 schedule allows, and otherwise on the HO-A gantry in the \atlas cavern.
The tracking modules would be produced from 2031 until the end of 2034, installed during LS4, and take data from the start of Run~5 in 2036.
The installation of the services and of the mechanical support system could already start in the YETS of Run~4, if this can be accommodated.

In parallel, the physics programme will be extended to further benchmarks, including the vector portal, LLPs produced in hadron decays with dedicated selections, and decays of heavy BSM states.
The background estimates will be validated with \proanubis data, and a full \geant simulation of the \anubis detector will be integrated into \setanubis.
The qualification of a low-GWP gas mixture, in line with the EU regulations on fluorinated gases, and the development of the front end and readout will be pursued in close collaboration with the \atlas RPC community and the DRD1 Collaboration.
The \anubis Collaboration, which currently comprises 38 authors from 15 institutes, will continue to grow to provide the effort needed for the construction of the detector.

Beyond the HL-LHC, the \anubis detector is a proof of concept for transverse LLP detectors at future $e^+e^-$ colliders, such as LAYCAST~\cite{laycast} and HECATE~\cite{hecate}, and therefore an important stepping stone towards the FCC~\cite{Benedikt:2928193}.
Should a future FCC-ee detector be hosted in the UX15 cavern, the \anubis detector could be upcycled and continue to take data at the FCC-ee.

In summary, the \anubis detector offers a cost-effective way to substantially extend the discovery potential of the HL-LHC for long-lived particles, building on proven technology and on a close integration with the \atlas experiment.